\documentclass[12pt]{article}

\usepackage[a4paper,text={16.8cm,22.4cm}]{geometry}
\usepackage{amsmath,amssymb,bm,psfrag,graphicx,feynarts,color}
\RequirePackage[sort&compress,square,comma,numbers]{natbib}
\allowdisplaybreaks 
\renewcommand{\arraystretch}{1.2}
\newcommand{\pslash}{\rlap{\hspace{0.3mm}/}{p}}
\newcommand{\delslash}{\rlap{\hspace{0.3mm}/}{\partial}}

\begin{document}

\begin{titlepage}

\begin{flushright}
\normalsize
MITP/13-01\\
March 22, 2013\\
Revised: 24 December 2013
% arXiv:1303.5702 (v1: 22 March 2013)
% arXiv:1303.5702 (v2: 24 December 2013)
\end{flushright}

\vspace{0.3cm}
\begin{center}
\Large\bf
5D Perspective on Higgs Production at the Boundary of a Warped Extra Dimension
\end{center}

\vspace{0.8cm}
\begin{center}
Raoul Malm, Matthias Neubert, Kristiane Novotny and Christoph Schmell\\
\vspace{0.7cm}
{\sl PRISMA Cluster of Excellence \& Mainz Institute for Theoretical Physics\\ 
Johannes Gutenberg University, 55099 Mainz, Germany}
\end{center}

\vspace{0.8cm}
\begin{abstract}
A comprehensive, five-dimensional calculation of Higgs-boson production in gluon fusion is performed for both the minimal and the custodially protected Randall-Sundrum (RS) model, with Standard Model fields propagating in the bulk and the scalar sector confined on or near the IR brane. For the first time, an exact expression for the $gg\to h$ amplitude in terms of the five-dimensional fermion propagator is derived, which includes the full dependence on the Higgs-boson mass. Various results in the literature are reconciled and shown to correspond to different incarnations of the RS model, in which the Higgs field is either localized on the IR brane or is described in terms of a narrow bulk state. The results in the two scenarios differ in a qualitative way: the $gg\to h$ amplitude is suppressed in models where the scalar sector is localized on the IR brane, while it tends to be enhanced in bulk Higgs models. In both cases, effects of higher-dimensional operators contributing to the $gg\to h$ amplitude at tree level are shown to be numerically suppressed under reasonable assumptions. There is no smooth cross-over between the two scenarios, since the effective field-theory description breaks down in the transition region. A detailed phenomenological analysis of Higgs production in various RS scenarios is presented, and for each scenario the regions of parameter space already excluded by LHC data are derived.
\end{abstract}
\vfil

\end{titlepage}
\newpage

\section{Introduction}

The discovery of a Higgs-like boson at the LHC \cite{ATLAS:2012gk,CMS:2012gu} marks the beginning of a new era in particle physics. The properties of the new particle appear to be close to those predicted for an elementary scalar with couplings as given by the Standard Model (SM). The hierarchy problem -- the question about the ultra-violet (UV) sensitivity of the scalar sector and the stability of the Higgs potential under quantum fluctuations -- is thus more pressing than ever. In extensions of the SM the scalar sector can be stabilized in various ways. The most popular solution to the hierarchy problem is low-scale supersymmetry, which protects the Higgs-boson mass by linking it to the masses of its fermionic partners. An interesting alternative is provided by models featuring a warped extra dimension \cite{Randall:1999ee}, in which the SM is embedded in a compact extra dimension of anti-de~Sitter space, while the scalar sector is localized on one of two branes bounding the fifth dimension. The fundamental UV cutoff of the model is the warped Planck scale, whose value near this ``infra-red (IR) brane'' lies in the TeV range. These models, introduced by Randall and Sundrum (RS), provide particularly attractive scenarios of TeV-scale new physics, since in addition to the hierarchy problem they also address the flavor puzzle and yield an attractive framework for understanding the hierarchies of fermion masses and mixing angles \cite{Grossman:1999ra,Gherghetta:2000qt,Huber:2000ie} and the smallness of flavor-changing neutral currents \cite{Agashe:2004ay,Agashe:2004cp,Csaki:2008zd,Casagrande:2008hr,Blanke:2008zb,Blanke:2008yr,Bauer:2009cf}.

Precision measurements of the Higgs-boson couplings to SM particles, which are accessible via studies of both the Higgs production cross sections and its decay rates into various final states, present unique opportunities to test the SM description of electroweak symmetry breaking and search for indirect hints of new physics. In the context of warped extra dimensions, Higgs physics has been studied by several authors \cite{Djouadi:2007fm,Falkowski:2007hz,Cacciapaglia:2009ky,Bhattacharyya:2009nb,Bouchart:2009vq,Casagrande:2010si,Azatov:2010pf,Azatov:2011qy,Goertz:2011hj,Carena:2012fk}. The effect on the $gg\to h$ amplitude caused by the heavy $b'$ state, the $SU(2)_R$ partner of the top quark predicted in RS models with custodial symmetry, was investigated in \cite{Djouadi:2007fm}. Models in which the Higgs scalar is a pseudo Nambu-Goldstone boson, such as warped gauge-Higgs unification scenarios, were studied in \cite{Falkowski:2007hz,Azatov:2011qy}. One finds that the result for the $gg\to h$ amplitude only depends on the fundamental parameter $v/f$ of these models, but that it is insensitive to the details about the spectrum of the Kaluza-Klein (KK) quarks. The authors of \cite{Cacciapaglia:2009ky,Bouchart:2009vq} have studied the effect of KK resonances on the loop-induced $hgg$ and $h\gamma\gamma$ couplings by working out the corrections to the top- and bottom-quark Yukawa couplings induced by their mixing with KK states. In these papers no significant contributions from the heavy KK quark states propagating in the loop were observed, because the Yukawa interactions coupling the Higgs to two $Z_2$-odd fermions (the second term in the last line of (\ref{gdef}) below) were implicitly assumed to be zero.\footnote{The fact that there are two towers of KK quark states for every massive SM quark, which is deeply connected to the finiteness of the 5D loop amplitude \cite{Carena:2012fk}, was overlooked in \cite{Bhattacharyya:2009nb}. In order to obtain a finite sum for the infinite KK tower, the authors made the approximation $m_{q_n}=\lambda_{q_n} v/\sqrt2$ with $\lambda_{q_n}\approx 1$ for the masses of the KK quarks, see eqs.~(8) and (10) of their paper, which is incorrect.} 
The possibly large effect on the Higgs-boson couplings induced by the shift of the Higgs vacuum expectation value (vev) relative to its SM value, which can arise in RS models with custodial symmetry, was emphasized in \cite{Bouchart:2009vq}. The first complete calculation of the $hgg$ and $h\gamma\gamma$ couplings, in which both types of Yukawa interactions in (\ref{gdef}) were included, was performed in \cite{Casagrande:2010si}. In this paper both the production of Higgs bosons in the gluon fusion process as well as the main decay channels were studied in an extended RS model with custodial symmetry. It was observed that the dominant corrections to the $hgg$ and $h\gamma\gamma$ couplings arise from the towers of KK quark states propagating in the loop, and that these effects are to a very good approximation independent of the masses of the corresponding SM quarks. The production rate was found to be suppressed in most regions of parameter space, while the branching fraction for the diphoton channel $h\to\gamma\gamma$ tends to be enhanced with respect to the SM. At about the same time, an independent analysis of the Higgs couplings to gluons and photons appeared \cite{Azatov:2010pf}, which reached the opposite conclusions. In a recent paper \cite{Carena:2012fk}, it was shown that the discrepancy between the two sets of results can be traced back to a subtlety in the calculation of the loop-induced Higgs couplings to gluons and photons. In order to compute the relevant overlap integrals of fermion wave functions with the brane-localized Higgs field, it is necessary to regularize the Higgs profile in an intermediate step and give it an infinitesimal width $\eta$ \cite{Azatov:2009na}. When the calculation of the gluon fusion amplitude is performed in a naive way, the limits of sending the regulator to zero ($\eta\to 0$) and including an infinite number of KK modes ($N\to\infty$) in the sum over virtual states do not commute. This ambiguity disappears once the loop calculation is performed in the presence of a consistent UV regulator, such as dimensional regularization with $d<4$ space-time dimensions. For the case of a brane-localized Higgs sector, one then obtains the results of \cite{Casagrande:2010si} no matter in which order the limits are taken. The same conclusion can be reached by using a hard UV momentum cutoff on the four-dimensional (4D) loop integral. The physical significance of the results found in \cite{Azatov:2010pf} was not fully elucidated in \cite{Carena:2012fk}, but the discussion in that paper suggests that they might refer to a certain limit of a model featuring a Higgs boson living in the bulk of the extra dimension. It was demonstrated that the gluon fusion amplitude receives an unsuppressed ``resonance contribution'' from high-mass KK states, which can resolve the wave function of the Higgs boson (see also \cite{Delaunay:2012cz}). This effect is absent for a brane-localized scalar sector.

In the present paper, we shed new light on these issues by performing the calculation of the $gg\to h$ amplitude as a five-dimensional (5D) loop calculation. In this way the very notion of KK states is avoided, the infinite sum over KK states is performed implicitly, and the only relevant limit to be considered is that of sending the regulator $\eta$ of the Higgs profile to zero. In the context of dimensional regularization, we find that this limit can be taken either before or after performing the loop integration. In both cases we confirm the results obtained in \cite{Casagrande:2010si,Carena:2012fk}. If the width of the Higgs profile is kept finite, in a way that will be specified more precisely below, we recover the findings of \cite{Azatov:2010pf}. They correspond to a model with a narrow bulk-Higgs field, whose shape along the extra dimension can be resolved by the high-momentum modes of the RS model. The 5D analysis highlights the relevance of different mass scales. In brane-Higgs models, these are the Higgs vev $v$, the KK mass scale $M_{\rm KK}$, and the physical UV cutoff $\Lambda_{\rm TeV}$ of the RS model near the IR brane. Models in which the Higgs boson is treated as a narrow bulk state contain, in addition, the scale $v/\eta\gg M_{\rm KK}$ (the inverse width of the Higgs profile). It makes an important difference whether this scale lies above or below the cutoff. The relevant loop integrand approaches a first plateau for Euclidean loop momenta $p_E\gg M_{\rm KK}$ and a second one for $p_E\gg v/\eta$ (see Figure~\ref{fig:T1gen} in Section~\ref{sec:analysis}). While in brane-Higgs models the second plateau is absent, in bulk-Higgs scenarios the $gg\to h$ amplitude receives an unsuppressed contribution from the high scale $v/\eta$, and it is thus sensitive to physics on distances shorter than $1/M_{\rm KK}$.

It is worth noting in this context that naive dimensional analysis (NDA) indicates that the $gg\to h$ amplitude is finite for the case of a bulk-Higgs field, but that it is logarithmically divergent by power counting if the Higgs sector is localized on the IR brane. As explained in \cite{Carena:2012fk}, however, systematic cancellations between the Yukawa couplings of the various fermion states within each KK level ensure the finiteness of the result also in the brane-localized Higgs scenario. Analogous cancellations were observed in \cite{Delaunay:2012cz} for the case of loop-induced dipole-operator contributions to flavor-changing processes. The 5D loop calculation performed in the present work confirms this observation and yields convergent results for both scenarios. We do not address the question whether the $gg\to h$ amplitude remains finite at two-loop order and beyond.

Our paper is structured as follows. In Section~\ref{sec:setup}, we define our setup and present some important remarks concerning the classification of the various RS models considered in our study. In Section~\ref{sec:Leff}, we derive an exact representation of the dimensionally-regularized gluon fusion amplitude in terms of an integral over the mixed-chirality components of the 5D quark propagator in the mixed momentum-position representation, including the contributions of the SM quarks and the full dependence on the Higgs-boson mass. To the best of our knowledge, such a result has not been presented before. Our expression holds for an arbitrary Higgs profile. The calculation of the 5D propagator for the case of a very narrow Higgs profile localized near the IR brane is performed in Section~\ref{sec:prop}, with technical details relegated to Appendix~\ref{app:details}. In Section~\ref{sec:analysis}, we use these results to evaluate the $gg\to h$ amplitude and show explicitly that taking the limit $\eta\to 0$ commutes with the integration over the 4D loop momentum. We prove a conjecture made in \cite{Carena:2012fk} for the analytic form of the contribution of the infinite tower of heavy KK quark states. We also present an alternative derivation of the same result by implementing the brane-localized Yukawa terms via appropriate boundary conditions in the field equations for the fermion mass eigenstates. In this approach, the notion of an infinitesimal regulator $\eta$ does not appear, and many of the subtleties related to the $\eta\to 0$ limit are avoided from the beginning. We also consider a generalization of the model in which two different Yukawa matrices enter in the 5D Yukawa interactions. We then discuss the changes that occur when the width of the Higgs profile is kept small but non-zero, corresponding to the case of a narrow bulk-Higgs field. In Section~\ref{sec:powerhgg}, we address the question of the numerical importance of power-suppressed operators, which contribute to the $gg\to h$ amplitude at tree level. They can arise because RS models are effective field theories valid below some cutoff. We argue that even if the UV completion of these models is strongly coupled, the corresponding power corrections are likely to be much smaller than the RS loop effects calculated in Section~\ref{sec:analysis}. While most of our discussion refers to the minimal RS model with the SM gauge group in the bulk, we generalize our results in Section~\ref{sec:custodial} to an extended RS model with a custodial symmetry protecting electroweak precision observables \cite{Agashe:2003zs,Csaki:2003zu,Agashe:2006at}. Contrary to the minimal RS scenario, this model allows for masses of KK excitations that are in reach of the LHC \cite{Carena:2006bn,Cacciapaglia:2006gp,Contino:2006qr,Carena:2007ua}. Phenomenological implications of our findings in the context of recent LHC data are discussed in Section~\ref{sec:pheno}, where we study the corrections to the Higgs-boson production cross section in three different versions of both the minimal and the custodially protected RS model. We illustrate the magnitude of the effects as a function of the mass of the lightest KK gluon state and the scale of the 5D Yukawa matrices, and derive the regions in parameter space that are already excluded by recent LHC measurements. Our main results are summarized in the conclusions. Some technical details of our calculations are collected in four appendices.

\section{Setup and classification of models}
\label{sec:setup}

Our focus in this work is on minimal RS models, in which the electroweak symmetry-breaking sector is localized on or near the IR brane. The extra dimension is taken to be an $S^1/Z_2$ orbifold, labeled by a coordinate $\phi\in[-\pi,\pi]$. Two branes are localized on the orbifold fixed-points $\phi=0$ (UV brane) and $|\phi|=\pi$ (IR brane). The size $r$ and curvature $k$ of the extra dimension are assumed to be of Planck size, $k\sim 1/r\sim M_{\rm Pl}$. The RS metric reads \cite{Randall:1999ee}
\begin{equation}
   ds^2 = e^{-2\sigma(\phi)}\,\eta_{\mu\nu}\,dx^\mu dx^\nu - r^2 d\phi^2
   = \frac{\epsilon^2}{t^2} \left( \eta_{\mu\nu}\,dx^\mu dx^\nu
    - \frac{1}{M_{\rm KK}^2}\,dt^2 \right) ,
\end{equation}
where $e^{-\sigma(\phi)}$ with $\sigma(\phi)=kr|\phi|$ is referred to as the warp factor. The quantity $L=\sigma(\pi)=kr\pi$ measures the size of the extra dimension. In the second equation above we have introduced a new coordinate $t=\epsilon\,e^{\sigma(\phi)}$, where $\epsilon=e^{-\sigma(\pi)}$ determines the hierarchy between the Planck scale and the TeV scale, and $M_{\rm KK}=k\epsilon$ sets the mass scale for the low-lying KK excitations of the SM particles.\footnote{The dimensionless variable $t$ is related to the conformal coordinate $z$ frequently used in the literature by the simple rescaling $z=t/M_{\rm KK}\equiv R'\,t$.} 
Our primary focus is on models where the scalar sector is localized on (or very near) the IR brane at $t=1$, in contrast to more complicated models, in which the Higgs boson is a 5D field propagating in the extended bulk of the extra dimension \cite{Contino:2003ve,Agashe:2004rs,Batell:2008me,Cabrer:2009we,Cabrer:2010si,Archer:2012qa}. While some of these extended models are rather appealing and deserve further investigation, also with regard to their Higgs phenomenology, we believe that the minimal models define important benchmark scenarios which should be explored first. This is not least because only in these cases analytic expressions for the production and decay amplitudes of the Higgs boson can be derived. A more detailed discussion of bulk-Higgs models can be found in Appendix~\ref{app:bulkHiggs}.

Before presenting our results, we find it useful to make a few comments concerning our definition of a brane-localized Higgs sector, which is general enough to allow for a non-zero width of the Higgs profile, as long as it cannot be resolved by the modes of the theory and hence does not affect any observables. Recall that RS models are effective field theories with an inherent, position-dependent UV cutoff given by the warped Planck scale \cite{Randall:2001gb,Pomarol:2000hp,Choi:2002wx,Goldberger:2002cz,Agashe:2002bx}
\begin{equation}\label{LamUVt}
   \Lambda_{\rm UV}(t) \sim M_{\rm Pl}\,e^{-\sigma(\phi)} 
   = M_{\rm Pl}\,\frac{\epsilon}{t} \equiv \frac{\Lambda_{\rm TeV}}{t} \,.
\end{equation}
This accounts for the fact that they do not provide a description of quantum gravity. The variation of the UV cutoff along the extra dimension is a crucial feature in order for RS models to provide a solution to the hierarchy problem. If the sector of electroweak symmetry breaking lives on or near the IR brane at $t=1$, then the effective UV cutoff regularizing quantum corrections to the scalar sector is of order $\Lambda_{\rm TeV}\sim 10\,M_{\rm KK}$. The little hierarchy problem is not addressed by RS models, because the theory must contain several KK modes (and hence the value of $\Lambda_{\rm TeV}$ must be in the multi-TeV range) in order to deserve the attribute ``extra dimensional''. As argued in \cite{Carena:2012fk}, the scale $\Lambda_{\rm TeV}$ also provides the effective UV cutoff in loop graphs involving Higgs bosons. The condition that the fermionic modes in the effective theory cannot resolve the width of the Higgs boson can be stated as
\begin{equation}\label{braneHiggs}
   \hspace{1.0cm} 
   \eta\ll\frac{v|Y_q|}{\Lambda_{\rm TeV}} \quad
   \mbox{(brane-localized Higgs),}
\end{equation}
where $|Y_q|$ sets the scale for the dimensionless, 5D Yukawa couplings of the model. Only if this condition is satisfied, the Higgs field can be regarded as being localized on the IR brane in the sense that any possible extension into the bulk does not give rise to observable effects. As shown in \cite{Carena:2012fk}, another consequence of condition (\ref{braneHiggs}) is that the results for the loop-induced $hgg$ and $h\gamma\gamma$ couplings can be well approximated by performing truncated sums over a small number of KK modes, whose individual Yukawa couplings are evaluated in the limit $\eta\to 0$. Relation (\ref{braneHiggs}) should be considered as a condition on the regulator $\eta$ at fixed, physical UV cutoff $\Lambda_{\rm TeV}$. For a brane-localized Higgs field one should take the limit $\eta\to 0$ wherever possible, but the above condition states that keeping $\eta$ finite but smaller than the bound on the right-hand side would not change the physics. 

A Higgs profile with a width $\eta>v|Y_q|/\Lambda_{\rm TeV}$ must be regarded as a bulk field. The features of the Higgs profile can then be resolved by the high-momentum states in the effective theory, and indeed one finds that high-mass KK fermions make sizable contributions. In the general case, the gluon fusion amplitude in an RS model with a bulk-Higgs field depends in a complicated way on the shapes of the Higgs and fermion profiles along the extra dimension (see \cite{Azatov:2010pf} for an approximate treatment; a more detailed analysis will be presented in \cite{inprep}). However, we find that for a narrow Higgs profile, defined by the relation
\begin{equation}\label{bulkHiggs}
   \hspace{1.0cm} 
   \frac{v|Y_q|}{\Lambda_{\rm TeV}} \ll \eta \ll \frac{v|Y_q|}{M_{\rm KK}} \quad
   \mbox{(narrow bulk Higgs),}
\end{equation}
a model-independent expression can be derived, which generalizes the findings of \cite{Azatov:2010pf}. Our results for the case of a narrow bulk-Higgs scenario are in full agreement with those obtained in \cite{Delaunay:2012cz} for the analogous case of loop-induced dipole-operator contributions to flavor-changing processes. Working under the assumption that the Higgs width is much larger than the inverse cutoff on the IR brane (i.e., $\eta\gg v|Y_q|/\Lambda_{\rm TeV}$), these authors find important contributions from high-mass KK states, which probe the ``bulky'' nature of the Higgs field.

We take an agnostic point of view regarding the question which kind of RS model is theoretically most appealing. The overwhelming majority of the RS literature has been based on models in which the scalar sector is localized on the IR brane. These models should therefore be included as a benchmark in any phenomenological study. Yet, having the Higgs as the only brane-localized field is somewhat peculiar, and after realizing that successful models of electroweak symmetry breaking can be constructed with a scalar sector in the bulk one may consider this to be a more appealing scenario. The fact that important one-loop amplitudes such as $gg\to h$ and $b\to s\gamma$ are convergent by naive power counting in bulk-Higgs models adds to their attractiveness. However, a bulk-Higgs model featuring a very narrow Higgs profile ($\eta\ll 1$) requires some fine-tuning. The most natural assumption would be that $\eta={\cal O}(1)$.

We will see that the results obtained under the two assumptions (\ref{braneHiggs}) and (\ref{bulkHiggs}) are rather different, both qualitatively and quantitatively. Indeed, one should consider the two scenarios as two different, distinguishable realizations of RS models. This fact has also been realized in \cite{Beneke:2012ie}. The situation resembles that encountered when one compares the original RS model, in which only gravity was allowed to propagate in the extra dimension while all SM fields were confined to the IR brane \cite{Randall:1999ee}, with the more popular models in which all matter and gauge fields live in the bulk \cite{Huber:2000ie}. While the original model only addressed the hierarchy problem, the latter models are qualitatively different in that they also provide successful theories of flavor. 

\begin{table}
\small
\begin{center}
\renewcommand{\arraystretch}{1.5}
\begin{tabular}{l|cc|c|c}
\hline\hline
Model & bulk Higgs & narrow bulk Higgs & transition region & brane Higgs \\
\hline
Higgs width & $\eta={\cal O}(1)$ & $\frac{v|Y_q|}{\Lambda_{\rm TeV}}\ll\eta\ll\frac{v|Y_q|}{M_{\rm KK}}$ & $\eta\sim\frac{v|Y_q|}{\Lambda_{\rm TeV}}$ & $\eta\ll\frac{v|Y_q|}{\Lambda_{\rm TeV}}$ \\
Power cors.\ & $\big(\frac{M_{\rm KK}}{\Lambda_{\rm TeV}}\big)^n$ & $\big(\frac{M_{\rm KK}}{\eta\Lambda_{\rm TeV}}\big)^n$ & $\big(\frac{M_{\rm KK}}{v|Y_q|}\big)^n$ & $\big(\frac{M_{\rm KK}}{\Lambda_{\rm TeV}}\big)^n$ \\
 & & $\!\!\frac{M_{\rm KK}}{\Lambda_{\rm TeV}} \frac{M_{\rm KK}}{v|Y_q|}\!\ll\!\frac{M_{\rm KK}}{\eta\Lambda_{\rm TeV}}\!\ll\!\frac{M_{\rm KK}}{v|Y_q|}\!\!$ & &  \\
Higgs profile & resolved by & resolved by & partially resolved by & not resolved \\[-3mm]
 & all modes & high-momentum modes & high-mom.\ modes & \\
${\cal A}(gg\to h)$ & enhanced & enhanced & not calculable & suppressed \\
Result & model-dependent & model-independent & unreliable & model-indep.\ \\
\hline\hline
\end{tabular}
\parbox{15.5cm}
{\vspace{3mm}
\caption{\label{tab:models} 
Comparison of the main features of various versions of RS model (see text for further explanation). The label ``model-independent result'' means that the corrections to the SM prediction for the Higgs production cross section can be calculated (to excellent approximation) without any reference to the Higgs and fermion bulk profiles.}}
\end{center}
\end{table}

While the width of the Higgs profile is a physical parameter, which in principle can be adjusted to take any desired value, the transition from the narrow bulk-Higgs scenario (\ref{bulkHiggs}) to the brane-Higgs scenario (\ref{braneHiggs}) cannot be described in a controlled analytical way. This fact can be understood by investigating the structures of the corresponding effective theories in more detail. Table~\ref{tab:models} summarizes the main features of the various models as defined by the size of the width parameter $\eta$. The second row in the table shows the scaling of power corrections, as represented by higher-dimensional operators in the effective Lagrangian of the RS model. Both in a generic bulk-Higgs model (with $\eta={\cal O}(1)$) and in models where the scalar sector is localized on the IR brane, effects of higher-dimensional operators in Higgs physics are suppressed by powers of the ratio $M_{\rm KK}/\Lambda_{\rm TeV}$, since as explained earlier the warped Planck scale $\Lambda_{\rm TeV}$ is the natural UV cutoff of these theories. The situation changes if one considers bulk-Higgs models, in which the width parameter $\eta$ is parametrically suppressed. Then the effective theory knows about an extra small parameter, and derivatives $\partial_t$ acting on the bulk scalar field can produce powers of $1/\eta$. As a result, there is a class of enhanced power corrections scaling like $\left(M_{\rm KK}/\eta\Lambda_{\rm TeV}\right)^n$. In the transition region between the narrow bulk-Higgs and brane-localized Higgs scenarios, these enhanced power corrections become of ${\cal O}(1)$ or larger, and hence the effective field-theory approach breaks down. In other words, because of the uncontrolled behavior of power-suppressed terms in the transition region, we lack the analytical control over the theory, which would be required to see how the results interpolate from the bulk-Higgs case to the brane-Higgs scenario as one reduces the value of $\eta$. In \cite{Azatov:2010pf}, the authors computed the $hgg$ amplitude in the context of a bulk-Higgs model and took the limit $\eta\to 0$ at the end of their calculation, stating that the answer corresponds to the case of a brane-localized Higgs. As we have just argued, such an approach gives the correct result in the model (\ref{bulkHiggs}), and we thus find it more appropriate to refer to it as a narrow bulk-Higgs scenario.

The above remarks referred to an idealized case, in which the electroweak scale $v|Y_q|$ and the KK mass scale $M_{\rm KK}$ are of comparable magnitude. In practice, due to the lack of KK modes below the TeV scale, there appears to be a little hierarchy between these scales, such that $v|Y_q|/M_{\rm KK}\lesssim 0.3$ or less. Then the power corrections in the transition region are even larger than ${\cal O}(1)$, and also in the narrow bulk-Higgs case the lower bound on $M_{\rm KK}/\eta\Lambda_{\rm TeV}$ cannot be much smaller than 1. In view of this fact, one must consider the results derived in this paper for the narrow bulk-Higgs case with some caution. A more reliable calculation should stay in a regime where $\eta={\cal O}(1)$ \cite{inprep}. This has the disadvantage that the results will depend in a complicated way on the shapes of the Higgs and fermion profiles. If it turns out that this dependence is weak, however, then the results obtained here for the narrow bulk-Higgs scenario might serve as reasonable approximations.

\section{5D analysis of the gluon fusion amplitude}
\label{sec:Leff}

We adopt the same definitions and notation as in the recent work \cite{Carena:2012fk}, in which  the gluon fusion process $gg\to h$ was analyzed in the context of an effective 4D theory, where it is understood as a sum over the contributions from an infinite tower of KK quarks propagating in the loop. Our goal is to repeat the calculation using 5D quark propagators instead, for which we adopt the mixed momentum-position representation \cite{Randall:2001gb,Puchwein:2003jq,Contino:2004vy,Carena:2004zn,Csaki:2010aj} (with $q=u,d$)
\begin{equation}\label{5Dprop}
\begin{aligned}
   i\bm{S}^q(t,t';p) &= \int d^4x\,e^{ip\cdot x}\,\langle\,0|\,T
    \big( {\cal Q}_L(t,x) + {\cal Q}_R(t,x) \big) 
    \big( \bar{\cal Q}_L(t',0) + \bar{\cal Q}_R(t',0) \big) |0\,\rangle \\
   &= \Big[ \bm{\Delta}_{LL}^q(t,t';-p^2)\,\pslash
    + \bm{\Delta}_{RL}^q(t,t';-p^2) \Big]\,P_R + (L\leftrightarrow R) \,,
\end{aligned}
\end{equation}
where $P_{R,L}=\frac12(1\pm\gamma_5)$, and the symbol $T$ denotes time ordering. We begin by considering the minimal RS model with the SM gauge group in the bulk. An extended model with a custodial symmetry will be discussed in Section~\ref{sec:custodial}. The minimal model contains an $SU(2)_L$ doublet quark field $Q(t,x)$ and two $SU(2)_L$ singlet fields $u(t,x)$ and $d(t,x)$ in the 5D Lagrangian, each of which are three-component vectors in generation space. The 5D fermion states can be described by four-component Dirac spinors \cite{Grossman:1999ra,Gherghetta:2000qt}. We use a compact notation, where we collect the left- and right-handed components of the up- and down-type states into six-component vectors ${\cal U}_A=(U_A,u_A)^T$ and ${\cal D}_A=(D_A,d_A)^T$ with $A=L,R$, which are collectively referred to as ${\cal Q}_{L,R}$ in the equation above. The Yukawa interactions of the Higgs boson with up- and down-type quarks are then given by\footnote{To keep the notation transparent, we do not use boldface symbols for unit and zero matrices.}
\begin{equation}\label{Lhqq}
\begin{aligned}
   {\cal L}_{hqq}(x) 
   &= - \sum_{q=u,d}\,\int_\epsilon^1\!dt\,\delta_h^\eta(t-1)\,h(x)\,\bar{\cal Q}_L(t,x)\,
    \frac{1}{\sqrt2}\,\bigg( \begin{array}{cc} 0 & \bm{Y}_q \\ 
                             \bm{Y}_q^\dagger & 0 \end{array} \bigg)\,
    {\cal Q}_R(t,x) + \mbox{h.c.} \\
   &= - \sum_{q=u,d} \sum_{m,n}\,g_{mn}^q\,h(x)\,\bar q_L^{(m)}(x)\,q_R^{(n)}(x)
    + \mbox{h.c.} \,,
\end{aligned}
\end{equation}
where the zeros in the diagonal blocks of the $6\times 6$ Yukawa matrices are required by gauge-invariance. The function $\delta_h^\eta(t-1)$ denotes the normalized Higgs profile along the extra dimension, which we take to be a regularized $\delta$-function (see below). In the second step we have decomposed the 5D fermion spinors into 4D KK modes,
\begin{equation}\label{KKdecomp}
   {\cal Q}_A(t,x) = \sum_n\,\,{\cal Q}_A^{(n)}(t)\,q_A^{(n)}(x) \,; \quad A=L,R \,.
\end{equation}
The superscript $n$ labels the different mass eigenstates in the 4D effective theory, such that $n=1,2,3$ refer to the SM quarks, while $n=4,\dots,9$ label the six fermion modes of the first KK level, and so on. The functions ${\cal Q}_{L,R}^{(n)}(t)$ denote the wave functions of the left- and right-handed components of the $n^{\rm th}$ KK mass eigenstate along the extra dimension. The Yukawa couplings $g_{mn}^q$ are given in terms of the overlap integrals \cite{Casagrande:2010si}
\begin{equation}\label{gdef}
\begin{aligned}
   g_{mn}^u 
   &= \frac{1}{\sqrt2} \int_\epsilon^1\!dt\,\delta_h^\eta(t-1)\,\,{\cal U}_L^{\dagger(m)}(t) 
    \bigg( \begin{array}{cc} 0 & \bm{Y}_u \\ 
           \bm{Y}_u^\dagger & 0 \end{array} \bigg)\,{\cal U}_R^{(n)}(t) \\
   &= \frac{\sqrt 2\pi}{L\epsilon} \int_\epsilon^1\!dt\,\delta_h^\eta(t-1)\,
    \Big[ a_m^{(U)\dagger}\,\bm{C}_m^{(Q)}(t)\,\bm{Y}_u\,\bm{C}_n^{(u)}(t)\,a_n^{(u)}
    + a_m^{(u)\dagger}\,\bm{S}_m^{(u)}(t)\,\bm{Y}_u^\dagger\,\bm{S}_n^{(Q)}(t)\,a_n^{(U)}
    \Big] \,,
\end{aligned}
\end{equation}
and likewise in the down-type quark sector. In the last step we have rewritten the answer in terms of the $Z_2$-even and $Z_2$-odd fermion profiles $\bm{C}_n^{(A)}(t)$ and $\bm{S}_n^{(A)}(t)$ introduced in \cite{Casagrande:2008hr}, which are diagonal $3\times 3$ matrices in generation space. These can be expressed in terms of combinations of Bessel functions, whose rank depends on the bulk mass parameters $\bm{c}_Q=\bm{M}_Q/k$ and $\bm{c}_{u,d}=-\bm{M}_{u,d}/k$ of the 5D fermion fields \cite{Grossman:1999ra,Gherghetta:2000qt}. Without loss of generality, we work in a basis where the $\bm{c}_i$ matrices are diagonal. The $SU(2)_L$ gauge symmetry in the bulk implies that the $SU(2)$-doublet quark fields have common $\bm{c}_Q$ parameters. The 3-component vectors $a_n^{(A)}$, on the other hand, describe the flavor mixings of the 5D interaction eigenstates into the 4D mass eigenstates, which are generated by the Yukawa interactions on the IR brane. Because of electroweak symmetry breaking, these vectors are different for $A=U,D,u,d$. For simplicity, from now on we use the generic notation $Q$ for $U,D$ and $q$ for $u,d$. The $3\times 3$ matrices $\bm{Y}_q$ contain the dimensionless Yukawa couplings of the 5D theory, which are obtained from the dimensionful Yukawa couplings $\bm{Y}_q^{\rm 5D}$ in the original 5D Lagrangian by the rescaling $\bm{Y}_q^{\rm 5D}=2\bm{Y}_q/k$ \cite{Grossman:1999ra,Gherghetta:2000qt} (see also the discussion of Yukawa interactions in Appendix~\ref{app:bulkHiggs}). Contrary to the SM, these matrices are assumed to have an anarchical structure, meaning that they are non-hierarchical matrices with ${\cal O}(1)$ complex elements. The hierarchies of the Yukawa matrices of the SM quarks in the effective 4D theory are explained in terms of a geometrical realization of the Froggatt-Nielsen mechanism in RS models \cite{Csaki:2008zd,Casagrande:2008hr,Blanke:2008zb,Huber:2003tu}. 

\begin{figure}
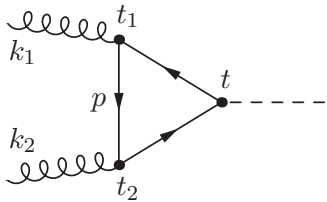

\begin{center}
\vspace{-8mm}
\mbox{\unitlength=0.3bp
\begin{feynartspicture}(432,504)(1,1)
\FADiagram{}
\FAProp(0.,15.)(7.,14.)(0.,){/Cycles}{0}
\FAProp(0.,5.)(7.,6.)(0.,){/Cycles}{0}
\FAProp(20.,10.)(13.5,10.)(0.,){/ScalarDash}{0}
\FAProp(7.,14.)(7.,6.)(0.,){/Straight}{1}
\FAProp(7.,14.)(13.5,10.)(0.,){/Straight}{-1}
\FAProp(7.,6.)(13.5,10.)(0.,){/Straight}{1}
\FALabel(6.2,10.)[r]{\small $p$}
\FALabel(1.8,13.2)[r]{\small $k_1$}
\FALabel(1.8,7.6)[r]{\small $k_2$}
\FALabel(14.1,11.5)[r]{\small $t$}
\FALabel(8.3,15.5)[r]{\small $t_1$}
\FALabel(8.3,4.5)[r]{\small $t_2$}
\FAVert(13.5,10.){0}
\FAVert(7.,14.){0}
\FAVert(7.,6.){0}
\end{feynartspicture}}
\vspace{-4mm}
\parbox{15.5cm}
{\vspace{-14mm}
\caption{\label{fig:graphs} 
Effective $hgg$ couplings induced by the exchange of 5D quark states. The positions of the vertices along the extra dimension are denoted by $t_{1,2}$ and $t$.}}
\end{center}
\end{figure}

The one-loop graph giving rise to the gluon fusion amplitude is shown in Figure~\ref{fig:graphs}, where at each vertex an integral over the fifth coordinate $t=e^{kr(|\phi|-\pi)}$ is implied, which varies between $\epsilon=e^{-kr\pi}\approx 10^{-15}$ on the UV brane and $t=1$ on the IR brane. We summarize the results of the calculation in terms of two coefficients $C_1$ and $C_5$ defined by the decomposition
\begin{equation}\label{Lhgg}
   {\cal A}(gg\to h)
   = C_1\,\frac{\alpha_s}{12\pi v}\,\langle \,0\,| G_{\mu\nu}^a\,G^{\mu\nu,a} |gg\rangle
    - C_5\,\frac{\alpha_s}{8\pi v}\,\langle \,0\,| G_{\mu\nu}^a\,\widetilde G^{\mu\nu,a} 
    |gg\rangle \,,
\end{equation}
where $\widetilde G^{\mu\nu,a}=-\frac12\epsilon^{\mu\nu\alpha\beta}\,G_{\alpha\beta}^a$ (with $\epsilon^{0123}=-1$) denotes the dual field-strength tensor. Note that, contrary to \cite{Carena:2012fk}, the Wilson coefficients $C_1$ and $C_5$ also include the contributions of the SM quarks. Throughout this paper, $v$ denotes the value of the Higgs vev in the RS model, which differs from the SM value $v_{\rm SM}\approx 246$\,GeV by a small amount~\cite{Bouchart:2009vq} (see Section~\ref{sec:pheno}). 

In order to perform the calculation of the gluon fusion amplitude at one-loop order consistently, it is necessary to introduce two different kinds of regulators. For a brane-localized scalar sector, the fermion profile functions are discontinuous on the IR brane, and hence their overlap integrals with a $\delta$-function type Higgs profile are ill defined. Before computing these integrals, it is important to regularize the Higgs profile by giving it a small but finite width $\eta\ll 1$ \cite{Azatov:2009na}. We therefore use the notation $\delta_h^\eta(t-1)$ in (\ref{Lhqq}) and (\ref{gdef}), where the regularized profile has unit area and support in the interval $1-\eta\le t\le 1$. Many of our results will be independent of the shape of the Higgs profile and would remain valid for the case of a general bulk-Higgs field, which we discuss in Appendix~\ref{app:bulkHiggs}. Only at the end of our analysis we will specialize to the case of a very narrow Higgs profile, with $\eta$ satisfying one of the conditions (\ref{braneHiggs}) or (\ref{bulkHiggs}). Note that we use the same Yukawa matrix $\bm{Y}_q$ in the two off-diagonal blocks in (\ref{Lhqq}). For a bulk-Higgs field, the equality of the two Yukawa matrices is a consequence of 5D Lorentz invariance. If the Higgs field is confined to the IR brane this argument no longer applies, and it would in principle be possible to allow for two different Yukawa matrices $\bm{Y}_q^C$ and $\bm{Y}_q^{S\dagger}$ in the two terms in the last line of (\ref{gdef}) \cite{Azatov:2010pf,Azatov:2009na}. This generalization is discussed in Appendix~\ref{app:Y1Y2}, and the corresponding results are summarized in Section~\ref{sec:2Yukawas}.

Secondly, as has been emphasized in \cite{Carena:2012fk}, it is important to introduce a consistent UV regulator in the calculation, even though the final answer for the gluon fusion amplitude is UV finite. This should not come as a surprise, as it is well known that even in the 4D case the introduction of a UV regulator is required in order to obtain a gauge-invariant answer. To see this, consider the loop diagram for a single KK mode, which naively is linearly divergent. Using invariance under $p\to -p$, a superficial logarithmic divergence remains. In dimensional regularization, one encounters the integral
\begin{equation}
   \int\frac{d^dp}{(2\pi)^d} \left[ \frac{4-d}{d}\,\frac{p^2}{\left(p^2-\Delta\right)^3}
    + \frac{\Delta}{\left(p^2-\Delta\right)^3} \right] \varepsilon(k_1)\cdot\varepsilon(k_2) \,,
\end{equation}
which identically vanishes for $d\ne 4$. Here $\Delta=m_{q_n}^2-xy(1-y)m_h^2$ arises after combining denominator using Feynman parameters. Note that if the calculation was performed naively in four dimensions, then only the second term would be present, and it would correspond to a gauge-dependent operator $A_\mu^a\,A^{\mu,a}$. In the 5D model, the UV regulator has the additional effect of regularizing the infinite sum over KK modes, which once again is superficially logarithmically divergent \cite{Carena:2012fk}. The relevant sum is of the form (recall that $n=4$ labels the lightest KK excitation)
\begin{equation}
   \lim_{N\to\infty,~\eta\to 0} \sum_{q=u,d} \sum_{n=4}^{3+6N} 
    \frac{v g_{nn}^q}{m_{q_n}}\,\bigg( \!\frac{\mu}{m_{q_n}}\! \bigg)^{4-d} ,
\end{equation} 
where $m_{q_n}$ are the masses of the KK quarks and $g_{nn}^q$ denote their effective 4D Yukawa couplings as defined in (\ref{gdef}). For $d=4$, one obtains different results depending on which of the two limits is evaluated first. However, in the presence of the dimensional regulator $d<4$ the order of limits becomes irrelevant, and one obtains a unique answer for the sum, which in the limit $d\to 4$ (taken at the end of the calculation) coincides with the result found in \cite{Casagrande:2010si}. Note that regularizing only the ordinary (4D) components of momentum space with a dimensional regulator is justified, since the warp factor and the presence of the branes break 5D Lorentz invariance, and because the integral over the compact interval $t\in[\epsilon,1]$ does not give rise to additional singularities. Introducing a UV cutoff in a way that respects the AdS$_5$ geometry leads to a warped 4D cutoff, as shown in (\ref{LamUVt}). Likewise, the scale $\mu$ of dimensional regularization should be replaced by $\mu_{\rm TeV}$ in the present case. 

With the regulators in place, the gluon fusion amplitude can be written in the form 
\begin{equation}\label{ampl}
\begin{aligned}
   {\cal A}(gg\to h)
   &= i g_s^2\,\delta^{ab} \sum_{q=u,d}\,\int\frac{d^dp}{(2\pi)^d}
    \int_\epsilon^1\!dt_1 \int_\epsilon^1\!dt_2 \int_\epsilon^1\!dt\,\delta_h^\eta(t-1) \\
   &\quad\mbox{}\times\mbox{Tr}\left[ \frac{1}{\sqrt2}\,
    \bigg( \begin{array}{cc} 0 & \bm{Y}_q \\ 
           \bm{Y}_q^\dagger & 0 \end{array} \bigg) \bm{S}^q(t,t_2;p-k_2)\,
    \rlap/\varepsilon(k_2)\,\bm{S}^q(t_2,t_1;p)\,\rlap/\varepsilon(k_1)\,
    \bm{S}^q(t_1,t;p+k_1) \right] ,
\end{aligned}
\end{equation}
where $k_i$ denote the incoming momenta of the external gluons, $a$ and $b$ their color indices, and $\varepsilon(k_i)$ their polarization vectors. We may now insert the decomposition of the 5D propagator given in (\ref{5Dprop}) and try to simplify the result. This task is made complicated by the fact that the propagator functions $\bm{\Delta}_{AB}$ are complicated functions of the 4-momentum $p$ and the coordinates $t$, $t'$. In order to simplify the calculation, it is convenient to use in intermediate steps their representations as sums over KK modes. Using the KK decomposition (\ref{KKdecomp}), it is straightforward to show that
\begin{equation}
\begin{aligned}
   \bm{\Delta}_{LL}^q(t,t';-p^2) 
   &= \sum_n\,\frac{1}{p^2-m_{q_n}^2}\,{\cal Q}_L^{(n)}(t)\,{\cal Q}_L^{(n)\dagger}(t') \,, \\
   \bm{\Delta}_{RL}^q(t,t';-p^2) 
   &= \sum_n\,\frac{m_{q_n}}{p^2-m_{q_n}^2}\,{\cal Q}_R^{(n)}(t)\,{\cal Q}_L^{(n)\dagger}(t') \,, 
\end{aligned}
\end{equation}
and similarly for the other two propagator functions. With the dimensional regulator in place, the 4D loop integral as well as the infinite sums over KK modes converge, and therefore the KK representations provide exact representations of the 5D propagator functions. The integrals over the coordinates $t_1$ and $t_2$ of the two external gluons can then be performed using the orthonormality relations \cite{Casagrande:2008hr}
\begin{equation}\label{ortho}
   \int_\epsilon^1\!dt\,{\cal Q}_A^{(m)\dagger}(t)\,{\cal Q}_A^{(n)}(t) = \delta_{mn} \,; 
   \quad A=L,R \,.
\end{equation}
After this is done, the 5D loop amplitude ${\cal A}$ in (\ref{ampl}) is expressed as a single sum over KK modes, and we find that it can be reduced to integrals of the regularized Higgs profile with traces of the mixed-chirality components of the 5D propagator evaluated at $t=t'$. We define
\begin{equation}\label{TRLdef}
\begin{aligned}
   T_+(p_E^2) 
   &= - \sum_{q=u,d} \frac{v}{\sqrt2} \int_\epsilon^1\!dt\,\delta_h^\eta(t-1)\,
    \mbox{Tr}\left[ \bigg( \begin{array}{cc} 0 & \bm{Y}_q \\ 
     \bm{Y}_q^\dagger & 0 \end{array} \bigg)\,
     \frac{\bm{\Delta}_{RL}^q(t,t;p_E^2)+\bm{\Delta}_{LR}^q(t,t;p_E^2)}{2} \right] , \\
   T_-(p_E^2) 
   &= - \sum_{q=u,d} \frac{v}{\sqrt2} \int_\epsilon^1\!dt\,\delta_h^\eta(t-1)\,
    \mbox{Tr}\left[ \bigg( \begin{array}{cc} 0 & \bm{Y}_q \\ 
     \bm{Y}_q^\dagger & 0 \end{array} \bigg)\,
     \frac{\bm{\Delta}_{RL}^q(t,t;p_E^2)-\bm{\Delta}_{LR}^q(t,t;p_E^2)}{2i} \right] ,
\end{aligned}     
\end{equation}
where $p_E^2\equiv-p^2$ denotes the square of the Euclidean loop momentum after the Wick rotation. Matching the resulting expression for the amplitude ${\cal A}$ with the two-gluon matrix elements in (\ref{Lhgg}), we obtain
\begin{equation}\label{C1C5res}
\begin{aligned}
   C_1 &= \frac32 \int_0^1\!dx \int_0^1\!dy\,\big( 1-4xy\bar y \big)\,I_+(xy\bar y\,m_h^2) 
    = \frac32 \int_0^1\!dz\,(1-z)\,f(z)\,I_+\Big(z\,\frac{m_h^2}{4}\Big) \,, \\
   C_5 &= \int_0^1\!dx \int_0^1\!dy\,\,I_-(xy\bar y\,m_h^2) 
    = \int_0^1\!dz\,f(z)\,I_-\Big(z\,\frac{m_h^2}{4}\Big) \,,
\end{aligned}
\end{equation}
where $m_h$ is the Higgs-boson mass, $x$ and $y$ are Feynman parameters, and we abbreviate $\bar y\equiv 1-y$ and $f(z)=\mbox{arctanh}\sqrt{1-z}$. The quantities
\begin{equation}\label{loopint}
\begin{aligned}
   I_\pm(m^2) 
   &= \frac{e^{\hat\epsilon\gamma_E}\mu^{2\hat\epsilon}}{\Gamma(2-\hat\epsilon)}
    \int_0^\infty\!dp_E^2\,p_E^{2(1-\hat\epsilon)} 
    \bigg( \frac{\partial}{\partial p_E^2} \bigg)^2 T_\pm\big(p_E^2-m^2-i0\big) \\
   &= - \frac{e^{\hat\epsilon\gamma_E}\mu^{2\hat\epsilon}}{\Gamma(1-\hat\epsilon)}
    \int_0^\infty\!dp_E\,p_E^{-2\hat\epsilon} \frac{\partial}{\partial p_E}\,
    T_\pm\big(p_E^2-m^2-i0\big)
\end{aligned}
\end{equation}
are the dimensionally regularized loop-momentum integrals (after Wick rotation) over the functions $T_\pm(p_E^2)$ in (\ref{TRLdef}), shifted by an amount $m^2$. We work in the $\overline{\rm MS}$ scheme with $d=4-2\hat\epsilon$ space-time dimensions. In the last step we have integrated by parts, which is justified as long as the quantity $p_E\,\partial T_\pm/\partial p_E$ vanishes at $p_E=0$ and at $p_E=\infty$. Our analysis in the following section confirms that these conditions are satisfied. 

In \cite{Carena:2012fk}, we have also explored a more intuitive regularization scheme based on using a hard UV momentum cutoff on the loop integral. This can be readily implemented once we have the answers in the form given above. Setting $\hat\epsilon=0$ and restricting the loop momentum to the range $0\le p_E\le\Lambda$, we obtain
\begin{equation}\label{hardLambda}
   I_\pm(m^2) = T_\pm(-m^2-i0) - T_\pm(\Lambda^2-m^2) 
    + \Lambda^2\,\frac{\partial}{\partial\Lambda^2}\,T_\pm(\Lambda^2-m^2) \,,
\end{equation}
where $\Lambda$ should be identified with the physical UV cutoff $\Lambda_{\rm TeV}$ of the RS model. 

The relations (\ref{C1C5res}) are one of our main results. They provide exact expressions for the Wilson coefficients corresponding to the 5D loop integral. The trick of using the KK representation in intermediate steps is legitimate and not different from similar techniques routinely used in 4D loop calculations. Note that in our analysis we have not taken the limit $m_h\to 0$, which is often adopted in discussions of the gluon fusion amplitude and provides a good approximation if the mass of the particle in the loop satisfies the inequality $m_{q_n}^2\gg m_h^2/4$. There would be no problem in using this approximation for the KK excitations, but for the light SM quarks (and to some extent even for the top quark) the Higgs mass must be kept in order to obtain a reliable result. The strategy adopted in \cite{Djouadi:2007fm,Falkowski:2007hz,Cacciapaglia:2009ky,Bhattacharyya:2009nb,Bouchart:2009vq,Casagrande:2010si,Azatov:2010pf,Azatov:2011qy,Goertz:2011hj,Carena:2012fk} was to first evaluate the gluon fusion amplitude in the limit $m_h\to 0$, then to subtract the contributions of the zero modes by hand, and finally to add back the contributions of the top and bottom quarks using the proper loop functions calculated with the physical value of the Higgs mass. Since in a 5D framework there is no distinction between zero modes and KK excitations, we are forced to keep the Higgs mass finite in order to include the SM contributions in the correct way. 

Our results (\ref{TRLdef}) and (\ref{C1C5res}) are valid for an arbitrary Higgs-boson profile along the extra dimension. As long as one succeeds in computing the mixed-chirality components of the 5D propagator in a generic bulk-Higgs model, one can use (\ref{C1C5res}) to compute the corresponding effective $ggh$ couplings. The limit of a brane-localized scalar sector corresponds to taking the limit $\eta\to 0$ in (\ref{TRLdef}). The calculation of the function $\bm{\Delta}_{RL}^q$ in that limit will be presented in the following section. It suffices to focus on one of the mixed-chirality components, since for space-like momenta the two components are related by $\bm{\Delta}_{LR}^q(t,t';p_E^2)=[\bm{\Delta}_{RL}^q(t',t;p_E^2)]^\dagger$.

\section{Calculation of the propagator functions $\bm{\Delta}_{LL}^q$ and $\bm{\Delta}_{RL}^q$}
\label{sec:prop}

We will now derive explicit expressions for the 5D fermion propagator in the mixed momentum-position representation (\ref{5Dprop}). Previous studies of the warped-space 5D fermion propagator have been presented in \cite{Contino:2004vy,Carena:2004zn,Csaki:2010aj}. We generalize these results by keeping for the first time the exact dependence on $v^2/M_{\rm KK}^2$ and the full three-generation flavor structure (see also \cite{Goertz:2011gk}), and by paying special attention to the effects of the regularized profile of the Higgs boson.

The profiles ${\cal Q}_{L,R}^{(n)}(t)$ form complete sets of functions on the interval $t\in[\epsilon,1]$, subject to the orthonormality conditions (\ref{ortho}). In our notation, the Dirac operator takes the form
\begin{equation}
   {\cal D} = \pslash - M_{\rm KK}\,\gamma_5\,\frac{\partial}{\partial t}
    - M_{\rm KK}\,{\cal M}_q(t) \,,
\end{equation}
where
\begin{equation}\label{Mqdef}
   {\cal M}_q(t) 
   = \frac{1}{t}\,\bigg(\! \begin{array}{cc} \bm{c}_Q & \,\,\,0 \\ 
                        0 & -\bm{c}_q \end{array} \!\bigg) 
    + \frac{v}{\sqrt 2 M_{\rm KK}}\,\delta_v^\eta(t-1)\,
    \bigg( \begin{array}{cc} 0 & \bm{Y}_q \\ 
           \bm{Y}_q^\dagger & 0 \end{array} \bigg) 
\end{equation}
is the generalized, hermitian mass matrix \cite{Carena:2012fk}. Here $\delta_v^\eta(t-1)$ denotes the normalized profile of the Higgs vev along the extra dimension. For a brane-localized scalar sector, we may without loss of generality assume that $\delta_v^\eta(t-1)=\delta_h^\eta(t-1)$ are given by the same regularized $\delta$-function. For the general case of a bulk-Higgs field, the two profiles differ, but as described in Appendix~\ref{app:bulkHiggs} these differences vanish in the limit of vanishing $\eta$.

Starting from the definition of the propagator in (\ref{5Dprop}), it is straightforward to show that
\begin{equation}\label{Dirac}
   {\cal D}\,\bm{S}^q(t,t';p) = \delta(t-t') \,,
\end{equation}
where we have used the completeness relations 
\begin{equation}
   \sum_n\,{\cal Q}_A^{(n)}(t)\,{\cal Q}_A^{(n)\dagger}(t') 
   = \delta(t-t') \,; \quad A=L,R
\end{equation}
for the bulk profiles. For the various propagator functions, this generalized Dirac equation implies the coupled system of equations
\begin{equation}\label{coupledeqs}
\begin{aligned}
   p^2 \bm{\Delta}_{LL}^q(t,t';-p^2)
    - M_{\rm KK} \left( \frac{\partial}{\partial t} + {\cal M}_q(t) \right) 
    \bm{\Delta}_{RL}^q(t,t';-p^2)
   &= \delta(t-t') \,, \\
   \bm{\Delta}_{RL}^q(t,t';-p^2)
    - M_{\rm KK} \left( - \frac{\partial}{\partial t} + {\cal M}_q(t) \right) 
    \bm{\Delta}_{LL}^q(t,t';-p^2)
   &= 0 \,, 
\end{aligned}
\end{equation}
and similarly for the other two functions.\footnote{For $p^2=0$, we recover the first-order differential equations for the mixed-chirality components derived in~\cite{Carena:2012fk}, once we identify $\bm{\Delta}_{RL}^q(t,t';0)\equiv -\bm{\Delta}_{RL}^q(t,t')$ and $\bm{\Delta}_{LR}^q(t,t';0)\equiv -\bm{\Delta}_{LR}^q(t,t')$.} 
Integrating these equations over an infinitesimal interval $t\in[t'-0,t'+0]$ at fixed $t'$ yields the jump conditions
\begin{equation}\label{jump}
\begin{aligned}
   \bm{\Delta}_{RL}^q(t'+0,t';-p^2) - \bm{\Delta}_{RL}^q(t'-0,t';-p^2) 
   &= - \frac{1}{M_{\rm KK}} \,, \\
   \bm{\Delta}_{LL}^q(t'+0,t';-p^2) - \bm{\Delta}_{LL}^q(t'-0,t';-p^2) &= 0 \,.
\end{aligned}
\end{equation}
We also need to specify appropriate boundary conditions on the UV and IR branes. In the presence of a regularized Higgs profile, they are
\begin{equation}\label{BCs}
   ( 0 \quad 1 )\,\bm{\Delta}_{LL}^q(t_i,t';-p^2) 
   = ( 1 \quad 0 )\,\bm{\Delta}_{RL}^q(t_i,t';-p^2) = 0 \,; \quad
   \mbox{for~~} t_i = \epsilon, 1 \,.
\end{equation}
This is nothing but the statement that the $Z_2$-odd fermion profiles obey Dirichlet boundary conditions on the two branes.

In order to solve the coupled equations (\ref{coupledeqs}), we first combine them to yield the second-order differential equation
\begin{equation}\label{2ndorder}
   \left[ \frac{\partial^2}{\partial t^2} - {\cal M}_q^2(t)
    - \frac{d{\cal M}_q(t)}{dt} - \hat p_E^2 \right] \bm{\Delta}_{LL}^q(t,t';-p^2)
   = \frac{1}{M_{\rm KK}^2}\,\delta(t-t') \,, 
\end{equation}
where $\hat p_E^2\equiv-p^2/M_{\rm KK}^2$. We then solve this equation assuming that $t\ne t'$, in which case the right-hand side vanishes. Next, we compute the function $\bm{\Delta}_{RL}^q(t,t';-p^2)$ from the second equation in (\ref{coupledeqs}). In the final step we determine the constants of integration by means of the jump conditions (\ref{jump}) and the boundary conditions (\ref{BCs}). The solution of the second-order differential equation involves as integration ``constants'' functions $\bm{C}_i(t')$ with $i=1,\dots,8$, which are $3\times 3$ matrices in generation space and whose values can differ depending on whether $t>t'$ or $t<t'$. In total, we thus have 16 functions $\bm{C}_i^>(t')$ and $\bm{C}_i^<(t')$. The jump conditions impose eight relations among these functions, and the boundary conditions give four conditions each on the UV and IR branes. Solving these relations determines the coefficient functions uniquely.

Up to this point our discussion is completely general and holds for an arbitrary bulk-Higgs field. Unfortunately, it is impossible to obtain a closed form of the solution for the general case of an arbitrary mass matrix ${\cal M}_q(t)$. Only for the special case where $p_E=0$ a formal solution in terms of an ordered exponential can be given \cite{Carena:2012fk}. To proceed, we exploit the fact that the result of the calculation must be regularization independent in the limit $\eta\to 0$. We therefore assume a particularly simple form of the regularized $\delta$-function for the profile of the Higgs vev, for which we take a square box of width $\eta$ and height $1/\eta$:
\begin{equation}\label{simplebox}
   \delta_v^\eta(t-1)\to \frac{1}{\eta}\,\theta(t-1+\eta) \,, \quad 
   \mbox{with~~} \eta \ll \frac{v|Y_q|}{M_{\rm KK}} \,.
\end{equation}
It then follows that for $t<1-\eta$, where the Higgs profile vanishes, we have 
\begin{equation}
   {\cal M}_q^2(t) + \frac{d{\cal M}_q(t)}{dt}
   = \frac{1}{t^2}\,\bigg(\! \begin{array}{cc}
    \bm{c}_Q \left( \bm{c}_Q - 1 \right) & 0 \\ 
    0 & \bm{c}_q \left( \bm{c}_q + 1 \right) \end{array} \!\bigg) \,, 
\end{equation}
while for $t>1-\eta$ we can approximate
\begin{equation}
   {\cal M}_q^2(t) + \frac{d{\cal M}_q(t)}{dt}
   = \frac{v^2}{2 M_{\rm KK}^2\eta^2} \left[ 
    \bigg( \begin{array}{cc} \bm{Y}_q \bm{Y}_q^\dagger & 0 \\ 
    0 & \bm{Y}_q^\dagger \bm{Y}_q \end{array} \bigg) 
    + {\cal O}\bigg( \frac{\eta M_{\rm KK}}{v|Y_q|} \bigg) \right] . 
\end{equation}
The omitted terms are suppressed, relative to the leading one, by at least a factor $\eta$. It will be useful to introduce the abbreviations
\begin{equation}\label{Xqdef}
   \bm{X}_q = \frac{v}{\sqrt2 M_{\rm KK}} \sqrt{\bm{Y}_q\bm{Y}_q^\dagger} \,, \qquad
   \bar{\bm{X}}_q = \frac{v}{\sqrt2 M_{\rm KK}} \sqrt{\bm{Y}_q^\dagger\bm{Y}_q}
\end{equation}
for the positive, hermitian $3\times 3$ matrices entering the leading term, which are given entirely in terms of the 5D anarchic Yukawa matrices. The general solution to (\ref{2ndorder}) in the region $t<1-\eta$ is given in terms of modified Bessel functions $I_\alpha(z)$. It can be written as
\begin{equation}\label{solregion1}
\begin{aligned}
   \bm{\Delta}_{LL}^q(t,t';-p^2)
   &= \hspace{3.5mm} \sqrt{t} \left( \begin{array}{cc}
    I_{\bm{c}_Q-\frac12}(\hat p_E t) & 0 \\ 
    0 & I_{\bm{c}_q+\frac12}(\hat p_E t) \end{array} \right)
    \left( \begin{array}{cc}
    \bm{C}_1(t') & \,\bm{C}_2(t') \\ 
    \bm{C}_3(t') & \,\bm{C}_4(t') \end{array} \right) \\
   &\quad\mbox{}+ \sqrt{t} \left( \begin{array}{cc}
    I_{-\bm{c}_Q+\frac12}(\hat p_E t) & 0 \\ 
    0 & I_{-\bm{c}_q-\frac12}(\hat p_E t) \end{array} \right)
    \left( \begin{array}{cc}
    \bm{C}_5(t') & \,\bm{C}_6(t') \\ 
    \bm{C}_7(t') & \,\bm{C}_8(t') \end{array} \right) .
\end{aligned}
\end{equation}
The general solution in the region $t>1-\eta$ can be expressed through hyperbolic trigonometric functions. It reads
\begin{equation}\label{sol2}
\begin{aligned}
   \bm{\Delta}_{LL}^q(t,t';-p^2)
   &= \hspace{3.5mm} \left( \begin{array}{cc}
    \cosh[\bm{S}_q\,\bar\theta^\eta(t-1)] & 0 \\ 
    0 & \cosh[\bar{\bm{S}}_q\,\bar\theta^\eta(t-1)] \end{array} \right)
    \left( \begin{array}{cc}
    \hat{\bm{C}}_1(t') & \,\hat{\bm{C}}_2(t') \\ 
    \hat{\bm{C}}_3(t') & \,\hat{\bm{C}}_4(t') \end{array} \right) \\
   &\quad\mbox{}+ \left( \begin{array}{cc}
    \sinh[\bm{S}_q\,\bar\theta^\eta(t-1)] & 0 \\ 
    0 & \sinh[\bar{\bm{S}}_q\,\bar\theta^\eta(t-1)] \end{array} \right)
    \left( \begin{array}{cc}
    \hat{\bm{C}}_5(t') & \,\hat{\bm{C}}_6(t') \\ 
    \hat{\bm{C}}_7(t') & \,\hat{\bm{C}}_8(t') \end{array} \right) ,
\end{aligned}
\end{equation}
where the dependence on the coordinate $t$ enters via the integral (for $t\ge 1-\eta$)
\begin{equation}\label{thetadef}
   \bar\theta^\eta(t-1) \equiv \int_t^1\!dt'\,\delta_v^\eta(t'-1)
   = \frac{1-t}{\eta} \,,
\end{equation}
and we have introduced the abbreviations
\begin{equation}\label{Sqdef}
   \bm{S}_q = \sqrt{\bm{X}_q^2+\eta^2\hat p_E^2} \,,
    \qquad
   \bar{\bm{S}}_q = \sqrt{\bar{\bm{X}}_q^2+\eta^2\hat p_E^2} \,.    
\end{equation}
Once again the coefficient functions $\hat{\bm{C}}_i(t')$ can take different values for $t>t'$ and $t<t'$. Requiring that the propagator functions $\bm{\Delta}_{LL}(t,t';-p^2)$ and $\bm{\Delta}_{RL}(t,t';-p^2)$ are continuous at $t=1-\eta$ gives eight conditions, which allow us to relate the coefficients $\hat{\bm{C}}_i(t')$ to $\bm{C}_i(t')$.

In working out the solutions we neglect the infinitesimal regularization parameter $\eta$ wherever possible, with two exceptions: First, like the profile of the Higgs vev itself, the $\bar\theta^\eta(t-1)$ functions vary rapidly over the interval $1-\eta\le t\le 1$, and hence $\eta$ appears in (\ref{thetadef}) in an essential way. Second, inside the quantities $\bm{S}_q$ and $\bar{\bm{S}_q}$ the regulator appears in the product $\eta\hat p_E$, and since in (\ref{loopint}) we integrate over all values of the loop momentum there might in principle be contributions from very large momenta, for which $\eta^2\hat p_E^2$ is comparable to the entries of $\bm{X}_q^2$ or larger. For the case of a brane-localized Higgs boson as defined in (\ref{braneHiggs}), such contributions are unphysical in view of the inherent UV cutoff of RS models, and we might therefore simply exclude them by hand. However, we find it more instructive to show their decoupling explicitly in the context of dimensional regularization.

Further details of the solution for the coefficient functions are described in Appendix~\ref{app:details}. In the following section we report our final expressions for the quantities $T_\pm(p_E^2)$ defined in (\ref{TRLdef}). The dependence on the Euclidean 4-momentum enters our results via the quantities $\bm{S}_q$ and $\bar{\bm{S}}_q$ introduced in (\ref{Sqdef}) and via the ratio of certain linear combinations of modified Bessel functions, which we define as
\begin{equation}\label{RAdef}
   \bm{R}_A(\hat p_E) 
   = \frac{I_{-\bm{c}_A-\frac12}(\epsilon\hat p_E)\,I_{\bm{c}_A-\frac12}(\hat p_E)
           - I_{\bm{c}_A+\frac12}(\epsilon\hat p_E)\,I_{-\bm{c}_A+\frac12}(\hat p_E)}%
          {I_{-\bm{c}_A-\frac12}(\epsilon\hat p_E)\,I_{\bm{c}_A+\frac12}(\hat p_E)
           - I_{\bm{c}_A+\frac12}(\epsilon\hat p_E)\,I_{-\bm{c}_A-\frac12}(\hat p_E)} \,;
    \quad A=Q,q \,.
\end{equation}
These quantities are diagonal matrices in generation space. A significant complication originates from the fact that they do not commute with the matrices $\bm{S}_q$ and $\bar{\bm{S}}_q$, giving rise to non-trivial matrix products. It will be important for our discussion to exploit the asymptotic behavior of the ratio $\bm{R}_A$ for large and small values of $\hat p_E$. Using the well-known properties of the modified Bessel functions $I_\alpha(z)$, we find that for $\mbox{Re}\,\hat p_E\gg 1$ 
\begin{equation}\label{RAlarge}
   \bm{R}_A(\hat p_E) 
   = 1 + \frac{\bm{c}_A}{\hat p_E} + \frac{\bm{c}_A\,(1+\bm{c}_A)}{2\hat p_E^2}
    + {\cal O}(\hat p_E^{-3}) \,,
\end{equation}
up to exponentially small ${\cal O}(e^{-2\hat p_E})$ terms. The asymptotic behavior for small values of $\hat p_E$ is
\begin{equation}\label{RAsmall}
   \bm{R}_A(\hat p_E) 
   = \frac{F^2(\bm{c}_A)}{\hat p_E} + \frac{\hat p_E}{1-2\bm{c}_A} \left[
    1 - F^2(\bm{c}_A) + \frac{F^4(\bm{c}_A)}{3+2\bm{c}_A} \right]
    + {\cal O}(\hat p_E^3) \,,
\end{equation}
where 
\begin{equation}\label{Fdef}
   F^2(c) = \frac{1+2c}{1-\epsilon^{1+2c}}
\end{equation}
denotes the squared value of the profile of a chiral component of a SM fermion on the IR brane~\cite{Grossman:1999ra,Gherghetta:2000qt}.

\section{Analysis of the loop amplitude}
\label{sec:analysis}

We now show how to calculate the loop integrals $I_\pm(m^2)$ in (\ref{loopint}) for the cases of a brane-localized Higgs boson and a narrow bulk-Higgs field, as defined in (\ref{braneHiggs}) and (\ref{bulkHiggs}). We perform the calculation in dimensional regularization, but we first motivate the results in the context of the more intuitive scheme in which a hard UV cutoff is used. We begin by collecting some general properties of the functions $T_\pm(p_E^2)$ defined in (\ref{TRLdef}), which are derived from the general solution to the differential equations discussed in the previous section and in Appendix~\ref{app:details}.

\subsection{\boldmath Properties of the functions $T_\pm(p_E^2)$}

In the region of small momenta ($|p_E|\ll M_{\rm KK}$), the functions $T_\pm(p_E^2)$ vary rapidly and in a way that is strongly dependent on the values of the bulk mass parameters $c_i$. This is expected, because in this momentum range their behavior is dominated by the contributions of the SM quarks. Remarkably, we find that at the special value $p_E=0$ the results are given by the very simple expressions
\begin{equation}\label{region1}
   T_+(0) = \sum_{q=u,d} \mbox{Tr}\,\big[ \bm{X}_q\coth\bm{X}_q \big] \equiv t_0 \,, \qquad
   T_-(0) = 0 \,,
\end{equation}
which only depend on the 5D Yukawa couplings, via the quantity $\bm{X}_q$ defined in (\ref{Xqdef}). In the neighborhood of this point the behavior is complicated and not described by a simple formula. For larger values of the Euclidean momentum, such that $p_E\gg M_{\rm KK}$, the function $T_+(p_E^2)$ converges towards a universal limiting value
\begin{eqnarray}\label{region2}
   T_+(p_E^2) 
   &=& \sum_{q=u,d} \mbox{Tr} \left\{ \bm{X}_q\tanh2\bm{X}_q 
    + \frac{1}{2\hat p_E} \left[ \frac{\bm{c}_Q\,\bm{X}_q\tanh2\bm{X}_q}{\cosh2\bm{X}_q}
    + \frac{\bm{c}_q\,\bar{\bm{X}}_q\tanh2\bar{\bm{X}}_q}{\cosh2\bar{\bm{X}}_q} \right] 
    + {\cal O}(\hat p_E^{-2}) \right\} \nonumber\\
   &\equiv& t_1 + \frac{t_2}{\hat p_E} + \dots \,, \qquad
    (M_{\rm KK}\ll p_E\ll v|Y_q|/\eta)
\end{eqnarray}
while $T_-(p_E^2)={\cal O}(\hat p_E^{-2})$ falls off more rapidly. To derive this result, we have taken the limit $\eta\hat p_E\to 0$ and used the asymptotic expansion in (\ref{RAlarge}). A dependence on the bulk mass parameters enters only at subleading order. Interestingly, there exists a third region of extremely large Euclidean momentum, $p_E\gg v|Y_q|/\eta$, for which the behavior changes once again, and the function $T_+(p_E^2)$ tends to zero according to 
\begin{equation}\label{region3}
   T_+(p_E^2) 
   = \frac{1}{\eta\hat p_E} \sum_{q=u,d} \mbox{Tr}\,\bm{X}_q^2 + {\cal O}(\hat p_E^{-2}) 
   \equiv \frac{t_3}{\eta\hat p_E} + \dots \,, \qquad
    (p_E\gg v|Y_q|/\eta)
\end{equation}
while still $T_-(p_E^2)={\cal O}(\hat p_E^{-2})$.
Note that in this region the loop momentum $p_E$ exceeds the value of the intrinsic UV cutoff of a consistent RS model with a brane-localized Higgs sector, because condition (\ref{braneHiggs}) implies $\Lambda_{\rm TeV}\ll v|Y_q|/\eta$. It can therefore only contribute if we consider a bulk-Higgs field as defined in (\ref{bulkHiggs}).

\begin{figure}
\begin{center}
\psfrag{x}[]{$p_E/M_{\rm KK}$}
\psfrag{y}[b]{$T_+^{\rm \,1\,gen}(p_E^2)$}
\psfrag{a}[r]{\footnotesize $\eta=0.1$~}
\psfrag{b}[r]{\footnotesize $\eta=0.005$}
\psfrag{c}[bl]{\footnotesize $\eta\to0$}
\includegraphics[width=0.6\textwidth]{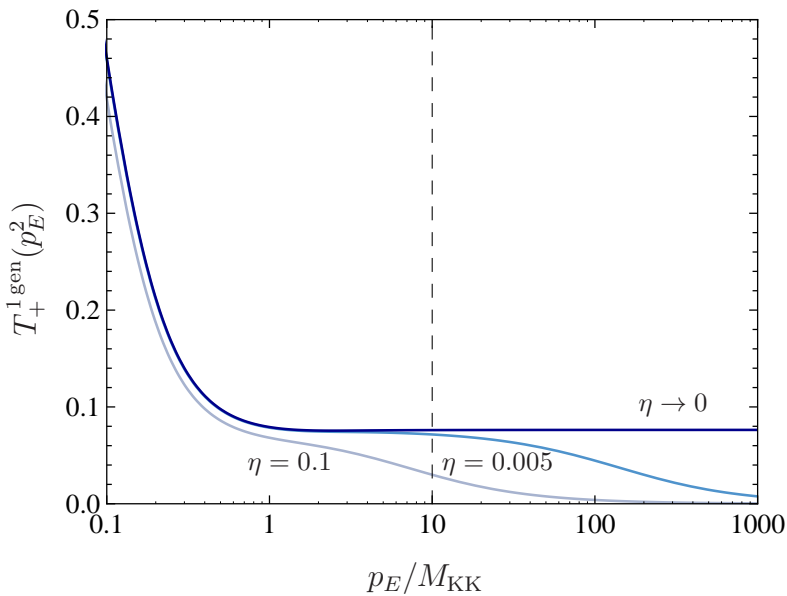}
\parbox{15.5cm}
{\vspace{2mm}
\caption{\label{fig:T1gen}
Momentum dependence of the propagator function $T_+(p_E^2)$ for the case of one fermion generation and parameters corresponding to the top quark. The three curves refer to different values of the regulator $\eta$, as indicated. The vertical dashed line indicates the value of the UV cutoff of the RS model (for $\Lambda_{\rm TeV}=10\,M_{\rm KK}$).}}
\end{center}
\end{figure}

It follows from this discussion that the functions $T_\pm(p_E^2)$ have all the properties required for the integration by parts in (\ref{loopint}). The exact momentum dependence of these functions is rather complicated, and we refrain from giving explicit expressions for the general case. We will instead discuss the simpler case of a single fermion generation, which exhibits all the relevant features mentioned above. In this case we have obtained the analytic expression
\begin{equation}\label{T1res}
   T_+^{\rm \,1\,gen}(p_E^2) = \sum_{q=u,d} \frac{X_q^2}{S_q}\,
    \frac{k_1(\hat p_E)\,S_q\sinh2S_q + k_2(\hat p_E)\,\eta\hat p_E
          \left( \cosh2S_q - \frac{\sinh2S_q}{2S_q} \right)}%
         {k_1(\hat p_E)\,S_q\,(\cosh2S_q-1) + k_2(\hat p_E)\,\eta\hat p_E \sinh2S_q + 2 S_q} \,,
\end{equation}
where $S_q$ has been defined in (\ref{Sqdef}), and
\begin{equation}
   k_1(\hat p_E) = 1 + R_q(\hat p_E)\,R_Q(\hat p_E) \,, \qquad
   k_2(\hat p_E) = R_q(\hat p_E) + R_Q(\hat p_E) \,.
\end{equation}
The function $T_-^{\rm \,1\,gen}(p_E^2)=0$ vanishes trivially. It is a simple exercise to derive from (\ref{T1res}) the various limiting behaviors shown in (\ref{region1})\,--\,(\ref{region3}), simplified to the one-generation case. Figure~\ref{fig:T1gen} shows the behavior of the result (\ref{T1res}) for the parameter choices $c_Q=-0.45$, $c_q=0.395$, and $|Y_q|=2.3$, which correspond to the physical mass $m_q=172.6$\,GeV of the top quark. We set the KK scale to $M_{\rm KK}=2$\,TeV, such that $X_q\approx 0.2$. The three curves correspond to different values of the regulator $\eta$. The three regions of Euclidean momenta mentioned above ($p_E/M_{\rm KK}\sim 1$, $p_E/M_{\rm KK}\gg 1$, and $p_E/M_{\rm KK}\gg X_q/\eta$) are clearly visible from the plot. The dark and light blue curves correspond to models for which $\Lambda_{\rm TeV}/M_{\rm KK}\ll X_q/\eta$, and hence condition (\ref{braneHiggs}) defining a brane-localized Higgs field holds. The gray curve corresponds to the case of a narrow bulk Higgs, as defined in (\ref{bulkHiggs}).

\subsection{\boldmath Analysis of the loop integrals $I_\pm(m^2)$}

Our final goal is to calculate the loop integrals $I_\pm(m^2)$ defined in (\ref{loopint}) in the dimensional regularization scheme. For simplicity, however, we first consider the integral $I_+(0)$ at the special point $m^2=0$ and work with a hard momentum cutoff $\Lambda=\Lambda_{\rm TeV}$. For the case of a brane-localized Higgs sector, defined according to condition (\ref{braneHiggs}), we obtain from (\ref{hardLambda})
\begin{equation}\label{resbrane}
   I_+(0) \big|_{\rm brane\;Higgs}
   = t_0 - t_1 - \frac{3t_2}{2}\,\frac{M_{\rm KK}}{\Lambda_{\rm TeV}} + \dots \,,
\end{equation}
with $t_0$ and $t_{1,2}$ as defined in (\ref{region1}) and (\ref{region2}), respectively. The last term is a small threshold correction (suppressed by the UV cutoff, which we assume to be much larger than the KK mass scale), which is present in a hard-cutoff scheme but will not be visible in the dimensional regularization scheme discussed below. Such power-suppressed terms can be included via higher-dimensional operators in the effective Lagrangian of the RS model. Their suppression $\sim M_{\rm KK}/\Lambda_{\rm TeV}$ is in accordance with Table~\ref{tab:models}.

The difference $(t_0-t_1)$ coincides with the expression for the quantity $\Sigma_q^{\rm (CGHNP)}$ (summed over $q=u,d$) derived in \cite{Carena:2012fk} for the case of a brane-localized Higgs sector. It corresponds to the numerical result first derived in \cite{Casagrande:2010si}. The same result would be obtained if one would take the limit $\eta\to 0$ before performing the integral over the loop momentum. For the opposite case of a narrow bulk-Higgs field, defined according to condition (\ref{bulkHiggs}), the UV cutoff is such that the quantity $T_+(\Lambda^2)$ in (\ref{hardLambda}) must be evaluated using (\ref{region3}), so that we obtain
\begin{equation}\label{resbulk}
   I_+(0) \big|_{\rm narrow\;bulk\;Higgs}
   = t_0 - \frac{3t_3}{2}\,\frac{M_{\rm KK}}{\eta\Lambda_{\rm TeV}} + \dots
\end{equation}
instead of (\ref{resbrane}). The two answers differ by an amount $t_1$ given by the first term on the right-hand side in (\ref{region2}). The term $t_0$ coincides with the expression for the quantity $\Sigma_q^{\rm (ATZ)}$ (summed over $q=u,d$) derived in \cite{Carena:2012fk}, which corresponds to the result first obtained in \cite{Azatov:2010pf}. We emphasize that the threshold corrections are enhanced by a factor $1/\eta$ in this case, which provides an example of the general behavior anticipated in Table~\ref{tab:models} for the case of a narrow bulk-Higgs field. We will comment more on the structure of power corrections and the role of higher-dimensional operators in Sections~\ref{sec:powercors} and \ref{sec:powerhgg}.

It is instructive to reproduce the above results in the less intuitive, but more consistent (from a mathematical point of view) dimensional regularization scheme. We will argue that also in this case the limit of a brane-localized Higgs sector can be taken without encountering any ambiguities. In order to demonstrate this, we should perform the integrals over $p_E$ in (\ref{loopint}) and then take the limit $\eta\to 0$, and show that this yields the same answer as first setting $\eta\to 0$ and then integrating over the loop momentum. However, our explicit result in (\ref{T1res}) and its generalization to three generations are so complicated that the dimensionally regularized integral cannot be evaluated in closed form. We will instead consider a toy model, which captures all important features of the exact result. To this end, we study the function
\begin{equation}\label{Tmodel}
   T_+^{\rm model}(p_E^2) = \frac{t_0-t_1-t_2}{1+\hat p_E^2}
    + \frac{t_2}{\sqrt{1+\hat p_E^2}} + \frac{t_3}{\sqrt{(t_3/t_1)^2+(\eta\hat p_E)^2}} \,,
\end{equation}
which exhibits the same asymptotic behavior in the three regions as the exact result. Evaluating the integrals in (\ref{loopint}) for this function, we obtain
\begin{equation}\label{resdimreg}
   I_+^{\rm model}(0) 
   = (t_0-t_1-t_2) \left(\! \frac{\mu}{M_{\rm KK}} \!\right)^{2\hat\epsilon}
    + t_2 \left(\! \frac{\mu}{2M_{\rm KK}} \!\right)^{2\hat\epsilon}
    + t_1 \left(\! \frac{t_1}{2t_3} \!\right)^{2\hat\epsilon} 
     \left(\! \frac{\mu\eta}{M_{\rm KK}} \!\right)^{2\hat\epsilon} 
    + {\cal O}(\hat\epsilon^2) \,,
\end{equation}
where $t_1/(2t_3)=1+{\cal O}(v^2/M_{\rm KK}^2)$. While the first two contributions are associated with the scale $M_{\rm KK}$, i.e.\ with low-lying KK modes, the third contribution is associated with the super-heavy scale $M_{\rm KK}/\eta$, which for a brane-localized Higgs sector is larger than the physical UV cutoff of the RS model. Note that in the limit $\eta\to 0$ this contribution tends to zero, leaving $I_+^{\rm model}(0)=(t_0-t_1)$ as the final result for the integral after the UV regulator $\hat\epsilon$ has been removed, in accordance with (\ref{resbrane}). The same result is obtained if the limit $\eta\to 0$ is taken in (\ref{Tmodel}) before the integral is evaluated. The last term in (\ref{Tmodel}) then reduces to a constant, which does not contribute to (\ref{loopint}). In the dimensional regularization scheme, the case of a narrow bulk Higgs, for which the loop momenta can resolve the shape of the Higgs profile, is obtained by removing the UV regulator $\hat\epsilon$ at small but finite value of $\eta$. In this case one finds $I_+^{\rm model}(0)=t_0$, in accordance with (\ref{resbulk}).

\subsection{Power corrections and higher-dimensional operators}
\label{sec:powercors}

Let us add some comments concerning the size of generic power corrections, which can be described in terms of higher-dimensional operators added to the Lagrangian of the RS model (with unknown coefficients). For example, what should one expect for the magnitude of the leading power corrections to the Yukawa interactions (\ref{Lhqq}) coupling the Higgs boson to bulk fermions? In general, higher-dimensional operators can be constructed by inserting one or more (covariant) derivatives acting on the fields.\footnote{Note that the 5D Lagrangian does not contain any small mass parameters, which could be used to construct non-derivative operators of higher dimension.} 
These operators are suppressed by the fundamental, physical UV cutoff of RS models, which is of order the Planck scale. The leading such operators involving a fermion bilinear contain a single derivative, possibly accompanied by a factor $\mbox{sgn}(\phi)$. We are thus led to study the object
\begin{equation}
   \frac{1}{M_{\rm Pl}}\,E_a^A\,iD_A\gamma^a
   = \frac{1}{M_{\rm Pl}} \left( e^{\sigma(\phi)}\,i\delslash
    - \frac{1}{r}\,\gamma_5\,\partial_\phi \right)
    + \mbox{terms containing gauge fields,}
\end{equation}
where $\gamma^a=\{\gamma^\mu,i\gamma_5\}$ are the 5D Dirac matrices and $E_a^A$ denotes the vielbein \cite{Grossman:1999ra,Gherghetta:2000qt}. From now on we focus on the derivative terms only. Changing variables from $\phi$ to $t$, and using the definition of the warped UV cutoff in (\ref{LamUVt}), we obtain
\begin{equation}\label{onederiv}
   \frac{1}{M_{\rm Pl}}\,E_a^A\,iD_A\gamma^a
   = \frac{1}{\Lambda_{\rm UV}(t)} 
    \left( i\delslash - \gamma_5\,M_{\rm KK}\,\partial_t \right) + \dots \,.
\end{equation}
Operators containing more than one derivative contain similar structures. For example, the 5D d'Alembertian can be written as 
\begin{equation}\label{box5}
   \frac{1}{M_{\rm Pl}^2}\,\Box_5
   = \frac{e^{2\sigma(\phi)}}{M_{\rm Pl}^2}
    \left( \Box_4 - \frac{e^{-2\sigma(\phi)}}{r^2}\,\partial_\phi^2 \right)
   = \frac{1}{\Lambda_{\rm UV}^2(t)} 
    \left( \Box_4 - M_{\rm KK}^2\,\frac{1}{t}\,\partial_t\,t\,\partial_t \right) .
\end{equation}

Several comments are in order. First, we note that higher-derivative operators in the effective Lagrangian are indeed suppressed by the position-dependent UV cutoff $\Lambda_{\rm UV}(t)$, as stated in the Introduction. If we consider power corrections to couplings involving the Higgs boson (no matter whether the Higgs field is localized on or near the IR brane), the corresponding cutoff scale is $\Lambda_{\rm TeV}$. The 4D derivatives contained in (\ref{onederiv}) and (\ref{box5}) will produce powers of external momenta or masses of the various fermion modes. The corresponding terms scale like $\left(M_{\rm KK}/\Lambda_{\rm TeV}\right)^n$. For models in which the Higgs field is a generic bulk scalar (with width $\eta\sim 1$) or a brane-localized field, derivatives $\partial_t$ acting on the fields near $t=1$ produce ${\cal O}(1)$ factors, since the wave functions are naturally expressed in terms of the $t$ variable, typically involving Bessel functions of argument $x_n t$ with $x_n=m_n/M_{\rm KK}$, or powers of $t$ in the case of the SM fermions. (For a brane-localized Higgs field, these derivatives must be evaluated at $t=1^-$, i.e., by approaching the IR brane from the left.) Hence, the $\partial_t$ terms in the derivative operators shown above also give rise to $\left(M_{\rm KK}/\Lambda_{\rm TeV}\right)^n$ corrections. The situation changes if we consider a limit of a bulk-Higgs model in which the width $\eta$ of the Higgs profile becomes parametrically suppressed. Then the Higgs profile itself, as well as the profiles of particles coupling to the Higgs field, change rapidly over a small interval of width $\eta$ near the IR brane. In such a scenario, a derivative $\partial_t$ acting on the Higgs field or any field coupling to the Higgs boson picks up a factor $1/\eta$, and hence the corresponding power corrections scale like $\left(M_{\rm KK}/\eta\Lambda_{\rm TeV}\right)^n$. We thus confirm the scaling of power corrections anticipated in Table~\ref{tab:models}.

\subsection{Final expressions for the loop integrals}

The above discussion shows that in the presence of the UV regulator, and for a brane-localized Higgs boson, it is possible to take the limit $\eta\to 0$ at the level of the functions $T_\pm(p_E^2)$, before the loop integral is performed. Using the results from Appendix~\ref{app:details}, we extended the form (\ref{T1res}) valid for one fermion generation to the general case of more than one fermion generations. For $\eta\to 0$, we find
\begin{equation}\label{T3gen}
\begin{aligned}
   T_+(p_E^2) 
   &= \sum_{q=u,d} \mbox{Tr} \left\{ \frac{2\bm{X}_q}{\sinh2\bm{X}_q} \left[ 
    \sinh^2\!\bm{X}_q + \frac12 \bigg( \frac{\bm{Z}_q(p_E^2)}{1 + \bm{Z}_q(p_E^2)}
    + \frac{\bm{Z}_q^\dagger(p_E^2)}{1 + \bm{Z}_q^\dagger(p_E^2)} \bigg) \right] 
    \right\} , \\
   T_-(p_E^2) 
   &= \sum_{q=u,d} \mbox{Tr} \left\{ \frac{2\bm{X}_q}{\sinh2\bm{X}_q} \left[ 
    \frac{1}{2i} \bigg( \frac{\bm{Z}_q(p_E^2)}{1 + \bm{Z}_q(p_E^2)}
    - \frac{\bm{Z}_q^\dagger(p_E^2)}{1 + \bm{Z}_q^\dagger(p_E^2)} \bigg) \right] 
    \right\} ,
\end{aligned}
\end{equation}
where the quantity
\begin{equation}\label{Zdef}
   \bm{Z}_q(p_E^2) 
   = \frac{v^2}{2M_{\rm KK}^2}\,\frac{\tanh\bm{X}_q}{\bm{X}_q}\,\bm{Y}_q\,
    \bm{R}_q(\hat p_E)\,\bm{Y}_q^\dagger\,
    \frac{\tanh\bm{X}_q}{\bm{X}_q}\,\bm{R}_Q(\hat p_E)
\end{equation}
involves a non-trivial product of matrix-valued functions. Note that we have removed any reference to the matrices $\bar{\bm X}_q$ in the final expressions by using the identities $\bm{Y}_q\,f(\bar{\bm X}_q)=f(\bm{X}_q)\,\bm{Y}_q$ and $f(\bar{\bm X}_q)\,\bm{Y}_q^\dagger=\bm{Y}_q^\dagger\,f(\bm{X}_q)$, which hold for an arbitrary function $f(\bm{X}_q)$ that has a non-singular expansion in powers of $\bm{X}_q^2$.

We are now ready to derive the final expressions for the loop integrals in (\ref{loopint}). 
 The quantities $T_\pm(-m^2-i0)$ computed using (\ref{T3gen}) replace the quantity $t_0$ in (\ref{resbrane}), (\ref{resbulk}), and (\ref{resdimreg}), while $t_1$ has already been given in (\ref{region2}). Removing the UV regulator after the integral over the loop momentum has been performed, we obtain
\begin{equation}\label{IplImi}
\begin{aligned}
   I_+(m^2) 
   &= \sum_{q=u,d} \left\{ \mbox{Tr}\,g(\bm{X}_q) 
    + \frac12\,\mbox{Tr} \left[ \frac{2\bm{X}_q}{\sinh2\bm{X}_q}\,
    \bigg( \frac{\bm{Z}_q(-m^2)}{1 + \bm{Z}_q(-m^2)}
    + \frac{\bm{Z}_q^\dagger(-m^2)}{1 + \bm{Z}_q^\dagger(-m^2)} \bigg) \right] \right\} , \\
   I_-(m^2) 
   &= \sum_{q=u,d}\,\frac{1}{2i}\,\mbox{Tr} \left[ \frac{2\bm{X}_q}{\sinh2\bm{X}_q}\,
    \bigg( \frac{\bm{Z}_q(-m^2)}{1 + \bm{Z}_q(-m^2)}
    - \frac{\bm{Z}_q^\dagger(-m^2)}{1 + \bm{Z}_q^\dagger(-m^2)} \bigg) \right] ,
\end{aligned}
\end{equation}
where $m^2\equiv m^2+i0$, and the function
\begin{equation}\label{gfundef}
   g(\bm{X}_q) \big|_{\rm brane\;Higgs} 
   = \bm{X}_q\tanh\bm{X}_q - \bm{X}_q\tanh2\bm{X}_q
   = - \frac{\bm{X}_q\tanh\bm{X}_q}{\cosh2\bm{X}_q} 
\end{equation}
obeys a non-singular series expansions in powers of $\bm{X}_q^2$. Note that due to the presence of strong-interaction phases arising from the analytic continuation from a Euclidean momentum $p_E^2$ to $-m^2-i0$, the functions $I_\pm (m^2)$ cannot simply be written in terms of the real and imaginary parts of a traces over matrices. If instead of the brane-localized Higgs boson we consider a narrow bulk-Higgs state, then the subtraction term $t_1$ is absent, see (\ref{resbrane}) and (\ref{resbulk}). The expressions in (\ref{IplImi}) remain valid also in this case, provided we use
\begin{equation}\label{gfundef2}
   g(\bm{X}_q) \big|_{\rm narrow\;bulk\;Higgs} = \bm{X}_q\tanh\bm{X}_q \,.
\end{equation}

The above equations are the main result of our paper. Up to some small corrections to be determined below, the first term on the right-hand side of the equation for $I_+(m^2)$ corresponds to the contribution of the infinite tower of KK quarks to the $ggh$ amplitude. The remaining terms describe the contributions of the SM quarks. For the case of a brane-localized Higgs sector, the function $g(\bm{X}_q)$ coincides with an expression first obtained in \cite{Carena:2012fk} by means of a conjecture. In the present work we have derived this form. For the case of a narrow bulk-Higgs field, the expansion of $g(\bm{X}_q)$ to ${\cal O}(\bm{X}_q^2)$ reproduces the result derived in \cite{Azatov:2010pf}. This demonstrates that the ``brane-Higgs limit'' considered in that paper really corresponds to the case of a narrow bulk scalar, as defined in (\ref{bulkHiggs}).

\subsection{Alternative derivation of the result for a brane Higgs}
\label{sec:4.5}

For the case of a brane-localized scalar sector, it has been shown in \cite{Carena:2012fk} that the fermion bulk profiles and the Yukawa couplings $g_{mn}^q$ to the fermion mass eigenstates defined in (\ref{gdef}) can also be derived in a different way, by solving the field equations for the fermion modes in the bulk and incorporating the effects of the Yukawa interactions by imposing appropriate boundary conditions on the IR brane. The Yukawa couplings are then derived by evaluating the fermion profiles in the limit $t\to 1^-$ (approached from the left), which defines their values on the IR brane by continuous extension. 

This method, which in \cite{Carena:2012fk} was established for individual fermion states, can also be applied to the infinite tower of KK modes, by imposing similar boundary conditions on the 5D propagator functions. Indeed, we find that with a brane-localized Higgs field the functions $T_\pm(p_E^2)$ defined in (\ref{TRLdef}) can also be computed as
\begin{equation}\label{TRLnaive}
   T_+(p_E^2) \big|_{\rm brane~Higgs}
   = - \sum_{q=u,d} \frac{v}{\sqrt2}\,
    \mbox{Tr}\left[ \bigg( \begin{array}{cc} 0 & \bm{Y}_q \\ 
     \bm{Y}_q^\dagger & 0 \end{array} \bigg)\,
     \frac{\bm{\Delta}_{RL}^q(1^-,1^-;p_E^2)+\bm{\Delta}_{LR}^q(1^-,1^-;p_E^2)}{2} \right] ,
\end{equation}
and similarly for $T_-(p_E^2)$. The propagator functions $\bm{\Delta}_{AB}^q$ are now computed by solving the coupled system of equations (\ref{coupledeqs}) {\em without\/} including the Yukawa term in the generalized mass matrix ${\cal M}_q(t)$ in (\ref{Mqdef}). Instead, one modifies the boundary conditions on the IR brane, such that 
\begin{equation}\label{eq56}
   \bigg( \frac{v\tilde{\bm{Y}}_q^\dagger}{\sqrt2 M_{\rm KK}} ~\quad 1 \bigg)\,
    \bm{\Delta}_{LL}^q(1^-,t';-p^2) 
   = \bigg( 1 \quad - \frac{v\tilde{\bm{Y}}_q}{\sqrt2 M_{\rm KK}} \bigg)\,
    \bm{\Delta}_{RL}^q(1^-,t';-p^2) = 0
\end{equation}
instead of condition (\ref{BCs}) with $t_i=1$. Here 
\begin{equation}\label{Ytildef}
   \tilde{\bm{Y}}_q = \frac{\tanh\bm{X}_q}{\bm{X}_q}\,\bm{Y}_q
\end{equation} 
are the modified Yukawa matrices introduced in \cite{Casagrande:2010si}. The boundary conditions on the UV brane (at $t_i=\epsilon$) and the jump conditions (\ref{jump}) remain unchanged. It is a straightforward exercise to derive the propagator functions from these equations, and in particular to determine the mixed-chirality components at $t=t'=1^-$. We have confirmed that inserting these results into (\ref{TRLnaive}) one reproduces the expressions given in (\ref{IplImi}). This method provides an independent derivation of the result for the brane-localized Higgs boson, in which the notion of a regulator $\eta$ never appears.

\subsection{Analysis of the zero-mode contributions}

We will now analyze the terms involving the matrices $\bm{Z}_q$ in (\ref{IplImi}), which include the contributions of the SM quarks, in more detail, using results derived in \cite{Casagrande:2008hr}. We first note that we can rewrite
\begin{equation}\label{Zdef1}
   \bm{Z}_q(p_E^2) 
   = \frac{v^2}{2M_{\rm KK}^2}\,\tilde{\bm{Y}}_q\,\bm{R}_q(\hat p_E)\,
    \tilde{\bm{Y}}_q^\dagger\,\bm{R}_Q(\hat p_E) \,,
\end{equation}
with $\tilde{\bm{Y}}_q$ as defined above. In terms of these quantities, the eigenvalue equation determining the KK masses reads
\begin{equation}\label{eigenvals}
   \mbox{det}\big[ 1 + \bm{Z}_q(-m_n^2) \big] = 0 \,.
\end{equation}
The asymptotic expansion for $\bm{R}_A$ in (\ref{RAsmall}) introduces the fermion profiles $F(\bm{c}_A)$ next to the modified Yukawa matrices. We can then reexpress the answer in terms of the effective Yukawa matrices defined as \cite{Casagrande:2010si}
\begin{equation}\label{Yqeff}
   \bm{Y}_q^{\rm eff} \equiv F(\bm{c}_Q)\,\tilde{\bm{Y}}_q\,F(\bm{c}_q)
   = \bm{U}_q\,\bm{\lambda}_q\,\bm{W}_q^\dagger \,,
    \qquad \mbox{where} \quad
   \bm{\lambda}_q = \frac{\sqrt2}{v}\,\bm{m}_{q,0}
\end{equation}
are diagonal, positive real matrices. The entries $m_{q_i,0}$ denote the zeroth-order values of the masses of the SM quarks. The unitary matrices $\bm{U}_q$ and $\bm{W}_q$ are defined by relation (\ref{Yqeff}). Including also the subleading terms in the expansion (\ref{RAsmall}), we obtain
\begin{equation}
   \bm{Z}_q(p_E^2)
   = F^{-1}(\bm{c}_Q)\,\bm{U}_q \left[ \frac{\bm{m}_{q,0}^2}{p_E^2}
    + \left( \bm{\delta}_Q + \bm{m}_{q,0}\,\bm{\delta}_q\,\bm{m}_{q,0}^{-1} \right) + \dots \right]
    \bm{U}_q^\dagger\,F(\bm{c}_Q) \,,
\end{equation}
where 
\begin{equation}\label{deltadef}
\begin{aligned}
   \bm{\delta}_Q &= \bm{x}_q\,\bm{W}_q^\dagger \left[
    \frac{1}{1-2\bm{c}_q} \left( \frac{1}{F^2(\bm{c}_q)} - 1 + \frac{F^2(\bm{c}_q)}{3+2\bm{c}_q} 
    \right) \right] \bm{W}_q\,\bm{x}_q \,, \\
   \bm{\delta}_q &= \bm{x}_q\,\bm{U}_q^\dagger \left[
    \frac{1}{1-2\bm{c}_Q} \left( \frac{1}{F^2(\bm{c}_Q)} - 1 + \frac{F^2(\bm{c}_Q)}{3+2\bm{c}_Q} 
    \right) \right] \bm{U}_q\,\bm{x}_q
\end{aligned}
\end{equation}
with $\bm{x}_q=\bm{m}_{q,0}/M_{\rm KK}$ are hermitian matrices giving rise to some small corrections of order $v^2/M_{\rm KK}^2$, which except for the two entries proportional to $m_{u_3}^2=m_t^2$ carry an additional strong chiral suppression \cite{Casagrande:2008hr}. Introducing the abbreviation $\bm{\varepsilon}_q=\bm{\delta}_Q+\bm{m}_{q,0}\,\bm{\delta}_q\,\bm{m}_{q,0}^{-1}$, and working to first order in $v^2/M_{\rm KK}^2$, we can rewrite the eigenvalue equation (\ref{eigenvals}) in the form
\begin{equation}\label{det2}
   \mbox{det}\big[ m_n^2 - \bm{m}_{q,0}^2 \left(1-\bm{\varepsilon}_q\right) + \dots \big] = 0 \,, 
\end{equation}
whereas
\begin{equation}\label{Zrat2}
   \frac{\bm{Z}_q(p_E^2)}{1+\bm{Z}_q(p_E^2)}
   = F^{-1}(\bm{c}_Q)\,\bm{U}_q \left[ \bm{\varepsilon}_q
    + \frac{\left(1-\bm{\varepsilon}_q\right) \bm{m}_{q,0}^2 \left(1-\bm{\varepsilon}_q\right)}%
           {p_E^2 + \bm{m}_{q,0}^2 \left(1-\bm{\varepsilon}_q\right)} + \dots \right]
    \bm{U}_q^\dagger\,F(\bm{c}_Q) \,.
\end{equation}
Only the diagonal elements of the matrices $\bm{\varepsilon}_q$ contribute when (\ref{det2}) and traces of (\ref{Zrat2}) are evaluated to first order in $v^2/M_{\rm KK}^2$. It is then not difficult to show that the masses of the SM quarks are given by
\begin{equation}
   m_{q_i}^2 = m_{q_i,0}^2 \left( 1 - \varepsilon_{q_i} + \dots \right) , \qquad
   \mbox{with} \quad
   \varepsilon_{q_i}\equiv \left( \bm{\varepsilon}_q \right)_{ii}
    = \left( \bm{\delta}_Q \right)_{ii} + \left( \bm{\delta}_q \right)_{ii} ,
\end{equation}
where the dots represent terms of order $v^4/M_{\rm KK}^4$ and higher. Moreover, we find
\begin{equation}\label{trZ}
   \sum_{q=u,d} \mbox{Tr} \left[ \frac{2\bm{X}_q}{\sinh2\bm{X}_q}\,
    \frac{\bm{Z}_q(p_E^2)}{1 + \bm{Z}_q(p_E^2)} \right]
   = \sum_i \left[ \kappa_{q_i}\,\frac{m_{q_i}^2}{m_{q_i}^2+p_E^2} + \varepsilon_{q_i} \right]
    + \dots \,,
\end{equation}
where 
\begin{equation}\label{kappadef}
   \kappa_{q_i} = 1 - \varepsilon_{q_i} - \frac23\,\Big[
    \bm{U}_q^\dagger\,F(\bm{c}_Q)\,\bm{X}_q^2\,F^{-1}(\bm{c}_Q)\,\bm{U}_q \Big]_{ii} \,.
\end{equation}
Note that while the parameters $\kappa_{q_i}$ are in general complex, the quantities $\varepsilon_{q_i}$ are real. The sum in (\ref{trZ}) extends over all six SM quarks. However, in practice the contributions of the light quarks can safely be neglected. For the third-generation quarks, we find that
\begin{equation}\label{kappat}
   \kappa_t = 1 - \varepsilon_t
    - \frac{v^2}{3M_{\rm KK}^2}\,
    \frac{\left( \bm{Y}_u \bm{Y}_u^\dagger \bm{Y}_u \right)_{33}}{\left( \bm{Y}_u \right)_{33}}
\end{equation}
up to chirally-suppressed terms, and a corresponding formula holds for $\kappa_b$. This expression coincides with the result derived in \cite{Carena:2012fk}. Explicit formulae for the matrix elements $\left( \bm{\delta}_A \right)_{33}$ can also be found in this reference.

It is now a simple exercise to evaluate the Wilson coefficients $C_{1,5}$ using (\ref{C1C5res}). We obtain
\begin{equation}\label{Ciresu}
\begin{aligned}
   C_1 &= \sum_{q=u,d}\,\mbox{Tr}\,\big[ g(\bm{X}_q) + \bm{\varepsilon}_q \Big]
    + \sum_i\,\mbox{Re}(\kappa_{q_i})\,A(\tau_i) + \dots \\
   &\approx \Bigg[ 1 - \frac{v^2}{3M_{\rm KK}^2}\,\mbox{Re}\,
    \frac{\left( \bm{Y}_u \bm{Y}_u^\dagger \bm{Y}_u \right)_{33}}{\left( \bm{Y}_u \right)_{33}}
    \Bigg]\,A(\tau_t) + A(\tau_b)
    + \mbox{Tr}\,g(\bm{X}_u) + \mbox{Tr}\,g(\bm{X}_d) \,, \\
   C_5 &= \sum_i\,\mbox{Im}(\kappa_{q_i})\,B(\tau_i) + \dots
    \approx - \frac{v^2}{3M_{\rm KK}^2}\,\mbox{Im} \Bigg[
    \frac{\left( \bm{Y}_u \bm{Y}_u^\dagger \bm{Y}_u \right)_{33}}{\left( \bm{Y}_u \right)_{33}}
    \Bigg]\,B(\tau_t) \,,
\end{aligned}
\end{equation}
where $\tau_i=4m_{q_i}^2/m_h^2-i0$, and the parameter integrals evaluate to \cite{Beneke:2002jn,Djouadi:2005gj}
\begin{equation}
   A(\tau) = \frac{3\tau}{2}\,\Big[ 1 + (1-\tau) \arctan^2\frac{1}{\sqrt{\tau-1}} \Big] \,,
    \qquad
   B(\tau) = \tau \arctan^2\frac{1}{\sqrt{\tau-1}} \,.
\end{equation}
For the light SM quarks, these functions must be analytically continued to $\tau<1$. In (\ref{Ciresu}), we first present expressions that are exact up to small corrections of order $v^4/M_{\rm KK}^4$, represented by the dots, which are numerically insignificant. The leading effects, which involve traces over functions of Yukawa matrices and thus increase with the number of fermion generations, are exact to all orders in $v^2/M_{\rm KK}^2$. The infinite sum over KK quark states contributes the trace term in the expression for $C_1$. The second term contains the sum over the contributions of the SM quarks, whose Yukawa interactions are modified with respect to the SM by factors $\kappa_{q_i}$. 

In the final, approximate expressions we have used the fact that all $\varepsilon_{q_i}$ parameters other than $\varepsilon_t$ can be neglected to a very good approximation, and that for the term proportional to $\varepsilon_t$ we can neglect the small deviation of the function $A(\tau_t)\approx 1.03$ from~1. Also, for the small $b$-quark contribution, it is safe to neglect the small deviation of $\kappa_b$ from~1. In this approximation, which is accurate to better than 1\% for $M_{\rm KK}\gtrsim 2$\,TeV, we observe that the Wilson coefficients $C_1$ and $C_5$ become independent of the bulk mass parameters $c_i$. They are entirely given in terms of the 5D Yukawa matrices of the RS model. In the SM, we have $C_1^{\rm SM}=A(\tau_t)+A(\tau_b)$ and $C_5^{\rm SM}=0$.

\subsection{Brane-localized Higgs sector with different Yukawa matrices}
\label{sec:2Yukawas}

Before closing this section, we return to the generalization of the RS model with a brane-localized Higgs sector in which one allows for different Yukawa matrices $\bm{Y}_q^C$ and $\bm{Y}_q^{S\dagger}$ in the two terms in the last line of (\ref{gdef}) \cite{Azatov:2010pf,Azatov:2009na}. We will refer to this model as ``type-II brane-Higgs'' scenario. As discussed in Appendix~\ref{app:Y1Y2}, we find that the expressions valid in this case can be obtained from the ones derived so far by means of some simple manipulations. Instead of the matrices $\bm{X}_q$ defined in (\ref{Xqdef}) and $\tilde{\bm{Y}}_q$ given in (\ref{Ytildef}), we must use
\begin{equation}\label{newdefs}
   \bm{X}_q = \frac{v}{\sqrt2 M_{\rm KK}} \sqrt{\bm{Y}_q^C\bm{Y}_q^{S\dagger}} \,, \qquad
   \tilde{\bm{Y}}_q = \frac{\tanh\bm{X}_q}{\bm{X}_q}\,\bm{Y}_q^C \,.
\end{equation}
It follows that instead of (\ref{Zdef}) we now have
\begin{equation}
   \bm{Z}_q(p_E^2) 
   = \frac{v^2}{2M_{\rm KK}^2}\,\frac{\tanh\bm{X}_q}{\bm{X}_q}\,\bm{Y}_q^C\,
    \bm{R}_q(\hat p_E)\,\bm{Y}_q^{C\dagger}\,
    \frac{\tanh\bm{X}_q^\dagger}{\bm{X}_q^\dagger}\,\bm{R}_Q(\hat p_E) \,.
\end{equation}
Also, the master formulae (\ref{IplImi}) must be generalized to read
{\small
\begin{equation}
\begin{aligned}
   I_+(m^2) 
   &= \sum_{q=u,d} \left\{ \mbox{Re}\,\mbox{Tr}\,g(\bm{X}_q,\tilde{\bm{Y}}_q) 
    + \frac12\,\mbox{Tr} \left[ 
    \frac{2\bm{X}_q}{\sinh2\bm{X}_q}\,\frac{\bm{Z}_q(-m^2)}{1 + \bm{Z}_q(-m^2)}
    + \frac{2\bm{X}_q^\dagger}{\sinh2\bm{X}_q^\dagger}\,
    \frac{\bm{Z}_q^\dagger(-m^2)}{1 + \bm{Z}_q^\dagger(-m^2)} \right] \right\} , \\
   I_-(m^2) 
   &= \sum_{q=u,d} \left\{ \mbox{Im}\,\mbox{Tr}\,g(\bm{X}_q,\tilde{\bm{Y}}_q) 
    + \frac{1}{2i}\,\mbox{Tr} \left[ 
    \frac{2\bm{X}_q}{\sinh2\bm{X}_q}\,\frac{\bm{Z}_q(-m^2)}{1 + \bm{Z}_q(-m^2)}
    - \frac{2\bm{X}_q^\dagger}{\sinh2\bm{X}_q^\dagger}\,
    \frac{\bm{Z}_q^\dagger(-m^2)}{1 + \bm{Z}_q^\dagger(-m^2)} \right] \right\} ,
\end{aligned}
\end{equation}
}
where
\begin{equation}\label{gtype2}
   g(\bm{X}_q,\tilde{\bm{Y}}_q) \big|_{\rm brane\;Higgs}^{\rm type-II}
   = - \frac{2\bm{X}_q}{\sinh2\bm{X}_q}\,
    \frac{\frac{v^2}{2M_{\rm KK}^2}\,\tilde{\bm{Y}}_q\tilde{\bm{Y}}_q^\dagger}%
    {1+\frac{v^2}{2M_{\rm KK}^2}\,\tilde{\bm{Y}}_q\tilde{\bm{Y}}_q^\dagger} 
   = - \frac{v^2}{2M_{\rm KK}^2}\,\bm{Y}_q^C\bm{Y}_q^{C\dagger} + \dots \,.
\end{equation}
Finally, in the formulae for $\kappa_t$ in (\ref{kappat}) one must replace the combination $\big(\bm{Y}_u\bm{Y}_u^\dagger\bm{Y}_u\big)_{33}/\big(\bm{Y}_u\big)_{33}$ by $\big(\bm{Y}_u^C\bm{Y}_u^{S\dagger}\bm{Y}_u^C\big)_{33}/\big(\bm{Y}_u^C\big)_{33}$. Note that because $\bm{X}_q$ is no longer a positive hermitian matrix, traces of $\bm{X}_q^n$ can now have arbitrary phases. However, at leading order in the expansion in $v^2/M_{\rm KK}^2$ the trace of the function $g(\bm{X}_q,\tilde{\bm{Y}}_q)$ is a negative real number. Indeed, at this order there is no difference between the result (\ref{gtype2}) and the original result in (\ref{gfundef}) valid for the brane-Higgs scenario with $\bm{Y}_q^C=\bm{Y}_q^S$. 

An interesting special case is that where $\bm{Y}_q^S=0$, meaning that the Yukawa couplings involving a product of two $Z_2$-odd fields, given by the second term in the last line of (\ref{gdef}), is put to zero. This choice was frequently adopted in the literature. It corresponds to taking the limit $\bm{X}_q\to 0$ in our results, in which case $\tilde{\bm{Y}}_q\to\bm{Y}_q^C$, and the quantities $\kappa_{q_i}$ in (\ref{kappadef}) reduce to $\kappa_{q_i}=1-\varepsilon_{q_i}$. It follows that in this particular model one obtains
\begin{equation}
\begin{aligned}
   C_1 &= \sum_{q=u,d}\,\mbox{Tr}\,\big[ g(0,\bm{Y}_q^C) + \bm{\varepsilon}_q \big]
    + \sum_i\,(1-\varepsilon_{q_i})\,A(\tau_i) + \dots \\
   &\approx C_1^{\rm SM}
    + \left[  1 - A(\tau_t) \right] \varepsilon_t + \varepsilon_b 
    - \frac{v^2}{2M_{\rm KK}^2}\,\mbox{Tr}\,\big[ \bm{Y}_u^C \bm{Y}_u^{C\dagger}
    + \bm{Y}_d^C \bm{Y}_d^{C\dagger} \big] \,, 
\end{aligned}
\end{equation}
whereas $C_5=0$. The first term in the first line is the result of the summation over the KK tower of quark states, while the second term gives the contributions of the SM quarks, whose Yukawa couplings are modified with respect to their values in the SM by factors $(1-\varepsilon_{q_i})$. It suffices for all practical purposes to keep only the terms shown in the second line. Apart from the last term, they agree with a corresponding result presented in \cite{Bouchart:2009vq}. The first two corrections to the SM result are numerically very small, because $1-A(\tau_t)\approx-0.03$ and the quantity $\varepsilon_b$ is chirally suppressed. The third correction, which arises from the infinite sum over KK states, gives the dominant contribution by far. This effect was not found in \cite{Azatov:2010pf}, because in this paper the brane-Higgs case was derived by taking a limit of a bulk-Higgs result. If one formally introduces two different Yukawa matrices in the narrow bulk-Higgs scenario, one indeed finds that $g(\bm{X}_q)$ defined in (\ref{gfundef2}) vanishes in the limit where $\bm{Y}_q^S\to 0$. However, in the context of a bulk Higgs model taking $\bm{Y}_q^S$ different from $\bm{Y}_q^C$ violates 5D Lorentz invariance, and moreover (as we have explained several times) the brane-Higgs case cannot be derived by taking a limit of the bulk-Higgs results.

In practice, we find that the corrections to the gluon fusion amplitude found in the type-II brane-Higgs scenario are numerically very similar to those obtained in the original brane-Higgs model. The main difference is a slightly larger spread of the distribution of points obtained when one scans the parameter space of the model. In our phenomenological analysis in Section~\ref{sec:pheno} we will therefore restrict ourselves to a study of the case where $\bm{Y}_q^C=\bm{Y}_q^S$.

\section{\boldmath Impact of higher-dimensional $|\Phi|^2 (G_{\mu\nu}^a)^2$ operators}
\label{sec:powerhgg}

We have argued in the introduction that RS models must be considered as effective field theories, valid below a (position-dependent) UV cutoff given by the warped Planck scale. The UV completion of these models is unknown. It may be strongly coupled, for instance due to effects of quantum gravity. Short-distance contributions from physics above the cutoff scale give rise to higher-dimensional operators, such as those studied briefly in Section~\ref{sec:powercors}. Two particularly interesting higher-dimensional operators relevant for Higgs production are $\Phi^\dagger\Phi\,{\cal G}_{MN}^a {\cal G}^{MN,a}$ and $\Phi^\dagger\Phi\,{\cal G}_{MN}^a {\cal \widetilde G}^{MN,a}$, which mediate effective $hgg$ couplings at tree level. Here ${\cal G}_{MN}^a$ is the 5D gluon field-strength tensor. We will now address the question how important the contributions of these operator are in the low-energy effective theory, focussing on the first operator for concreteness.

In the RS model with the scalar sector localized on the IR brane, the relevant effective action is
\begin{equation}\label{Shhgg}
   S_{\rm eff} = \int d^4x \int_{-r\pi}^{r\pi}\!dx_5\,c_{\rm eff}\,\delta(|x_5|-r\pi)\,
    \frac{\Phi^\dagger\Phi}{\Lambda_{\rm TeV}^2}\,
    \frac{g_{s,5}^2}{4}\,{\cal G}_{\mu\nu}^a\,{\cal G}^{\mu\nu,a} + \dots \,,
\end{equation} 
where we do not bother to write down terms involving ${\cal G}_{\mu 5}^a$. Here $g_{s,5}$ is the five-dimensional strong coupling, which is related to the coupling $g_s$ of the SM by $g_{s,5}=\sqrt{2\pi r}\,g_s$ \cite{Davoudiasl:1999tf}. The natural UV cutoff governing the suppression of the brane-localized higher-dimensional operator is $\Lambda_{\rm TeV}$. NDA suggests that the dimensionless coupling $c_{\rm eff}$ could be as large as ${\cal O}(1)$ if the UV completion above the cutoff of the RS model is strongly coupled. In the absence of a complete model, it is impossible to say how $c_{\rm eff}$ might depend on other parameters, such as the Yukawa couplings or the number of fermion generations. Even in a strongly coupled theory, it is possible that $c_{\rm eff}$ could be significantly smaller than~1,\footnote{An example is provided by the $\pi^0\to\gamma\gamma$ decay amplitude, which is loop suppressed in the SM despite the fact that QCD is strongly coupled in the low-energy regime.}% 
{} for instance because the effective degrees of freedom coupling the Higgs boson to two gluons can only be pair produced, or because they have suppressed couplings to the operators $\Phi^\dagger\Phi$ or ${\cal G}_{\mu\nu}^a\,{\cal G}^{\mu\nu,a}$. Following common practice, we shall assume that taking $c_{\rm eff}={\cal O}(1)$ provides a conservative upper bound for the effect of the brane-localized operators on the gluon fusion amplitude.
 
Using the KK decomposition of the gluon field, 
\begin{equation}
   {\cal G}_{\mu\nu}^a(x,\phi) = \frac{1}{\sqrt r} \sum_n G_{\mu\nu}^{(n)\,a}(x)\,\chi_n^G(\phi)
   = \frac{1}{\sqrt{2\pi r}}\,G_{\mu\nu}^a(x) + \mbox{KK modes} \,,
\end{equation} 
where the zero mode (the SM gluon $G_{\mu\nu}^a\equiv G_{\mu\nu}^{(0)\,a}$) has a flat profile along the extra dimension, and writing the scalar doublet in the standard form
\begin{equation}
   \Phi(x) = \begin{pmatrix} - i\varphi^+(x) \\ 
     \frac{1}{\sqrt2} \big[ v + h(x) + i\varphi_3(x) \big] \end{pmatrix} ,
\end{equation}
we find that the relevant terms in the action (\ref{Shhgg}) gives rise to the effective Lagrangian
\begin{equation}\label{Leff}
   {\cal L}_{\rm eff} = \frac{c_{\rm eff}}{\Lambda_{\rm TeV}^2}\,{\cal O}_{\rm eff} \,, 
\end{equation} 
where
\begin{equation}
   {\cal O}_{\rm eff} = \Phi^\dagger\Phi\,\frac{g_s^2}{4}\,G_{\mu\nu}^a\,G^{\mu\nu,a} 
   \ni \frac{g_s^2 v^2}{8} \left( 1 + \frac{h(x)}{v} \right)^2 G_{\mu\nu}^a\,G^{\mu\nu,a} \,.
\end{equation} 

We now repeat this analysis for an RS model in which the Higgs field lives in the bulk of the extra dimension. A detailed discussion of the properties of a bulk-Higgs field and its vev is presented in Appendix~\ref{app:bulkHiggs}. In this case the higher-dimensional operator can be localized on both the IR and UV branes, or it can live in the bulk. We thus consider the action
\begin{equation}
   S_{\rm eff} = \int d^4x \int_{-r\pi}^{r\pi}\!dx_5\,
    \Big[ c_1 + c_2\,\delta(|x_5|-r\pi) + c_3\,\delta(x_5) \Big]\,
    \frac{\Phi^\dagger\Phi}{M_{\rm Pl}^2}\,\frac{g_{s,5}^2}{4}\,
    {\cal G}_{\mu\nu}^a\,{\cal G}^{\mu\nu,a} + \dots \,,
\end{equation} 
where the coupling $c_1$ is dimensionless, while $c_{2,3}\sim 1/M_{\rm Pl}$. Since all fields live in the bulk, the natural cutoff suppressing the operator is set by the Planck scale. Also, the scalar field now takes the form shown in relation (\ref{bulkPhi}) in Appendix~\ref{app:bulkHiggs}. Using the KK decomposition of the Higgs field given in (\ref{hKKdec}), we find that
\begin{equation}
\begin{aligned}
   S_{\rm eff} 
   &= \int d^4x\,\frac{g_s^2}{4}\,G_{\mu\nu}^a(x)\,G^{\mu\nu,a}(x)\,
    \frac{2\pi}{L} \int_\epsilon^1\!\frac{dt}{t}\,\frac{v^2(t)}{2\Lambda_{\rm UV}^2(t)}
    \left( 1 + h(x)\,\frac{\chi_0(t)}{v(t)} \right)^2 \\
   &\quad\times \left\{ c_1 + \frac{k}{2}\,\big[ c_2\,\delta(t-1)
    + \epsilon\,c_3\,\delta(t-\epsilon) \big] \right\} + \dots \,,
\end{aligned}
\end{equation} 
where $\Lambda_{\rm UV}(t)=M_{\rm Pl}\,\epsilon/t$ is the warped Planck scale as introduced in (\ref{LamUVt}), and $v(t)$ and $\chi_0(t)$ are the profiles of the Higgs vev and the physical SM Higgs boson along the extra dimension. We now use the explicit form of the profile of the Higgs vev given in (\ref{vtfinal}), as well as the fact that according to (\ref{chi0fin}) we have $\chi_0(t)/v(t)=1/v$ up to corrections of order $m_h^2/M_{\rm KK}^2$, which we neglect here. Here $v\approx 246$\,GeV denotes the SM value of the Higgs vev. It is then straightforward to perform the integration over $t$ in the above result. Matching the answer onto the effective Lagrangian given in (\ref{Leff}), we obtain
\begin{equation}\label{cfinbulk}
   c_{\rm eff} = \frac{1+\beta}{2+\beta}\,c_1 + (1+\beta)\,k c_2 
   \quad \stackrel{\beta\gg 1}{\longrightarrow} \quad
   c_1 + |\mu| c_2 \,,
\end{equation}
where the parameter $\beta\sim 1/\eta$ is related to the width of the profile of the scalar field (see Appendix~\ref{app:bulkHiggs}). NDA suggests that $c_1$ and $k c_2$ can be as large as ${\cal O}(1)$ if the UV completion of the RS model is strongly coupled. The contribution of the operator localized on the UV brane is of ${\cal O}(\epsilon^{4+2\beta})\,c_3$ and thus entirely negligible. This suppression results from a factor $1/M_{\rm Pl}^2$ times $v^2(\epsilon)\sim\epsilon^{2+2\beta}$ reflecting the smallness of the Higgs vev profile on the UV brane. Note that in the limit of a very narrow bulk-Higgs field, corresponding to $\beta\gg 1$ (or $\eta\ll 1$), the largest mass scale in the model is the Higgs mass parameter $|\mu|\approx\beta k={\cal O}(M_{\rm Pl})$ in (\ref{eqn:HSaction}) and (\ref{eqn:diffv}), and hence it is more appropriate to assume that $c_2\sim 1/|\mu|\sim 1/M_{\rm Pl}$. Once again, this leads to $c_{\rm eff}={\cal O}(1)$. The structure of the result (\ref{cfinbulk}) is completely analogous to the corresponding expression in (\ref{Leff}) valid for a brane-localized Higgs boson. In both cases the results for $c_{\rm eff}$, and hence the magnitude of the contributions of higher-dimensional operators, are expected to be of the same order. 

The effective Lagrangian (\ref{Leff}) yields a contribution to the Wilson coefficient $C_1$ in (\ref{Lhgg}) given by
\begin{equation}\label{DeltaC1}
   \Delta C_1 = \frac{3 c_{\rm eff}}{4} \left( \frac{4\pi v}{\Lambda_{\rm TeV}} \right)^2 
   \approx c_{\rm eff} \left( \frac{2.7\,\mbox{TeV}}{\Lambda_{\rm TeV}} \right)^2 . 
\end{equation}
In order for this contribution to be much smaller than the SM value $C_1=1$, we need to assume that either the cutoff scale is much larger than about 3\,TeV or that $|c_{\rm eff}|\ll 1$ for some reason. With $\Lambda_{\rm TeV}\sim 10\,M_{\rm KK}\sim 20\,\mbox{--}\,50$\,TeV, the first criterion is satisfied in realistic RS models even if $c_{\rm eff}={\cal O}(1)$. The expected contribution to the Wilson coefficient $C_1$ is then in the percent range, which is negligible in view of the current experimental uncertainty in the measurements of the Higgs-boson couplings. Another interesting question is under which assumptions the contribution (\ref{DeltaC1}) is much smaller than the corrections to the SM result $C_1=1$ which we have obtained from loop effects in the RS model, which are approximately given by 
\begin{equation}
   |C_1-1|
   \approx \frac{v^2}{2M_{\rm KK}^2} \sum_{q=u,d} \mbox{Tr}\left( \bm{Y}_q\bm{Y}_q^\dagger \right)
   \approx \frac{v^2}{2M_{\rm KK}^2}\,2 N_g^2\,|Y_q|^2 \,,
\end{equation}
where $N_g=3$ is the number of fermion generations, and $|Y_q|$ is the typical size of an element of the anarchic 5D Yukawa matrices, defined by
\begin{equation}\label{eq87}
   |Y_q|^2 \equiv \langle|(\bm{Y}_q)_{ij}|^2\rangle = \frac{y_*^2}{2} \,.
\end{equation}
We work with anarchic 5D Yukawa matrices and assume that the entries $(\bm{Y}_q)_{ij}$ are random complex numbers, which with equal probability can take any value in the complex plane inside a circle of radius $y_*$. Throughout our paper, we will assume that $y_*$ is an ${\cal O}(1)$ parameter. This is natural, since we have obtained the matrices $\bm{Y}_q$ by multiplying the underlying, dimensionful Yukawa couplings $\bm{Y}_q^{\rm 5D}$ of the original 5D Lagrangian by the AdS curvature $k$, which sets the natural scale for dimensionful quantities in the RS model. It follows that the power-suppressed contribution (\ref{DeltaC1}) can be neglected as long as
\begin{equation}\label{ceffesti}
   c_{\rm eff} \left( \frac{M_{\rm KK}}{\Lambda_{\rm TeV}} \right)^2
   \ll \frac{N_g^2\,y_*^2}{24\pi^2} \,,
\end{equation}
which for $\Lambda_{\rm TeV}\approx 10 M_{\rm KK}$ can be rewritten as $c_{\rm eff}\ll 3.8\,y_*^2$. In the custodial RS model studied in the next section, the expression on the right-hand side of this relation is multiplied by~4, yielding the weaker condition $c_{\rm eff}\ll 15.2\,y_*^2$. In our phenomenological analysis in Section~\ref{sec:pheno} we will consider values of $y_*$ between 3 and 0.5. In order to neglect the power-suppressed contributions for $y_*=0.5$ in the minimal RS model, we would need to rely on the assumption that $|c_{\rm eff}|\ll 1$. 

Relation (\ref{ceffesti}) makes it clear that, in comparing the contributions from higher-dimensional operators with the contribution from virtual KK states, we are comparing a power-suppressed effect with a loop effect. Since we treat the dimensionless Yukawa couplings as ${\cal O}(1)$ random complex parameters, it would follow that in the formal limit $\Lambda_{\rm TeV}\to\infty$ the higher-dimensional operator contribution tends to zero, while the loop contribution remains as the dominant effect.\footnote{Since for too large values of the cutoff the Yukawa sector becomes strongly coupled (see below), our result in such an academic limit could at best be taken as a rough estimate of the KK loop contributions.} 
However, since by construction the RS model is free of large hierarchies, the ratio $M_{\rm KK}/\Lambda_{\rm TeV}$ cannot be made arbitrarily small. We therefore do not expect a strong hierarchy between the contribution from virtual KK states and those from higher-dimensional operators. In practice, which of the effect wins is more of a numerical question than a parametric one. In our phenomenological analysis in Section~\ref{sec:pheno}, we include the contribution $\Delta C_1$ in (\ref{DeltaC1}) by treating $c_{\rm eff}$ as a random number with magnitude less than~1.

For our loop calculation to be trustable, we should impose an upper bound on the size of $y_*$ by requiring that the Yukawa interactions remain perturbative up to the cutoff of the RS model under consideration (see e.g.\ \cite{Csaki:2008zd,Cacciapaglia:2006mz}). Following common practice, we will assume that $y_*<y_{\rm max}\approx 3$. A detailed discussion of different estimates of the perturbativity bound $y_{\rm max}$ is presented in Appendix~\ref{app:perturbativity}.

\section{Extension to the RS model with custodial symmetry}
\label{sec:custodial}

We will now present the generalization of the above results to an extended version of the RS model, in which large corrections to electroweak precision observables are avoided by means of an enlarged gauge symmetry in the bulk of the extra dimension. Electroweak precision tests are then no longer in conflict with having the masses of the lightest KK states lie in the range of a few TeV, in reach for direct production of these particles at the LHC \cite{Carena:2006bn,Cacciapaglia:2006gp,Contino:2006qr,Carena:2007ua}. Specifically, we consider an RS model based on the gauge symmetry $SU(3)_C\times SU(2)_L\times SU(2)_R\times U(1)_X\times P_{LR}$. On the IR brane, the symmetry-breaking pattern $SU(2)_L\times SU(2)_R\to SU(2)_V$ provides a custodial symmetry, which protects the $T$ parameter from receiving excessively large contributions \cite{Agashe:2003zs,Csaki:2003zu}. This symmetry breaking is accomplished by means of a Higgs field transforming as a bi-doublet under the two $SU(2)$ symmetries. The additional $P_{LR}$ symmetry, which interchanges the two $SU(2)$ groups, protects the left-handed $Zb\bar b$ coupling from receiving large modifications \cite{Agashe:2006at}. On the UV brane, the symmetry breaking $SU(2)_R\times U(1)_X\to U(1)_Y$ generates the SM gauge group. The symmetry breaking to $U(1)_{\rm EM}$ is implemented by means of an interplay of the UV and IR boundary conditions. Thorough discussions of this model containing many technical details have been presented in \cite{Casagrande:2010si,Albrecht:2009xr}, and we will adopt the notations of the first paper throughout our analysis.

The fermion representations we adopt are chosen such that they can be embedded into complete $SO(5)$ multiplets used in the context of models with gauge-Higgs unification \cite{Contino:2006qr,Carena:2007ua,Medina:2007hz}. As a consequence of the discrete $P_{LR}$ symmetry, which is instrumental in protecting the left-handed $Zb\bar b$ coupling \cite{Agashe:2006at} and its flavor-changing counterparts \cite{Blanke:2008zb}, the left-handed bottom quark has to be embedded in a $SU(2)_L\times SU(2)_R$ bi-doublet with isospin quantum numbers $T_L^3=-T_R^3=-1/2$. This fixes the quantum numbers of the other fields uniquely. In particular, the right-handed down-type quarks have to be embedded in an $SU(2)_R$ triplet in order to obtain an $U(1)_X$-invariant Yukawa coupling. One arrives at the following multiplet structure for the quark fields with even $Z_2$ parity:
\begin{equation}
\begin{aligned}
   Q_L &= \left( \begin{array}{cc}
     {u_L^{(+)}}_{\frac 23} & {\lambda_L^{(-)}}_{\frac 53} \\
     {d_L^{(+)}}_{-\frac 13} & {u_L^{\prime\,(-)}}_{\frac 23}
    \end{array} \right)_{\!\!\frac 23} , \qquad 
    u_R^c = \left( {u_R^{c\,(+)}}_{\frac 23} \right)_{\frac 23} , \\
    {\cal T}_R = {\cal T}_{1R}\oplus{\cal T}_{2R}
    &= \left( \begin{array}{c}
      {\Lambda_R^{\prime\,(-)}}_{\frac 53} \\
      {U_R^{\prime\,(-)}}_{\frac 23} \\
      {D_R^{\prime\,(-)}}_{-\frac 13}
     \end{array} \right)_{\!\!\frac 23}
     \oplus \left( {D_R^{(+)}}_{-\frac 13} \quad {U_R^{(-)}}_{\frac 23} \quad 
      {\Lambda_R^{(-)}}_{\frac 53} \right)_{\frac 23} .
\end{aligned}
\end{equation}
$Q_L$ is a bi-doublet under $SU(2)_L\times SU(2)_R$, while ${\cal T}_R$ transforms as $(\bm{3},\bm{1})\oplus(\bm{1},\bm{3})$. The fields with odd $Z_2$ parity have the opposite chirality. Their profiles are related to those of the $Z_2$-even fields by the field equations. The inner and outer subscripts on the various fields denote their $U(1)_{\rm EM}$ and $U(1)_X$ charges, respectively, which are connected through the relations $Y=-T_R^3+Q_X$ and $Q=T_L^3+Y$. 

The superscripts on the fields specify the type of boundary conditions they obey on the UV boundary. Fields with superscript $(+)$ obey the usual mixed boundary conditions allowing for a light zero mode, meaning that we impose the Dirichlet condition $\bm{S}_n^{A(+)}(\epsilon)=0$ on the profile functions of the corresponding $Z_2$-odd fields. These zero modes correspond to the SM quarks.\footnote{Note that the notation $u_L$, $d_L$, $u_R^c$, $D_R$ for these fields adopted here differs from the notation $U_L$, $D_L$, $u_R$, $d_R$ we used for the minimal RS model.} 
Fields with superscripts $(-)$ correspond to heavy, exotic fermions with no counterparts in the SM. For these states, the Dirichlet boundary condition is imposed on the $Z_2$-even fields (this means imposing the conditions $\bm{C}_n^{A(-)}(\epsilon)=0$ on the profile functions) in order to avoid the presence of a zero mode. The UV boundary conditions for the profiles $\bm{S}_n^{A(-)}(t)$ and $\bm{C}_n^{A(+)}(t)$ are of mixed type and follow from the field equations. We do not explicitly show the boundary conditions on the IR brane, which in the presence of a regularized Higgs profile are of Dirichlet type for all fields, $\bm{S}_n^{A(\pm)}(1^-)=0$. 

Note that we have chosen the same $SU(2)_L \times SU(2)_R$ representations for all three quark generations, which is necessary if one wants to consistently incorporate quark mixing in the fully anarchic approach to flavor in warped extra dimensions. The chosen representations also play a crucial role in the suppression of flavor-changing, left-handed $Z$-boson couplings \cite{Blanke:2008zb,Casagrande:2010si}. Altogether, there are fifteen different quark states in the up sector and nine in the down sector. The boundary conditions give rise to three light modes in each sector, which are identified with the SM quarks. These are accompanied by KK towers consisting of groups of fifteen and nine modes of similar masses in the up and down sectors, respectively. In addition, there is a KK tower of exotic fermion states with electric charge 5/3, which exhibits nine excitations in each KK level. 

In order to simplify the notation as much as possible, it is convenient to introduce the vectors
\begin{equation}
  \vec U = \bigg( \begin{array}{c} u \\ u' \end{array} \bigg) \,, \hspace{7mm} 
  \vec u = \left( \begin{array}{c} u^c \\ U' \\ U \end{array} \right) , \hspace{7mm}  
  \vec D = d \,, \hspace{7mm} 
  \vec d = \bigg( \begin{array}{c} D \\ D' \end{array} \bigg) \,, \hspace{7mm} 
  \vec\Lambda = \lambda \,, \hspace{7mm} 
  \vec\lambda = \bigg( \begin{array}{c} \Lambda' \\ \Lambda \end{array} \bigg) \,,
\end{equation}
which collect the fields with same electric charges (2/3, $-1/3$, and 5/3). Upper-case (lower-case) symbols denote fields whose left-handed (right-handed) components are $Z_2$ even. The corresponding matrices of bulk mass parameters are
\begin{equation}\label{ciparams}
\begin{aligned}
   \bm{c}_{\vec U} 
   &= \mbox{diag}\big( \bm{c}_Q ,\, \bm{c}_Q \big) \,, &\,\quad
   \bm{c}_{\vec D} &= \bm{c}_Q \,, &\,\quad
   \bm{c}_{\vec\Lambda} &= \bm{c}_Q \,, \\
   \bm{c}_{\vec u}
   &= \mbox{diag}\big( \bm{c}_{u^c} ,\, \bm{c}_{\tau_1} ,\, \bm{c}_{\tau_2} \big) \,, &\,\quad
   \bm{c}_{\vec d}
   &= \mbox{diag}\big( \bm{c}_{\tau_2} ,\, \bm{c}_{\tau_1} \big) \,, &\,\quad
  \bm{c}_{\vec\lambda}
   &= \mbox{diag}\big( \bm{c}_{\tau_1} ,\, \bm{c}_{\tau_2} \big) \,,
\end{aligned}    
\end{equation}
where each entry is a $3\times 3$ diagonal matrix in generation space. Note that the fields $\vec U$, $\vec D$, and $\vec\Lambda$ are governed by the same bulk mass matrix $\bm{c}_Q$, while $\vec u$, $\vec d$, and $\vec\lambda$ are associated with three different mass matrices $\bm{c}_{u^c}$, $\bm{c}_{\tau_2}$, and $\bm{c}_{\tau_1}$. The first two of them, $\bm{c}_{u^c}\equiv\bm{c}_u$ and $\bm{c}_{\tau_2}\equiv\bm{c}_d$, can be identified with the mass matrices appearing in the minimal RS model. The three new parameters contained in the matrix $\bm{c}_{\tau_1}$ can be related to the other ones by extending the $P_{LR}$ symmetry to the part of the quark sector that mixes with the left-handed down-type zero modes, by requiring that the action be invariant under the exchange of the fields $D'$ and $D$ \cite{Casagrande:2010si}. This extended version of the $P_{LR}$ symmetry implies
\begin{equation}\label{extended}
   \bm{c}_{\tau_1} = \bm{c}_{\tau_2} \,,
\end{equation}
and hence the number of independent bulk mass parameters is reduced to the same number as in the minimal RS model. Whether or not this equation holds will turn out to be largely irrelevant to our discussion.

In generalization of (\ref{Lhqq}), we now collect all left- and right-handed fields in the up, down, and exotic sectors into the 15-component vectors $(\vec U_A,\vec u_A)^T$ and the 9-component vectors $(\vec D_A,\vec d_A)^T$ and $(\vec\Lambda_A,\vec\lambda_A)^T$ (with $A=L,R$), to which we will  collectively refer as ${\cal Q}_{L,R}$. The Yukawa couplings of the Higgs boson to these fields can then be written in the form
\begin{equation}\label{Yuknew}
   {\cal L}_{hqq}(x) 
   = - \sum_{q=u,d,\lambda}\,\int_\epsilon^1\!dt\,\delta_h^\eta(t-1)\,
    h(x)\,\bar{\cal Q}_L(t,x)\,
    \frac{1}{\sqrt2}\,\bigg( \begin{array}{cc} 0 & \bm{Y}_{\vec q} \\ 
                             \bm{Y}_{\vec q}^\dagger & 0 \end{array} \bigg)\,
    {\cal Q}_R(t,x) + \mbox{h.c.} \,,
\end{equation}
where
\begin{equation}
   \bm{Y}_{\vec u} = \left( \begin{array}{ccc}
    \bm{Y}_u \,&\, \frac{1}{\sqrt2}\,\bm{Y}_d & \frac{1}{\sqrt2}\,\bm{Y}_d \\
    \bm{Y}_u \,&\, - \frac{1}{\sqrt2}\,\bm{Y}_d \,&\, - \frac{1}{\sqrt2}\,\bm{Y}_d 
    \end{array} \right) , \qquad
   \bm{Y}_{\vec d} = \bm{Y}_{\vec\lambda} 
   = \big( \bm{Y}_d \quad \bm{Y}_d \big)
\end{equation}
denote the corresponding $6\times 9$ and $3\times 6$ Yukawa matrices. The $3\times 3$ block matrices $\bm{Y}_q$ entering these expressions are the same as in the minimal RS model. Even though the extended RS model with custodial symmetry has a much richer structure than the minimal model, it thus features the same number of parameters in the fermion sector, once relation (\ref{extended}) is imposed.
 
With all the notation in place, we are now ready to generalize the analysis presented in the previous sections to the extended RS model with custodial symmetry. Since the Yukawa interactions (\ref{Yuknew}) have the same structure as in (\ref{Lhqq}), and since the boundary conditions on the IR brane are the same as in the minimal model, the only difference in the solution of the differential equations (\ref{coupledeqs}) concerns the UV boundary conditions imposed on the propagator functions. While the boundary conditions for fields with superscript $(+)$ give rise to the particular combination of Bessel functions defined in (\ref{RAdef}), $\bm{R}_A^{(+)}(\hat p_E)\equiv\bm{R}_A(\hat p_E)$, the corresponding linear combination for fields with superscript $(-)$ is given by
\begin{equation}\label{Rminus}
   \bm{R}_A^{(-)}(\hat p_E)
   = \frac{I_{\bm{c}_A-\frac12}(\epsilon\hat p_E)\,I_{-\bm{c}_A+\frac12}(\hat p_E)
           - I_{-\bm{c}_A+\frac12}(\epsilon\hat p_E)\,I_{\bm{c}_A-\frac12}(\hat p_E)}%
           {I_{\bm{c}_A-\frac12}(\epsilon\hat p_E)\,I_{-\bm{c}_A-\frac12}(\hat p_E)
           - I_{-\bm{c}_A+\frac12}(\epsilon\hat p_E)\,I_{\bm{c}_A+\frac12}(\hat p_E)}
   = \frac{1}{\bm{R}_A^{(+)}(\hat p_E)}\,\bigg|_{\bm{c}_A\to -\bm{c}_A} \,.
\end{equation}
Apart from this effect, we find that the central results (\ref{IplImi}) remain valid if we extend the sum over flavors appropriately, i.e.\
\begin{equation}\label{Iplcust}
   I_+(m^2) = \sum_{q=u,d,\lambda} \left\{ \mbox{Tr}\,g(\bm{X}_{\vec q}) 
    + \frac12\,\mbox{Tr} \left[ \frac{2\bm{X}_{\vec q}}{\sinh2\bm{X}_{\vec q}}\,
    \bigg( \frac{\bm{Z}_{\vec q}(-m^2)}{1 + \bm{Z}_{\vec q}(-m^2)}
    + \frac{\bm{Z}_{\vec q}^\dagger(-m^2)}{1 + \bm{Z}_{\vec q}^\dagger(-m^2)} \bigg) \right] 
    \right\} ,
\end{equation}
and similarly for $I_-(m^2)$. In analogy to (\ref{Zdef}), the matrices $\bm{Z}_{\vec q}(p_E^2)$ are given by
\begin{equation}\label{Zdefcust}
   \bm{Z}_{\vec q}(p_E^2) 
   = \frac{v^2}{2M_{\rm KK}^2}\,\frac{\tanh\bm{X}_{\vec q}}{\bm{X}_{\vec q}}\,
    \bm{Y}_{\vec q}\,\bm{R}_{\vec q}(\hat p_E)\,\bm{Y}_{\vec q}^\dagger\,
    \frac{\tanh\bm{X}_{\vec q}}{\bm{X}_{\vec q}}\,\bm{R}_{\vec Q}(\hat p_E) \,,
\end{equation}
where
\begin{equation}\label{eq85}
\begin{aligned}
   \bm{R}_{\vec U}
   &= \mbox{diag}\big( \bm{R}_Q^{(+)} ,\, \bm{R}_Q^{(-)} \big) \,, &\,\quad
   \bm{R}_{\vec D}
   &= \bm{R}_Q^{(+)} \,, &\,\quad
   \bm{R}_{\vec\Lambda}
   &= \bm{R}_Q^{(-)} \,, \\   
   \bm{R}_{\vec u}
   &= \mbox{diag}\big( \bm{R}_{u^c}^{(+)} ,\, \bm{R}_{\tau_1}^{(-)} ,\, \bm{R}_{\tau_2}^{(-)}
    \big) \,, &\,\quad
   \bm{R}_{\vec d}
   &= \mbox{diag}\big( \bm{R}_{\tau_2}^{(+)} ,\, \bm{R}_{\tau_1}^{(-)} \big) \,, &\,\quad
   \bm{R}_{\vec\lambda}
   &= \mbox{diag}\big( \bm{R}_{\tau_1}^{(-)} ,\, \bm{R}_{\tau_2}^{(-)} \big) \,,
\end{aligned}    
\end{equation}
which resembles the structure of the bulk mass matrices in (\ref{ciparams}). For simplicity of notation, we have omitted the argument $\hat p_E$ of the various $\bm{R}_A^{(\pm)}$ matrices. 

The relevant squared Yukawa matrices entering the quantities $\bm{X}_{\vec q}$ in (\ref{Iplcust}) and (\ref{Zdefcust}), which are defined in analogy with (\ref{Xqdef}), are given by the $6\times 6$ matrix
\begin{equation}
\begin{aligned}
   \bm{Y}_{\vec u}^{} \bm{Y}_{\vec u}^\dagger
   &= \left( \begin{array}{cc}
    \bm{Y}_u \bm{Y}_u^\dagger \!+\! \bm{Y}_d \bm{Y}_d^\dagger 
    ~&~ \bm{Y}_u \bm{Y}_u^\dagger \!-\! \bm{Y}_d \bm{Y}_d^\dagger \\
    \bm{Y}_u \bm{Y}_u^\dagger \!-\! \bm{Y}_d \bm{Y}_d^\dagger 
    ~&~ \bm{Y}_u \bm{Y}_u^\dagger \!+\! \bm{Y}_d \bm{Y}_d^\dagger
    \end{array} \right) 
   = \bm{V} \left( \begin{array}{cc}
    2\bm{Y}_d \bm{Y}_d^\dagger \,&\, 0 \\
    0 \,&\, 2\bm{Y}_u \bm{Y}_u^\dagger 
    \end{array} \right) \bm{V}^\dagger \,, \\
   &\hspace{25mm}\mbox{with} \quad
   \bm{V} = \bm{V}^\dagger =
    \frac{1}{\sqrt2}\,\bigg( \begin{array}{cc}
    -1 \,&\, 1 \\ 1 \,&\, 1 
    \end{array} \bigg) \,,
\end{aligned}    
\end{equation}
and the $3\times 3$ matrices $\bm{Y}_{\vec d}^{}\bm{Y}_{\vec d}^\dagger=\bm{Y}_{\vec\lambda}^{}\bm{Y}_{\vec\lambda}^\dagger=2\bm{Y}_d\bm{Y}_d^\dagger$. It follows that
\begin{equation}\label{gsumcust}
   \sum_{q=u,d,\lambda}\,\mbox{Tr}\,g(\bm{X}_{\vec q}) 
   = \mbox{Tr}\,g\big(\sqrt2\bm{X}_u\big) + 3\,\mbox{Tr}\,g\big(\sqrt2\bm{X}_d\big) \,,
\end{equation}
where the final answer is now expressed in terms of traces over the same $3\times 3$ matrices $\bm{X}_q$ as in the minimal RS model. 

Our next task is to reduce also the second term in (\ref{Iplcust}) to traces over $3\times 3$ matrices. From the definition (\ref{Zdefcust}), it is straightforward to derive that
\begin{equation}\label{Zvecs}
\begin{aligned}
   \bm{Z}_{\vec u}(p_E^2) 
   &= \frac{v^2}{2M_{\rm KK}^2}\,\bm{V}
    \left( \begin{array}{cc}
    \tilde{\bm{Y}}_d\,\big[ \bm{R}_{\tau_1}^{(-)} 
    + \bm{R}_{\tau_2}^{(-)} \big]\,\tilde{\bm{Y}}_d^\dagger & 0 \\
    0 & 2\tilde{\bm{Y}}_u\,\bm{R}_{u^c}^{(+)}\,\tilde{\bm{Y}}_u^\dagger
    \end{array} \right) \bm{V}^\dagger 
    \left( \begin{array}{cc}
    \bm{R}_Q^{(+)} & 0 \\
    0 & \bm{R}_Q^{(-)}
    \end{array} \right) , \\
   \bm{Z}_{\vec d}(p_E^2) 
   &= \frac{v^2}{2M_{\rm KK}^2}\,\tilde{\bm{Y}}_d \left[ 
    \bm{R}_{\tau_2}^{(+)} + \bm{R}_{\tau_1}^{(-)} \right]
    \tilde{\bm{Y}}_d^\dagger\,\bm{R}_Q^{(+)} \,, \\
   \bm{Z}_{\vec\lambda}(p_E^2) 
   &= \frac{v^2}{2M_{\rm KK}^2}\,\tilde{\bm{Y}}_d \left[ 
    \bm{R}_{\tau_1}^{(-)} + \bm{R}_{\tau_2}^{(-)} \right]
    \tilde{\bm{Y}}_d^\dagger\,\bm{R}_Q^{(-)} \,,
\end{aligned}
\end{equation}
where again we have omitted the argument $\hat p_E$ of the $\bm{R}_A^{(\pm)}$ matrices on the right-hand side of the equations. In the custodial model, the modified Yukawa matrices are defined as $\tilde{\bm{Y}}_q=\big[\!\tanh(\sqrt2\bm{X}_q)/(\sqrt2\bm{X}_q)\big]\,\bm{Y}_q$ \cite{Casagrande:2010si}, with an extra factor of $\sqrt2$ inserted compared with the minimal model. In (\ref{C1C5res}), we need to evaluate the result (\ref{Iplcust}) for values $|p_E^2|\ll M_{\rm KK}^2$. Using the expansion in (\ref{RAsmall}), we obtain after a straightforward calculation (again with $\bm{x}_q=\bm{m}_{q,0}/M_{\rm KK}$)
\begin{equation}
\begin{aligned}
   \bm{V}^\dagger\bm{Z}_{\vec u}(p_E^2)\,\bm{V} 
   &= F^{-1}(\bm{c}_Q)\,\bm{U}_u\,\Bigg\{ 
    \left[ \frac{\bm{m}_{u,0}^2}{p_E^2}
    + \left( \bm{\Phi}_U + \bm{m}_{u,0}\,\bm{\Phi}_u\,\bm{m}_{u,0}^{-1} \right) + \dots \right]
    \bigg( \begin{array}{rc} 0 ~&\, 0 \\ \!-1 ~&\, 1 \end{array} \bigg) \\
   &\quad\mbox{}+ \bm{V}_{\rm CKM}\,\bm{x}_d\,\bm{W}_d^\dagger\,
    \frac{1}{2F^2(\bm{c}_{\tau_2})} \left[ \frac{1}{F^2(-\bm{c}_{\tau_1})} 
    + \frac{1}{F^2(-\bm{c}_{\tau_2})} \right]
    \bm{W}_d\,\bm{x}_d\,\bm{V}_{\rm CKM}^\dagger\,
    \bigg( \begin{array}{cc} 1 \,&\, -1 \\ 1 \,&\, -1 \end{array} \bigg) \\
   &\quad\mbox{}+ \bm{x}_u^2\,\bm{U}_u^\dagger\,
    \frac{2}{F^2(\bm{c}_Q)\,F^2(-\bm{c}_Q)}\,\bm{U}_u\,
    \bigg( \begin{array}{cc} 0 \,&\, 0 \\ 1 \,&\, 0 \end{array} \bigg) 
    + \dots \Bigg\}\,\bm{U}_u^\dagger\,F(\bm{c}_Q) \,, \\
   \bm{Z}_{\vec d}(p_E^2) 
   &= F^{-1}(\bm{c}_Q)\,\bm{U}_d \left[ \frac{\bm{m}_{d,0}^2}{p_E^2}
    + \left( \bm{\Phi}_D + \bm{m}_{d,0}\,\bm{\Phi}_d\,\bm{m}_{d,0}^{-1} \right) 
    + \dots \right] \bm{U}_d^\dagger\,F(\bm{c}_Q) \,,
\end{aligned}
\end{equation}
where $\bm{V}_{\rm CKM}=\bm{U}_u^\dagger\,\bm{U}_d$ is the CKM mixing matrix. The terms shown explicitly above are of leading and subleading order in $v^2/M_{\rm KK}^2$. To this order, the quantity $\bm{Z}_{\vec\lambda}(p_E^2)$ vanishes. The quantities $\bm{\Phi}_A$ are generalizations of the matrices $\bm{\delta}_A$ given in (\ref{deltadef}). They are defined as \cite{Casagrande:2010si}
\begin{equation}\label{Phidef}
\begin{aligned}
   \bm{\Phi}_U &= \bm{x}_u\,\bm{W}_u^\dagger \left[
    \frac{1}{1-2\bm{c}_u} \left( \frac{1}{F^2(\bm{c}_u)} 
    - 1 + \frac{F^2(\bm{c}_u)}{3+2\bm{c}_u} \right) \right] \bm{W}_u\,\bm{x}_u \\
   &\quad\mbox{}+ \bm{V}_{\rm CKM}\,\bm{x}_d\,\bm{W}_d^\dagger\,
    \frac{1}{2F^2(\bm{c}_{\tau_2})} \left[ \frac{1}{F^2(-\bm{c}_{\tau_1})} 
    + \frac{1}{F^2(-\bm{c}_{\tau_2})} \right] \bm{W}_d\,\bm{x}_d\,
    \bm{V}_{\rm CKM}^\dagger \,, \\
   \bm{\Phi}_u &= \bm{x}_u\,\bm{U}_u^\dagger \left[
    \frac{1}{1-2\bm{c}_Q} \left( \frac{1}{F^2(\bm{c}_Q)} 
    \left[ 1 + \frac{1-2\bm{c}_Q}{F^2(-\bm{c}_Q)} \right]    
    - 1 + \frac{F^2(\bm{c}_Q)}{3+2\bm{c}_Q} \right) \right] 
    \bm{U}_u\,\bm{x}_u \,, \\
   \bm{\Phi}_D &= \bm{x}_d\,\bm{W}_d^\dagger \left[
    \frac{1}{1-2\bm{c}_{\tau_2}} \left( \frac{1}{F^2(\bm{c}_{\tau_2})} 
    \left[ 1 + \frac{1-2\bm{c}_{\tau_2}}{F^2(-\bm{c}_{\tau_1})} \right]
    - 1 + \frac{F^2(\bm{c}_{\tau_2})}{3+2\bm{c}_{\tau_2}} \right) \right] 
    \bm{W}_d\,\bm{x}_d \,, \\
   \bm{\Phi}_d &= \bm{x}_d\,\bm{U}_d^\dagger \left[
    \frac{1}{1-2\bm{c}_Q} \left( \frac{1}{F^2(\bm{c}_Q)} - 1 
    + \frac{F^2(\bm{c}_Q)}{3+2\bm{c}_Q} \right) \right] 
    \bm{U}_d\,\bm{x}_d \,.  
\end{aligned}
\end{equation}
After a lengthy calculation, we find that in analogy with (\ref{trZ})
\begin{equation}
   \sum_{q=u,d,\lambda} \mbox{Tr} \left[ \frac{2\bm{X}_{\vec q}}{\sinh2\bm{X}_{\vec q}}\,
    \frac{\bm{Z}_{\vec q}(p_E^2)}{1 + \bm{Z}_{\vec q}(p_E^2)} \right]
   = \sum_i\,\bigg[ \kappa_{q_i}\,\frac{m_{q_i}^2}{m_{q_i}^2+p_E^2} + \varepsilon_{q_i} \bigg] 
    + \dots \,,
\end{equation}
where
\begin{equation}\label{level42}
   \kappa_{q_i} = 1 - \varepsilon_{q_i} - \frac23\,\Big[
    \bm{U}_q^\dagger\,F(\bm{c}_Q)\,2\bm{X}_q^2\,F^{-1}(\bm{c}_Q)\,\bm{U}_q \Big]_{ii}
\end{equation}
now contains an extra factor of 2 in the last term compared with the result (\ref{kappadef}) for the minimal model, while
\begin{equation}
   \varepsilon_{q_i}
   = \left( \bm{\Phi}_Q \right)_{ii} + \left( \bm{\Phi}_q \right)_{ii} .
\end{equation}

We are now ready to present our final expressions for the Wilson coefficients $C_1$ and $C_5$ in the RS model with custodial symmetry. To an excellent approximation, we obtain instead of~(\ref{Ciresu})
\begin{equation}\label{C1C5custo}
\begin{aligned}
   C_1 &\approx \Bigg[ 1 - \frac{2v^2}{3M_{\rm KK}^2}\,\mbox{Re}\,
    \frac{\left( \bm{Y}_u \bm{Y}_u^\dagger \bm{Y}_u \right)_{33}}{\left( \bm{Y}_u \right)_{33}}
    \Bigg]\,A(\tau_t) + A(\tau_b)
    + \mbox{Tr}\,g\big(\sqrt2\bm{X}_u\big) + 3\,\mbox{Tr}\,g\big(\sqrt2\bm{X}_d\big) \,, \\
   C_5 &\approx - \frac{2v^2}{3M_{\rm KK}^2}\,\mbox{Im} \Bigg[
    \frac{\left( \bm{Y}_u \bm{Y}_u^\dagger \bm{Y}_u \right)_{33}}{\left( \bm{Y}_u \right)_{33}}
    \Bigg]\,B(\tau_t) \,,
\end{aligned}
\end{equation}
which once again is independent of the bulk mass parameters $c_i$. We find that this approximation is accurate to better than 2\% for $M_{\rm KK}\gtrsim 2$\,TeV. Whereas the small corrections parameterized by $\kappa_{q_i}$ and $\varepsilon_{q_i}$ have only a minor impact, the main difference between the minimal and the custodial RS models consists in the different multiplicity factors in the trace terms in (\ref{Ciresu}) and (\ref{C1C5custo}). Since the functions $g(\bm{X}_q)$ start with a quadratic term, we must compare $\bm{X}_u^2+\bm{X}_d^2$ in the minimal model with the combination $2\bm{X}_u^2+6\bm{X}_d^2$ in the custodial model. Since we assume that the 5D Yukawa matrices in the up- and down-type quark sectors are random matrices of similar magnitude, it follows that the effect of the KK modes in the custodial model is approximately four times as large as in the minimal model.\footnote{Based on a naive counting of degrees of freedom, this factor was estimated as 11/4 (instead of 4) in \cite{Goertz:2011hj}.}

\section{Phenomenological implications}
\label{sec:pheno}

We now present a numerical study of our results for both the minimal RS model and its extension with custodial symmetry. In each case, we distinguish the two cases of a brane-localized scalar sector and a narrow bulk-Higgs scenario. At the end of this section, we also discuss the generalization of the brane-Higgs scenario with two different Yukawa matrices, which was discussed in Section~\ref{sec:2Yukawas}. For the purposes of our discussion here, we will include the possible effects of the power-suppressed, higher-dimensional $|\Phi|^2 (G_{\mu\nu}^a)^2$ operators, which give rise to the effective Lagrangian (\ref{Leff}), by treating the coefficient $c_{\rm eff}$ as a random variable, whose value is scanned between $-1$ and 1 using a flat distribution. As discussed in detail in Section~\ref{sec:powerhgg}, the numerical impact of such operators is very small provided that $c_{\rm eff}\ll 3.8\,y_*^2$ in the minimal RS model and $c_{\rm eff}\ll 15.2\,y_*^2$ in the custodial RS model, where $y_*$ is the upper bound on the magnitudes of the complex entries of the random 5D Yukawa matrices, see (\ref{eq87}). 

Based on the expressions obtained in Sections~\ref{sec:analysis} and \ref{sec:custodial}, we evaluate the Higgs-boson production cross section via gluon fusion relative to the SM cross section \cite{Carena:2012fk},
\begin{equation}\label{eqn:ratio_h}
   R_h = \frac{\sigma(gg\to h)_{\rm RS}}{\sigma(gg\to h)_{\rm SM}} 
   = \frac{|\kappa_g|^2+|\kappa_{g5}|^2}{\kappa_v^2} \,,
\end{equation}
where $\kappa_g$ and $\kappa_{g5}$ parametrize the values of the Wilson coefficients, normalized to the SM value $C_1^{\rm SM}=A(\tau_t)+A(\tau_b)$, such that $\kappa_g=C_1/C_1^{\rm SM}$ and $\kappa_{g5}=\frac{3}{2}\,C_5/C_1^{\rm SM}$. The quantity $\kappa_v$ in (\ref{eqn:ratio_h}) denotes the shift of the Higgs vev $v$ in the RS model relative to the value $v_{\rm SM}$ of the SM \cite{Bouchart:2009vq}. We determine $\kappa_v$ from the shift to the Fermi constant derived in the RS model by considering (at tree level) the effect of the exchange of the infinite tower of KK gauge bosons on the rate for muon decay.\footnote{If one uses instead the shift on the value of the $W$-boson mass, one finds some additional contributions not enhanced by a factor of $L$, which are numerically insignificant \cite{Casagrande:2008hr}.} 
Using the definition $v_{\rm SM}=(\sqrt2 G_F)^{-1/2}$ along with results derived in \cite{Casagrande:2008hr}, we then obtain to first order in $v^2/M_{\rm KK}^2$
\begin{equation}\label{kappav}
   \kappa_v \big|_{\rm minimal~RS}= \frac{v}{v_{\rm SM}}
   \approx 1 + \frac{L m_W^2}{4 M_{\rm KK}^2} \,, \qquad
   \kappa_v \big|_{\rm custodial~RS} = \frac{v}{v_{\rm SM}}
   \approx 1 + \frac{L m_W^2}{2 M_{\rm KK}^2} \,, 
\end{equation}
where $m_W=gv/2$ is the lowest-order expression for the mass of the $W$ boson, and $L=-\ln\epsilon=\ln(M_{\rm Pl}/\Lambda_{\rm TeV})\sim 33\,\mbox{--}\,34$ for $\Lambda_{\rm TeV}\sim 20\,\mbox{--}\,50$\,TeV. In the custodial RS model with $P_{LR}$ symmetry, the correction is twice as large as in the minimal model.

Concerning the contributions from the infinite towers of KK quarks to $C_1$ and $C_5$, we need to evaluate the traces of the functions $g(\bm{X}_q)$ defined in (\ref{gfundef}) and (\ref{gfundef2}), which can be expanded in a power series in the positive matrix $\bm{X}_q^2={\cal O}(v^2/M_{\rm KK}^2)$. Keeping only the first term in this expansion, one encounters the quantity 
\begin{equation}\label{centrallimit}
   \mbox{Tr}\,\bm{X}_q^2 
   = \frac{v^2}{2M_{\rm KK}^2}\,\sum_{i,j=1}^{N_g} \left|(\bm{Y}_q)_{ij}\right|^2 
   \approx \frac{v^2}{2M_{\rm KK}^2}\,\frac{N_g^2\,y_*^2}{2} \,,
\end{equation}
where $N_g=3$ is the number of quark generations. In the last step we have used relation (\ref{eq87}), which states that on average $\langle\left|(\bm{Y}_q)_{ij}\right|^2\rangle=y_*^2/2$ for a complex random number.\footnote{In \cite{Carena:2012fk} the modulus and phase of the elements of the Yukawa matrices were chosen as random variables, in which case $\langle|(\bm{Y}_q)_{ij}|^2\rangle=y_*^2/3$.} 
It follows that, to good approximation, the effect of the KK tower of quark states scales proportional to the number of quark generations squared. While each entry of the Yukawa matrices $\bm{Y}_q$ is a randomly distributed complex number, the central limit theorem implies that the sum over the $N_g^2=9$ positive numbers in (\ref{centrallimit}) is (approximately) normally distributed about the average value shown in the equation. It is this fact which allows us to predict the Higgs-boson production rate to good accuracy in terms of only the two parameters $M_{\rm KK}$ and $y_*$ (see also \cite{Goertz:2011hj,Carena:2012fk}). This observation has an important implication, though. If we were interested in an observable depending on a {\em single\/} Yukawa coupling $|(\bm{Y}_q)_{ij}|$ (for some particular choice of $i$ and $j$), then scanning this parameter over all allowed values between 0 and the perturbativity bound $y_{\rm max}$ would cover the range of all possible results for the observable. We would not introduce a bias by allowing $|(\bm{Y}_q)_{ij}|$ to take values close to the upper bound. The situation is different in our case. Scanning $N_g^2=9$ random numbers $(\bm{Y}_q)_{ij}$ in the complex plane within a radius set by $y_*$, the sum of their absolute squares in (\ref{centrallimit}) will be described by a narrow gaussian distribution centered at $N_g^2\,y_*^2/2$. Even though it is mathematically allowed that this sum takes a value much smaller or much larger than this result (any value between 0 and $N_g^2\,y_*^2$ is possible), this will almost never happen in practice. It is thus necessary that we consider sets of model predictions for several different values of $y_*$, some close to the perturbativity bound $y_{\rm max}$ and some significantly smaller than it. For the numerical analysis, we generate three sets of 5000 random and anarchic 5D Yukawa matrices, whose entries satisfy $|(\bm{Y}_q)_{ij}|\le y_*$ with $y_*=0.5$, 1.5, and 3. As a further constraint, we impose that these matrices correctly reproduce the Wolfenstein parameters $\rho$ and $\eta$ of the unitarity triangle (see \cite{Casagrande:2008hr} for explicit formulae). This requirement helps to eliminate some outliers in the plots presented below. We also require that, with appropriately chosen bulk mass parameters $c_i$, one can reproduce the correct values for the masses of the SM quarks; however, imposing this condition only has a minor impact on our results.

\begin{figure}
\begin{center}
\psfrag{x}[]{\small $M_{g^{(1)}}~{\rm [TeV]}$}
\psfrag{y}[b]{\small $R_h$}
\psfrag{z}[]{\small \begin{tabular}{l} minimal RS model \\[-1mm] brane Higgs \end{tabular}}
\psfrag{w}[]{\small \begin{tabular}{l} minimal RS model \\[-1mm] bulk Higgs \end{tabular}}
\includegraphics[width=1.03\textwidth]{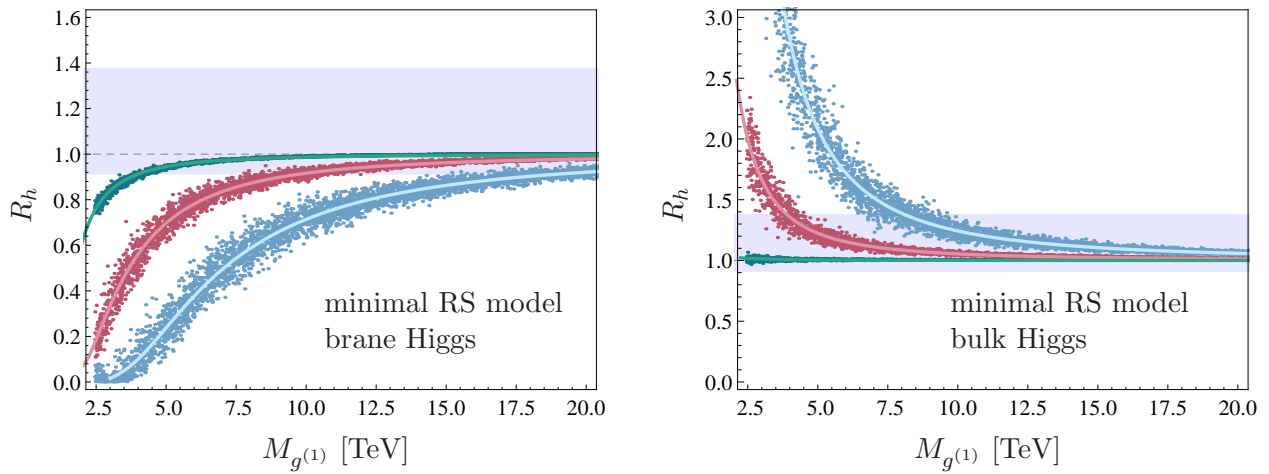}
\parbox{15.5cm}
{\caption{\label{fig:MRS}
Predictions for the ratio $R_h$ in the minimal RS model, for the cases of a brane-localized Higgs boson (left) and a narrow bulk-Higgs field (right). The green, red, and blue scatter points correspond to model points obtained using $y_*=0.5$, 1.5, and 3, respectively. The overlaid lines show fits to the various distributions. The area colored in blue represents the experimental $1\sigma$ band. See text for further explanation.}}
\end{center}
\end{figure}

Figure~\ref{fig:MRS} shows the results for the ratio $R_h$ defined in (\ref{eqn:ratio_h}) in the minimal RS model for the scenarios with a brane-localized Higgs boson (left) and a narrow bulk-Higgs field (right), in dependence of the mass $M_{g^{(1)}}$ of the lightest KK gluon state. We use the mass of the first excited gluon state as a reference, because it is more physical than the KK scale $M_{\rm KK}$, and because its value $M_{g^{(1)}}\approx 2.45\,M_{\rm KK}$ is a model-independent prediction of the RS models considered in this work. The green, red, and blue scatter points refer to the three different values of $y_*$. They have been obtained using the approximate expressions for the Wilson coefficients given in (\ref{Ciresu}), but at the scale of the plots they are indistinguishable from the results one would obtain using the exact expressions in (\ref{C1C5res}) and (\ref{IplImi}). We use $m_h=126$\,GeV for the mass of the Higgs boson, and $m_t=172.6$\,GeV and $m_b=\overline{m}_b(m_h)=2.9$\,GeV for the masses of the third-generation quarks. While for the heavy top-quark it is appropriate to use the pole mass, a running mass should be used for the $b$-quark. We observe that the ratio $R_h$ is strictly below~1 for the case of a brane-localized Higgs sector, while it is larger than 1 for the case of a narrow bulk-Higgs state. This observation allows for a clear distinction between the two scenarios. Only for very small $y_*$, a few points exist for which $R_h$ lies slightly below~1. This effect is due to the modification of the Higgs vev in the RS model, which always gives rise to a negative contribution.

In order to compare our predictions with experiment, we consider the cross section for the process $pp\to h\to Z Z^{(*)}\to 4\ell$ measured at the LHC. Since $gg\to h$ is the dominant production channel, accounting for about 90\% of the events in the SM, and because corrections to the $hZZ$ coupling in RS models are in general very small \cite{Casagrande:2010si,Goertz:2011hj}, we assume that any deviation of the rate for this process from its SM value can be traced back to new-physics contributions to the gluon fusion amplitude. The ATLAS and CMS collaborations have recently reported updated results for the ratio $\mu_{ZZ}=\sigma(pp\to h\to Z Z^{(*)})/\sigma_{\rm SM}(pp\to h\to Z Z^{(*)})$, which were obtained using the full data set collected up to the end of 2012 (approximately 25~fb$^{-1}$). The observed values are $\mu_{ZZ}^{\rm ATLAS}=1.7\,_{-0.4}^{+0.5}$ (at $m_h=124.3$\,GeV) and $\mu_{ZZ}^{\rm CMS}=0.91\,_{-0.24}^{+0.30}$ (at $m_h=125.8$\,GeV) \cite{Moriond_results},\footnote{At $m_h=125.5$\,GeV, the ATLAS result is shifted to $\mu_{ZZ}^{\rm ATLAS}=1.5\pm 0.4$, which is closer to the CMS value and gives rise to the average result $\mu_{ZZ}=1.09\,_{-0.21}^{+0.24}$.} 
which we naively average to obtain $\mu_{ZZ}=1.12\,_{-0.21}^{+0.26}$. The $1\sigma$ range corresponding to this result is shown by the blue band in the two plots. In our analysis we will assume that $\mu_{ZZ}\approx R_h$, i.e., that any possible deviation from 1 is due to a modification of the production cross section of the Higgs boson in gluon fusion. Model points falling outside these bands are excluded at the 68\% confidence level (CL). While for small $y_*=0.5$ most model points are in agreement with the data, it is interesting to observe that for larger $y_*$ the data already disfavor KK gluon masses in the low TeV range. The discrepancies between theory and experiment are stronger for the brane-Higgs model, because the mild tendency of an enhanced production rate seen in the data is in conflict with the suppression of the cross section predicted in this case. 

The overlaid, solid lines in Figure~\ref{fig:MRS} show fits to the various distributions of model points. In regions of parameter space where the deviations of $R_h$ from 1 are modest enough in order to be compatible with the data, a good approximation to these curves can be obtained by approximating the functions $g(\bm{X}_q)$ in (\ref{gfundef}) and (\ref{gfundef2}) by the first terms in their Taylor expansions and exploiting the anarchy of the 5D Yukawa matrices. In this way we find
\begin{equation}\label{eqn:Rh_Min}
   R_h\approx  1 - \frac{v^2}{2 M_{\rm KK}^2} \left[
    \left( \pm 4 N_g^2 + \frac83\,N_g - \frac43 \right) \langle |(\bm{Y}_q)_{ij}|^2 \rangle 
    + \frac{L m_W^2}{v^2} \right] ,
\end{equation}
where the upper sign corresponds to the brane-localized Higgs sector and the lower sign to the narrow bulk-Higgs scenario. For randomly chosen complex elements of the Yukawa matrices, it follows that $\langle\left|(\bm{Y}_q)_{ij}\right|^2\rangle=y_*^2/2$. The terms in brackets then evaluate to approximately $[21.3\,y_*^2+3.6]$ for the RS model with a brane-localized Higgs, and $[-14.7\,y_*^2+3.6]$ for the model with a narrow bulk Higgs (with $L=33.5$). Relation (\ref{eqn:Rh_Min}) exhibits the quadratic dependency on the number of quark generations $N_g$ and on the maximum absolute value $y_*$ imposed on the entries of the random Yukawa matrices. 

\begin{figure}
\begin{center}
\psfrag{x}[]{\small $M_{g^{(1)}}~{\rm [TeV]}$}
\psfrag{y}[b]{\small $y_*$}
\psfrag{z}[]{\small \begin{tabular}{l} \\[-1mm] \hspace{-3mm}minimal RS model \\[-1mm] \hspace{-3mm}brane Higgs \end{tabular}}
\psfrag{w}[]{\small \begin{tabular}{l} minimal RS model \\[-1mm] bulk Higgs \end{tabular}}
\includegraphics[width=1.03\textwidth]{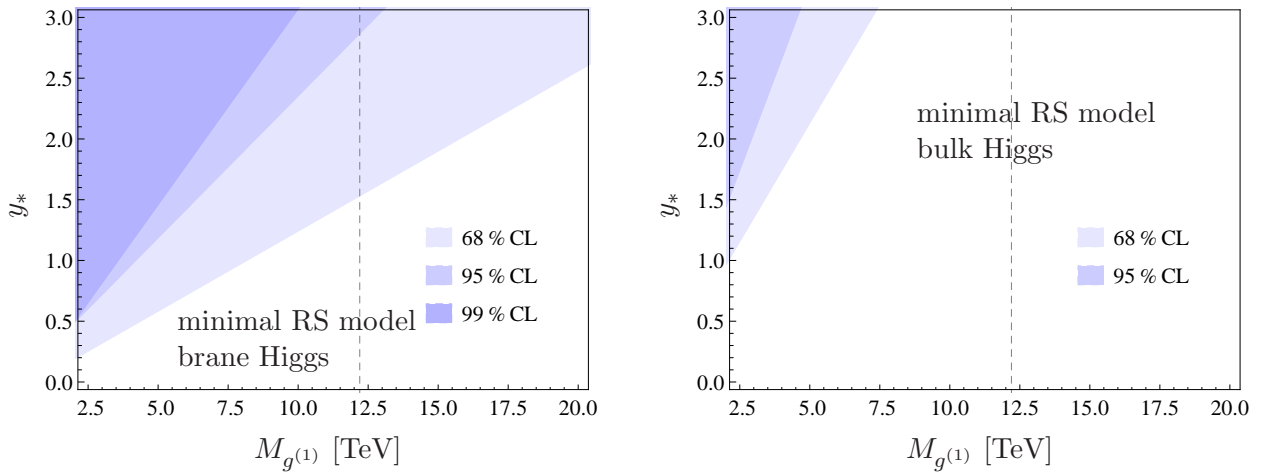}
\parbox{15.5cm}
{\caption{\label{fig:MRSexcl}
Excluded regions of parameter space in the minimal RS model, for the cases of a brane-localized Higgs boson (left) and a narrow bulk-Higgs field (right). The vertical dashed line shows the lower bound on $M_{g^{(1)}}$ obtained from a tree-level analysis of electroweak precision observables.}}
\end{center}
\end{figure}

Even at the present level of precision, the existing measurements of the Higgs-boson production cross section already provide highly non-trivial constraints on the parameter space of RS models. 
In Figure~\ref{fig:MRSexcl}, we show the regions in the $M_{g^{(1)}}$\,--\,$y_*$ parameter space which are already excluded by the current experimental data at various confidence levels. To obtain these regions, we first fit an approximately gaussian distribution to the model points shown in Figure~\ref{fig:MRS} for each pair of $M_{g^{(1)}}$ and $y_*$, and extract from it our theoretical prediction $R_h^{\rm th}$ and uncertainty $\Delta R_h^{\rm th}$ for these parameters. We then take the ratio $R_h^{\rm th}/R_h^{\rm exp}$, combine the theoretical and experimental errors in quadrature, and test at which confidence level this ratio is compatible with 1. In both versions of the RS model, the data exclude significant portions of the model parameter space. With the conventional choice $y_*=3$, for example, one finds $M_{g^{(1)}}>13$\,TeV for the brane-Higgs model and $M_{g^{(1)}}>4.5$\,TeV for the bulk-Higgs scenario, both at 95\% CL. Weaker constraints are obtained for smaller values of $y_*$. These bounds may be compared with those derived from the analysis of electroweak precision observables. The strongest constraint arises from the $S$ and $T$ parameters \cite{Peskin:1991sw}, whose present values are $S=0.03\pm 0.10$ and $T=0.05\pm 0.12$, with a correlation coefficient $\rho=0.89$ \cite{Baak:2012kk}. In the minimal RS model, one obtains at tree level \cite{Carena:2003fx}
\begin{equation}\label{SandT}
   S = \frac{2\pi v^2}{M_{\rm KK}^2} \left( 1 - \frac{1}{L} \right) , \qquad
   T = \frac{\pi v^2}{2\cos^2\theta_W\,M_{\rm KK}^2} \left( L - \frac{1}{2L} \right) .
\end{equation}
Requiring that these corrections are compatible with the experimental data, we find that $M_{g^{(1)}}>12$\,TeV at 95\% CL. This strong bound, which is indicated by the dashed line in Figure~\ref{fig:MRSexcl}, may however be weakened in several ways, for instance by including loop corrections, by reducing the size $L$ of the extra dimension (so-called ``little RS models'') \cite{Davoudiasl:2008hx}, or by introducing large brane-localized kinetic terms in the RS Lagrangian \cite{Carena:2003fx}. We note that for $M_{g^{(1)}}>12$\,TeV there is no significant flavor problem of the minimal RS model, as the tightest constraint from the $\epsilon_K$ parameter in $K$\,--\,$\bar K$ mixing \cite{Csaki:2008zd} can be satisfied with a modest 25\% fine-tuning \cite{Bauer:2011ah}.

\begin{figure}
\begin{center}
\psfrag{x}[]{\small $M_{g^{(1)}}~{\rm [TeV]}$}
\psfrag{y}[b]{\small $R_h$}
\psfrag{z}[]{\small \begin{tabular}{l} \hspace{-2mm}custodial model \\[-1mm] \hspace{-2mm}brane Higgs \end{tabular}}
\psfrag{w}[]{\small \begin{tabular}{l} \hspace{-2mm}custodial model \\[-1mm] \hspace{-2mm}bulk Higgs \end{tabular}}
\includegraphics[width=1.03\textwidth]{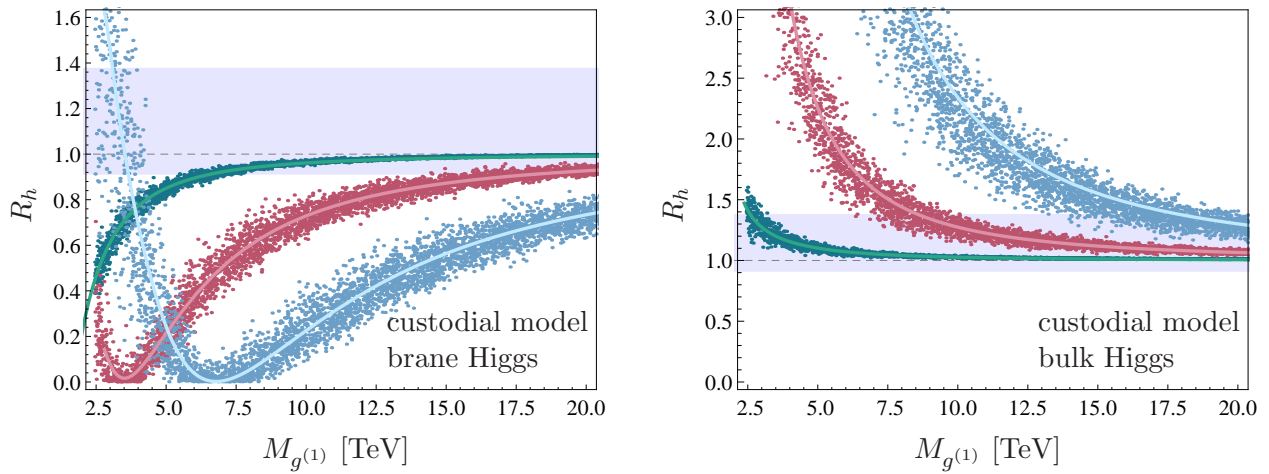}
\parbox{15.5cm}
{\caption{\label{fig:CRS}
Predictions for the ratio $R_h$ in the custodial RS model, for the cases of a brane-localized Higgs boson (left) and a narrow bulk-Higgs field (right). The meaning of the colors and curves is the same as in Figure~\ref{fig:MRS}.}}
\end{center}
\end{figure}

Softening the constraints from electroweak precision tests by means of a symmetry is the main motivation for extending the RS model by enlarging the gauge group in the bulk \cite{Agashe:2003zs,Csaki:2003zu,Agashe:2006at}. We now proceed to study the RS model with custodial symmetry, in which the Wilson coefficients $C_1$ and $C_5$ are given in (\ref{C1C5custo}). The corresponding numerical results are shown in Figure~\ref{fig:CRS}. For large masses $M_{g^{(1)}}$ we can derive analogously to (\ref{eqn:Rh_Min}) a formula for $R_h$ depending explicitly on $N_g$ and $y_*$, which in the present case reads
\begin{equation}\label{eq100}
   R_h\approx  1 - \frac{v^2}{2 M_{\rm KK}^2} \left[
    \left( \pm 16 N_g^2 + \frac{16}{3}\,N_g - \frac83 \right) \langle |(\bm{Y}_q)_{ij}|^2 \rangle 
    + \frac{2L m_W^2}{v^2} \right] .
\end{equation}
Note that the leading terms proportional to $N_g^2$ are enhanced by a factor 4 compared with the minimal model, reflecting the larger multiplicity of KK quark states. The remaining terms are enhanced by a factor 2, as can be seen from (\ref{level42}) and (\ref{kappav}). As a result, in the custodial RS model one finds significantly larger corrections to the SM prediction $R_h=1$ than in the minimal model \cite{Goertz:2011hj}. The terms in brackets then evaluate to approximately $[78.7\,y_*^2+7.1]$ for the RS model with a brane-localized Higgs, and $[-65.3\,y_*^2+7.1]$ for the model with a narrow bulk Higgs. For the same reason, the relative effect of higher-dimensional operators is suppressed compared with the minimal RS model. In relation~(\ref{ceffesti}), the right-hand side must be multiplied by a factor~4. 

\begin{figure}
\begin{center}
\psfrag{x}[]{\small $M_{g^{(1)}}~{\rm [TeV]}$}
\psfrag{y}[b]{\small $y_*$}
\psfrag{z}[]{\small \begin{tabular}{l} \\[-2mm] custodial model \\[-1mm] brane Higgs \\[-2mm] 
             \phantom{x} \end{tabular}}
\psfrag{w}[]{\small \begin{tabular}{l} custodial model \\[-1mm] bulk Higgs \end{tabular}}
\includegraphics[width=1.03\textwidth]{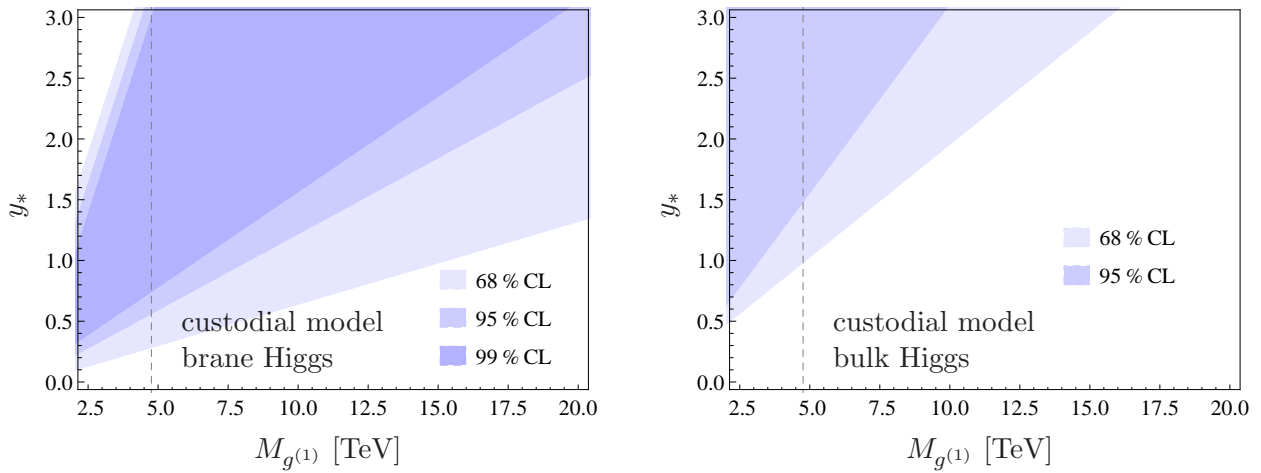}
\parbox{15.5cm}
{\caption{\label{fig:CRSexcl}
Excluded regions of parameter space in the custodial RS model, for the cases of a brane-localized Higgs boson (left) and a narrow bulk-Higgs field (right). The vertical dashed line shows the lower bound on $M_{g^{(1)}}$ obtained from a tree-level analysis of electroweak precision observables.}}
\end{center}
\end{figure}

Figure~\ref{fig:CRS} confirms the fact that the corrections to the Higgs-boson production rate are much enhanced compared with the case of the minimal RS model. Correspondingly, we obtain significantly larger exclusion regions than for the minimal model. This is shown in Figure~\ref{fig:CRSexcl}. In the brane-Higgs scenario, we obtain the exclusion range $4.5\,\mbox{TeV}<M_{g^{(1)}}<19$\,TeV for $y_*=3$ at 99\% CL, while in the bulk-Higgs model we find the lower bound $M_{g^{(1)}}>9.5$\,TeV at 95\% CL. Note that the allowed region in the upper left corner (at small $M_{g^{(1)}}$ and large $y_*$) of the first plot in the figure is one in which the new-physics contribution to the gluon fusion amplitude is larger than the SM contribution by about a factor 2 and interferes destructively, which appears somewhat unnatural. Moreover, it has been argued that most models in which the gluon fusion amplitude has the opposite sign than in the SM have problems with fine-tuning and vacuum stability \cite{Reece:2012gi}. The bounds on the RS parameter space that can be derived from Figure~\ref{fig:CRS} are stronger than those derived from the analysis of electroweak precision observables. In the custodial model the formula for the $S$ parameter shown in (\ref{SandT}) is left unchanged, while the custodial protection removes the leading term proportional to $L$ in the expression for the $T$ parameter, such that $T=-\pi v^2/(4L\cos^2\theta_W M_{\rm KK}^2)$ \cite{Agashe:2003zs}. Requiring that these corrections are compatible with the experimental data, we find that $M_{g^{(1)}}>4.7$\,TeV at 95\% CL. As indicated by the dashed line in Figure~\ref{fig:CRSexcl}, this lower bound is generally much weaker than the constraints implied by Higgs physics, except for regions in parameter space where $y_*$ is very small. Note that for such small values of the KK mass scale but $y_*\approx 3$, the RS flavor problem for the $\epsilon_K$ parameter can be solved by a fine-tuning of 5\,--\,10\%, or alternatively by enlarging the strong-interaction gauge group in the bulk \cite{Bauer:2011ah}. 

We may also read the exclusion plots in a different way. If we would like to have the first KK excitations in the reach for direct production at the LHC, then this imposes a strong upper bound on the maximum allowed values of the elements of the 5D Yukawa matrices. For instance, assuming that $M_{g^{(1)}}=5$\,TeV, we find that $y_*<0.6$ in the brane-Higgs model, and $y_*<1.5$ in the bulk-Higgs scenario (both at 95\% CL). Too small Yukawa couplings would however give rise to enhanced corrections to $\epsilon_K$ \cite{Csaki:2008zd}, and hence they would reinforce the RS flavor problem. 

The above analysis shows that Higgs physics, and in particular the Higgs-boson production rate in gluon fusion, provide sensitive probes of the virtual effects of KK excitations in the context of various RS scenarios. While models with a brane-localized scalar sector predict a suppression of the gluon fusion rate, this rate tends to be enhanced in scenarios with a bulk-Higgs field. The two classes of models can thus easily be distinguished in their signatures. The bounds on the model parameters obtained from Higgs physics are complementary to and sometimes stronger than those derived from the analysis of electroweak precision observables and rare flavor-changing processes. In models with a custodial protection of electroweak precision observables, the indirect effects of KK states on the Higgs-boson production rate are strongly enhanced compared with minimal RS models, and hence Higgs physics provides the strongest constraints in this case. Even under the pessimistic (but not unrealistic) assumption that the direct detection of KK excitations is out of the reach of the LHC, one may still see sizable modifications of the Higgs-boson production cross section. For example, even with $M_{g^{(1)}}=10$\,TeV or even 15\,TeV, Figures~\ref{fig:MRS} and \ref{fig:CRS} show that virtual effects of KK particles can have significant effects on the Higgs-boson production cross section, provided that the 5D Yukawa couplings are not too small. We also note that different implementations of warped extra-dimension models, such as little RS models in which the bounds from electroweak precision measurements and flavor physics are relaxed by reducing the size $L$ of the extra dimension \cite{Davoudiasl:2008hx}, give rise to very similar Higgs phenomenology, because the $L$-dependent corrections in (\ref{eqn:Rh_Min}) and (\ref{eq100}) have only a minor impact.

\section{Conclusions}
\label{sec:concl}

The discovery of a Higgs-like boson at the LHC \cite{ATLAS:2012gk,CMS:2012gu} has raised the demand for an explanation of the hierarchy problem. Precise measurements of the Higgs-boson couplings to various SM particles can provide valuable tools to distinguish between different new-physics models addressing this problem. Such measurements can elucidate the mechanism of electroweak symmetry breaking and probe for indirect hints of new particles. Of particular importance are loop-induced processes, such as the Higgs-boson production process $gg\to h$ and the radiative decay $h\to\gamma\gamma$, since possible new heavy resonances can lead to sizable deviations from the SM expectations.

In this paper, we have focused on the gluon fusion cross section in various incarnations of RS models, in which the scalar sector is localized on or near the IR boundary of a warped extra dimension. We have derived an exact expression for the $gg\to h$ amplitude in terms of an integral of the mixed-chirality components of the 5D fermion propagator with the Higgs-boson profile along the extra dimension. This expression can be used to calculate the effective CP-even and CP-odd $ggh$ couplings, as long as one succeeds in deriving an explicit expression for the propagator. In contrast to the procedure commonly used in the literature, all our calculations have been performed by keeping the exact dependence on the Higgs-boson mass. Moreover, working in a 5D framework we have avoided the notion of KK modes from the beginning, which means that the infinite sum over fermionic KK modes is performed implicitly. Only at the end of the calculation we have been able to identify the contributions of the SM particles and their KK towers to the effective $ggh$ couplings. This removes any ambiguities as to the order in which one should perform the limits $N\to\infty$ and $\eta\to 0$ \cite{Carena:2012fk}. The 5D analysis also elucidates the relevance of different mass scales. While in models with a brane-localized Higgs sector the gluon fusion amplitude receives the dominant new-physics contributions from states with masses of order several times $M_{\rm KK}$, in a narrow bulk-Higgs scenario there is another equally important contribution arising from states with masses of order $v/\eta$, which can resolve the ``bulky nature'' of the Higgs boson. 

In Table~\ref{tab:models}, we have classified different versions of RS models according to the parametric relation of the characteristic width $\eta$ of the Higgs-boson profile with respect to the two ratios $v|Y_q|/M_{\rm KK}$ and $v|Y_q|/\Lambda_{\rm TeV}$, where $\Lambda_{\rm TeV}$ is the value of the inherent UV cutoff near the IR brane. We have shown that it is possible to obtain explicit analytic expressions for the 5D propagator for both a brane-localized scalar sector and a scalar sector that lives very near the IR brane (narrow bulk-Higgs scenario). To an excellent approximation, the effective $ggh$ couplings in these cases only depend on the 5D Yukawa matrices $\bm{Y}_q$ and the ratio $v^2/M_{\rm KK}^2$, see e.g.\ (\ref{Ciresu}) and (\ref{C1C5custo}). On the contrary, the results for a generic bulk-Higgs model, in which the width of the Higgs profile is not parametrically small, depend in a complicated way on the 5D fermion masses and the shape of the Higgs profile \cite{inprep}. 

Importantly, we have pointed out that there is no controllable interpolation between bulk-Higgs and brane-Higgs models. In RS models in which the scalar sector is localized on the IR brane, one finds that the gluon fusion cross section is reduced compared with its SM value, in accordance with the findings of \cite{Casagrande:2010si,Goertz:2011hj,Carena:2012fk}. In this context, we have proved a conjecture made in \cite{Carena:2012fk} for the analytic form of the contribution from virtual KK states. On the other hand, in models in which the Higgs-boson is described in terms of a narrow bulk field localized near the IR brane, the cross section is enhanced (apart from regions in parameter space in which the 5D Yukawa matrices have very small entries). This result confirms the calculations performed in \cite{Azatov:2010pf}. The qualitative difference between the predictions obtained in the two types of scenarios provides an opportunity to distinguish between the two classes of models, provided that a deviation of the gluon fusion rate from its SM value is observed in the future. When one tries to interpolate between the bulk-Higgs and brane-Higgs scenarios, for instance by considering the limit $\eta\to 0$ in the context of a bulk-Higgs model, one enters a transition region with $\eta\sim v|Y_q|/\Lambda_{\rm TeV}$, in which the contributions from certain higher-dimensional operators involving additional derivatives in the RS Lagrangian become unsuppressed, so that the effective field-theory approach breaks down.

We have furthermore addressed the question of the numerical impact of power-suppressed $|\Phi|^2 (G_{\mu\nu}^a)^2$ operators, which contribute to the $gg\to h$ amplitude at tree level. They can be induced because RS models are effective field theories valid below some cutoff. We have shown that, irrespective of whether the Higgs sector is localized on the IR brane or lives in the bulk, one expects power corrections of similar size, as described by the effective Lagrangian in (\ref{Leff}) with a coefficient $c_{\rm eff}$ that can be of ${\cal O}(1)$ if the UV completion of the RS model is strongly coupled. We have argued that the resulting power corrections are likely to be numerically smaller than the RS loop effects calculated in our paper.

For most of our discussion we have focused on the minimal RS model with gauge symmetry $SU(3)_C\times SU(2)_L\times U(1)_Y$ in the bulk. However, in Section~\ref{sec:custodial} we have considered an extension with a custodial symmetry, based on the gauge group $SU(3)_C\times SU(2)_L\times SU(2)_R\times U(1)_X\times P_{LR}$. We have succeeded in deriving analytical expressions for the effective $ggh$ couplings in terms of the same input parameters that appear in the minimal model. Due to the higher multiplicity of particles running in the loop, the contribution from the infinite KK tower of virtual quark states turns out to be four times larger than in the minimal model. 

Investigating the phenomenological implications of our results, we have focused on the ratio $R_h$ representing the gluon fusion cross section in the various RS models normalized to its SM value. We have distinguished between a brane-localized and a narrow bulk-Higgs field for both the minimal and the custodial RS model. We have pointed out the fact that the KK contribution to $R_h$ does not only depend strongly on the number of quark generations ($N_g=3$), but also on the maximal value $y_*$ one imposes on the magnitudes of the individual entries of the anarchic 5D Yukawa matrices, which are assumed to be random complex numbers. To a good approximation, our results can be parameterized in terms of $y_*$ and the mass $M_{g^{(1)}}\approx 2.45 M_{\rm KK}$ of the lightest KK gluon state. Provided that the value of $y_*$ is not too small, we have shown that quite generically the new-physics effects in RS models can lead to significant deviations of $R_h$ from~1, even for KK masses that are not in the reach of the LHC. This is evident from Figures~\ref{fig:MRS} and \ref{fig:CRS}, which show that significant corrections can be obtained even for KK gluon masses in the range of 10\,--\,20\,TeV. For RS models with a custodial symmetry, whose original motivation was to lower the KK scales via a protection of the $T$ parameter and the $Zb\bar b$ vertex, the effects are even more pronounced. As mentioned earlier, $R_h$ is strictly less than 1 in RS models with a brane-localized scalar sector, whereas it exceeds~1 in models with a bulk-Higgs field for almost all points in parameter space. 

Comparing our predictions with the latest ATLAS and CMS data \cite{Moriond_results}, we have derived exclusion regions in the $M_{g^{(1)}}$\,--\,$y_*$ parameter space of the various models. The corresponding results shown in Figures~\ref{fig:MRSexcl} and \ref{fig:CRSexcl} demonstrate the new-physics reach of Higgs-boson observables such as $R_h$ in an impressive way. In the minimal RS model and at 95\% CL, one can exclude KK gluon masses lighter than 12.8\,TeV$\times(y_*/3)$ for the brane-Higgs case and 4.4\,TeV$\times(y_*/3)$ for the case of a narrow bulk-Higgs field. In custodially protected RS models, these bounds increase to 24.4\,TeV$\times(y_*/3)$ and 9.6\,TeV$\times(y_*/3)$, respectively. Especially in this latter case, the bounds derived from Higgs physics are already much stronger than those obtained from electroweak precision tests. A possible way to weaken these bounds is to assume that $y_*$ is significantly smaller than the value commonly adopted in the literature ($y_*\approx 3$). However, this would create a tension with other observables, such as the parameter $\epsilon_K$ in the neutral-kaon system,  which in the context of RS models receives corrections scaling like $1/y_*^2$ \cite{Csaki:2008zd}. 

The methods developed in this work can be extended to analyze the loop-mediated decay $h\to\gamma\gamma$ as well as other decay modes of the Higgs boson. As the experimental precision on the extracted Higgs couplings increases, it will be exciting to confront the theoretical predictions obtained in various RS models with the data. One might hope that, perhaps, one day this could provide a first hint of the possible existence of a warped extra dimension, even if no KK excitations of SM particles are to be discovered at the LHC.

\vspace{3mm}
\noindent
{\bf Note added:\/}
While this paper was in writing, the work \cite{Frank:2013un} appeared, in which similar questions as in the present work were addressed. While we have no objections to the analytical calculations presented in that paper, we disagree with the interpretation of the results obtained by these authors. In particular, the argument that higher-derivative operators in the RS Lagrangian would ``dress'' the brane-localized Higgs to make it look like a bulk field is incompatible with our findings. Rather, these operators dress the bulk Higgs as its profile is made narrower, and they are responsible for the transition from an enhanced $gg\to h$ amplitude (bulk Higgs) to a suppressed one (brane-localized Higgs), see Table~\ref{tab:models}.

\vspace{3mm}
{\em Acknowledgements:\/}
We are grateful to Kaustubh Agashe, Paul Archer and Florian Goertz for useful discussions. The research of M.N.\ is supported by the Advanced Grant EFT4LHC of the European Research Council (ERC), the Cluster of Excellence {\em Precision Physics, Fundamental Interactions and Structure of Matter\/} (PRISMA -- EXC 1098), grant 05H12UME of the German Federal Ministry for Education and Research (BMBF), and the Rhineland-Palatinate Research Center {\em Elementary Forces and Mathematical Foundations}. R.M., K.N., and C.S.\ are supported by the DFG Graduate School GRK~1581 {\em Symmetry Breaking in Fundamental Interactions}.

\newpage
\begin{appendix}

\section{Details of the solution for the propagator functions}
\label{app:details}
\renewcommand{\theequation}{A.\arabic{equation}}
\setcounter{equation}{0}

Here we present in more detail the derivations needed to calculate the functions $T_\pm(p_E^2)$ defined in (\ref{TRLdef}). Since ultimately we only need the mixed-chirality components of the 5D fermion propagator evaluated with $t=t'$ and convoluted with the profile of the Higgs boson along the extra dimension, we can from the beginning assume that $1-\eta\le t'\le 1$, but we allow $t$ to take any value. 

\subsubsection*{Calculation of the propagator functions $\bm{\Delta}_{LL}$ and $\bm{\Delta}_{RL}$}

For $t<1-\eta$ in the bulk, the most general solutions for the propagator functions are superpositions of modified Bessel functions, as shown in (\ref{solregion1}) for the case of $\bm{\Delta}_{LL}$. The function $\bm{\Delta}_{RL}$ then follows from the second equation in (\ref{coupledeqs}). Imposing the boundary conditions (\ref{BCs}) on the UV brane (at $t=\epsilon$), one finds four relations among the right coefficients $\bm{C}_i(t')$. Rescaling these coefficients appropriately, we write the solutions in the form
\begin{equation}\label{Kidef}
\begin{aligned}
   \bm{\Delta}_{LL}^q(t,t';-p^2)
   &= \sqrt{t} \left( \begin{array}{cc}
    \bm{D}_1^Q(\hat p_E,t) & 0 \\ 
    0 & \bm{D}_2^q(\hat p_E,t) \end{array} \right)
    \left( \begin{array}{cc}
    \bm{K}_1(t') & \,\bm{K}_2(t') \\ 
    \bm{K}_3(t') & \,\bm{K}_4(t') \end{array} \right) , \\
   \bm{\Delta}_{RL}^q(t,t';-p^2)
   &= - M_{\rm KK}\,\hat p_E \sqrt{t} \left( \begin{array}{cc}
    \bm{D}_2^Q(\hat p_E,t) & 0 \\ 
    0 & \bm{D}_1^q(\hat p_E,t) \end{array} \right)
    \left( \begin{array}{cc}
    \bm{K}_1(t') & \,\bm{K}_2(t') \\ 
    \bm{K}_3(t') & \,\bm{K}_4(t') \end{array} \right) ,
\end{aligned}
\end{equation}
where (with $A=Q,q$)
\begin{equation}
\begin{aligned}
   \bm{D}_1^A(\hat p_E,t) 
   &= I_{-\bm{c}_A-\frac12}(\epsilon\hat p_E)\,I_{\bm{c}_A-\frac12}(\hat p_E t)
     - I_{\bm{c}_A+\frac12}(\epsilon\hat p_E)\,I_{-\bm{c}_A+\frac12}(\hat p_E t) \,, \\ 
   \bm{D}_2^A(\hat p_E,t) 
   &= I_{-\bm{c}_A-\frac12}(\epsilon\hat p_E)\,I_{\bm{c}_A+\frac12}(\hat p_E t)
     - I_{\bm{c}_A+\frac12}(\epsilon\hat p_E)\,I_{-\bm{c}_A-\frac12}(\hat p_E t) 
\end{aligned}
\end{equation}
are diagonal matrices, and $\bm{D}_2^A(\hat p_E,\epsilon)=0$. 

In the region $t>1-\eta$ very near the IR brane, the general solution for $\bm{\Delta}_{LL}^q(t,t';-p^2)$ has been given in (\ref{sol2}), while the second equation in (\ref{coupledeqs}) yields
\begin{equation}\label{2ndeq}
\begin{aligned}
   \bm{\Delta}_{RL}^q(t,t';-p^2)
   &= \hspace{3.5mm} \frac{M_{\rm KK}}{\eta} \left( \begin{array}{cc}
    \bm{S}_q\,{\cal S}(t) & \varrho\bm{Y}_q\,\bar{\cal C}(t) \\ 
    \varrho\bm{Y}_q^\dagger\,{\cal C}(t) & \bar{\bm{S}}_q\,\bar{\cal S}(t) \end{array} \right)
    \left( \begin{array}{cc}
    \hat{\bm{C}}_1(t') & \,\hat{\bm{C}}_2(t') \\ 
    \hat{\bm{C}}_3(t') & \,\hat{\bm{C}}_4(t') \end{array} \right) \\
   &\quad\mbox{}+  \frac{M_{\rm KK}}{\eta} \left( \begin{array}{cc}
    \bm{S}_q\,{\cal C}(t) & \varrho\bm{Y}_q\,\bar{\cal S}(t) \\ 
    \varrho\bm{Y}_q^\dagger\,{\cal S}(t) & \bar{\bm{S}}_q\,\bar{\cal C}(t) \end{array} \right)
    \left( \begin{array}{cc}
    \hat{\bm{C}}_5(t') & \,\hat{\bm{C}}_6(t') \\ 
    \hat{\bm{C}}_7(t') & \,\hat{\bm{C}}_8(t') \end{array} \right) ,
\end{aligned}
\end{equation}
where $\varrho=v/(\sqrt2 M_{\rm KK})$, and we have used the abbreviations
\begin{equation}
   {\cal C}(t) = \cosh[\bm{S}_q\,\bar\theta^\eta(t-1)] \,, \qquad
   {\cal S}(t) = \sinh[\bm{S}_q\,\bar\theta^\eta(t-1)] \,,
\end{equation}
and similarly $\bar{\cal C}(t)$ and $\bar{\cal S}(t)$ defined with $\bar{\bm{S}}_q$ instead of $\bm{S}_q$. Because of the discontinuity at $t=t'$, we must distinguish the cases where $t>t'$ and $t<t'$. We indicate this by means of a superscript on the coefficient functions, using the notation $\hat{\bm{C}}_i^>(t')$ for $t>t'$, and $\hat{\bm{C}}_i^<(t')$ for $t<t'$. Imposing the boundary conditions (\ref{BCs}) on the IR brane (at $t=1$), and satisfying the jump conditions (\ref{jump}), it is straightforward to show that
\begin{equation}
\begin{aligned}
   \hat{\bm{C}}_1^>(t') 
   &= \hat{\bm{C}}_1^<(t') + \frac{\eta}{M_{\rm KK}^2}\,\frac{1}{\bm{S}_q}\,{\cal S}(t') \,, &\qquad
   \hat{\bm{C}}_5^<(t') &= \frac{\eta}{M_{\rm KK}^2}\,\frac{1}{\bm{S}_q}\,{\cal C}(t') \,, \\
   \hat{\bm{C}}_2^>(t') &= \hat{\bm{C}}_2^<(t') \,, &\qquad
   \hat{\bm{C}}_7^>(t') &= \hat{\bm{C}}_7^<(t') \,, \\
   \hat{\bm{C}}_4^<(t') &= - \frac{\eta}{M_{\rm KK}^2}\,\frac{1}{\bar{\bm{S}}_q}\,\bar{\cal S}(t')
    \,, &\qquad
   \hat{\bm{C}}_8^>(t') 
   &= \hat{\bm{C}}_8^<(t') - \frac{\eta}{M_{\rm KK}^2}\,\frac{1}{\bar{\bm{S}}_q}\,\bar{\cal C}(t') \,,
\end{aligned}    
\end{equation}
while all other coefficients vanish. These relations allow us to express the solution in terms of the four functions $\hat{\bm{C}}_i^<(t')$ with $i=1,2,7,8$.

The remaining eight coefficients are determined by requiring that the solutions for the propagator functions be continuous at $t=1-\eta$. Continuity of $\bm{\Delta}_{LL}$ yields the conditions
\begin{equation}
\begin{aligned}
   \bm{D}_1^Q(\hat p_E,1-\eta)\,\bm{K}_1(t')
   &= \cosh\bm{S}_q\,\hat{\bm{C}}_1^<(t')
    + \frac{\eta}{M_{\rm KK}^2}\,\frac{\sinh\bm{S}_q}{\bm{S}_q}\,{\cal C}(t') \,, \\
   \bm{D}_1^Q(\hat p_E,1-\eta)\,\bm{K}_2(t')
   &= \cosh\bm{S}_q\,\hat{\bm{C}}_2^<(t') \,, \\
   \bm{D}_2^q(\hat p_E,1-\eta)\,\bm{K}_3(t')
   &= \sinh\bar{\bm{S}}_q\,\hat{\bm{C}}_7^<(t') \,, \\
   \bm{D}_2^q(\hat p_E,1-\eta)\,\bm{K}_4(t')
   &= \sinh\bar{\bm{S}}_q\,\hat{\bm{C}}_8^<(t')
    - \frac{\eta}{M_{\rm KK}^2}\,\frac{\cosh\bar{\bm{S}}_q}{\bar{\bm{S}}_q}\,\bar{\cal S}(t') \,,
\end{aligned}    
\end{equation}
which can be used to eliminate the coefficients $\bm{K}_i(t')$. Note that on the left-hand sides of these equations we can take the limit $\eta\to 0$ without difficulty. When the solutions are inserted into the expression for $\bm{\Delta}_{RL}$ in (\ref{Kidef}), we then encounter the ratios $\bm{R}_A(\hat p_E)=\bm{D}_1^A(\hat p_E,1)/\bm{D}_2^A(\hat p_E,1)$ defined in (\ref{RAdef}). The remaining four coefficients $\hat{\bm{C}}_i^<(t')$ can be derived by requiring that the propagator function $\bm{\Delta}_{RL}$ is continuous at $t=1-\eta$. To express the answers in a compact form, we introduce the definitions
\begin{equation}
\begin{aligned}
   \bm{N}_q^{\eta,1}(p_E^2)
   &= 1 + \bm{Z}_q^{\eta,1}(p_E^2)
    + \eta\hat p_E \left[ 1 + \bm{R}_Q^{-1}(\hat p_E)\,(\bm{Y}_q^\dagger)^{-1}\,
    \bm{R}_q(\hat p_E)\,\bm{Y}_q^\dagger \right]
    \frac{\tanh\bm{S}_q}{\bm{S}_q}\,\bm{R}_Q(\hat p_E) \,, \\
   \bm{N}_q^{\eta,2}(p_E^2)
   &= 1 + \bm{Z}_q^{\eta,2}(p_E^2)
    + \eta\hat p_E\,\frac{\tanh\bm{S}_q}{\bm{S}_q}
    \left[ \bm{R}_Q(\hat p_E) + \bm{Y}_q\,\bm{R}_q(\hat p_E)\,\bm{Y}_q^{-1} \right] , 
\end{aligned}
\end{equation}
where
\begin{equation}
\begin{aligned}
   \bm{Z}_q^{\eta,1}(p_E^2) 
   &= \frac{v^2}{2M_{\rm KK}^2}\,\frac{\bm{S}_q\tanh\bm{S}_q}{\bm{X}_q^2}\,\bm{Y}_q\,
    \bm{R}_q(\hat p_E)\,\bm{Y}_q^\dagger\,\frac{\tanh\bm{S}_q}{\bm{S}_q}\,\bm{R}_Q(\hat p_E) \,, \\
   \bm{Z}_q^{\eta,2}(p_E^2) 
   &= \frac{v^2}{2M_{\rm KK}^2}\,\frac{\tanh\bm{S}_q}{\bm{S}_q}\,\bm{Y}_q\,
    \bm{R}_q(\hat p_E)\,\bm{Y}_q^\dagger\,\frac{\bm{S}_q\tanh\bm{S}_q}{\bm{X}_q^2}\,
    \bm{R}_Q(\hat p_E) \,.
\end{aligned}
\end{equation}
In the limit $\eta\to 0$, the quantities $\bm{N}_q^{\eta,i}(p_E^2)$ approach $1+\bm{Z}_q(p_E^2)$ with $\bm{Z}_q(p_E^2)$ as defined in (\ref{Zdef}), while the quantities $\bm{Z}_q^{\eta,i}(p_E^2)$ approach $\bm{Z}_q(p_E^2)$. After some lengthy algebra, we now obtain
\begin{equation}\label{Asliver}
\begin{aligned}
   \hat{\bm{C}}_1^<(t') &= - \frac{\eta}{M_{\rm KK}^2}\,\frac{1}{\bm{S}_q\sinh\bm{S}_q} \\
   &\quad\times \left[
    \sinh^2\bm{S}_q + \bm{Z}_q^{\eta,1}(p_E^2)\,\frac{1}{\bm{N}_q^{\eta,1}(p_E^2)}
    + \frac{\bm{S}_q\tanh\bm{S}_q}{\eta\hat p_E}\,\bm{R}_Q(\hat p_E)\,
    \frac{1}{\bm{N}_q^{\eta,1}(p_E^2)} \right] \frac{{\cal C}(t')}{\cosh\bm{S}_q} \,, \\
   \hat{\bm{C}}_2^<(t') &= \frac{1}{p_E M_{\rm KK}}\,\frac{1}{\cosh\bm{S}_q}\,\bm{R}_Q(\hat p_E)\,
    \frac{1}{\bm{N}_q^{\eta,2}(p_E^2)}\,\frac{{\cal S}(t')}{\bm{S}_q\cosh\bm{S}_q}\,
    \varrho\bm{Y}_q \,, \\
   \hat{\bm{C}}_7^<(t')
   &= \frac{\varrho\bm{Y}_q^\dagger}{p_E M_{\rm  KK}}\,\frac{1}{\bm{S}_q\cosh\bm{S}_q}\,
    \bm{R}_Q(\hat p_E)\,\frac{1}{\bm{N}_q^{\eta,1}(p_E^2)}\,\frac{{\cal C}(t')}{\cosh\bm{S}_q} \,, \\ 
   \hat{\bm{C}}_8^<(t') 
   &= \frac{\eta\varrho\bm{Y}_q^\dagger}{M_{\rm KK}^2}\,\frac{1}{\bm{X}_q^2\sinh\bm{S}_q} \\
   &\quad\times \left[ 
    \sinh^2\bm{S}_q + \frac{\bm{N}_q^{\eta,2}(p_E^2)-1}{\bm{N}_q^{\eta,2}(p_E^2)}
    - \frac{\bm{S}_q\tanh\bm{S}_q}{\eta\hat p_E}\,\bm{R}_Q(\hat p_E)\,
    \frac{1}{\bm{N}_q^{\eta,2}(p_E^2)} \right] \frac{{\cal S}(t')}{\bm{S}_q\cosh\bm{S}_q}\,
    \varrho\bm{Y}_q \,.
\end{aligned}
\end{equation}   

\subsubsection*{Calculation of the functions $T_\pm(p_E^2)$}

Equipped with all required coefficients, we can now derive explicit expressions for the quantities $T_{\pm}(p_E^2)$ defined in (\ref{TRLdef}). Using (\ref{2ndeq}), we find that 
\begin{equation}\label{beforetrace}
\begin{aligned}
   &\frac{v}{\sqrt2}\,\bigg( \begin{array}{cc} 0 & \bm{Y}_q \\ 
     \bm{Y}_q^\dagger & 0 \end{array} \bigg)\,
     \frac{\bm{\Delta}_{RL}^q(t,t;p_E^2)+\bm{\Delta}_{LR}^q(t,t;p_E^2)}{2} \\
   &= \frac{M_{\rm KK}^2}{2\eta}\,\bigg\{
    \bm{X}_q^2 \left[ {\cal C}(t)\,\hat{\bm{C}}_1^<(t)
    + \frac{\eta}{M_{\rm KK}^2}\,\frac{{\cal S}(t)\,{\cal C}(t)}{\bm{S}_q} \right]
    + \varrho\bm{Y}_q^\dagger\,\bm{S}_q\,{\cal S}(t)\,\hat{\bm{C}}_2^<(t) \\
   &\hspace{1.4cm}\mbox{}+ \bar{\bm{X}}_q^2 \left[ \bar{\cal S}(t)\,\hat{\bm{C}}_8^<(t)
    - \frac{\eta}{M_{\rm KK}^2}\,\frac{\bar{\cal S}(t)\,\bar{\cal C}(t)}{\bar{\bm{S}}_q} \right]  
    + \varrho\bm{Y}_q\,\bar{\bm{S}}_q\,\bar{\cal C}(t)\,\hat{\bm{C}}_7^<(t) 
    + \mbox{h.c.} \bigg\} \,,
\end{aligned}     
\end{equation}
where the contribution from $\bm{\Delta}_{LR}^q$ is the hermitian conjugate of that from $\bm{\Delta}_{RL}^q$ (assuming $p_E^2>0$ for now). Upon taking the trace in (\ref{TRLdef}), the two terms proportional to $\eta$ in the square brackets cancel each other. Next, using the explicit expressions for the coefficients in (\ref{Asliver}), we find that the contribution involving the terms proportional to $1/\eta$ in square brackets in the expression for $\hat{\bm{C}}_1^<$ cancel against the contribution from $\hat{\bm{C}}_7^<$ in (\ref{beforetrace}), and likewise for the terms involving $\hat{\bm{C}}_2^<$ and $\hat{\bm{C}}_8^<$. After the dust settles, we obtain
\begin{equation}\label{Tplgeneral}
\begin{aligned}
   T_+(p_E^2) 
   &= \sum_{q=u,d} \int_\epsilon^1\!dt\,\delta_h^\eta(t-1)\,\mbox{Tr}\,\bigg\{
    \frac{\bm{X}_q^2}{\bm{S}_q\sinh2\bm{S}_q} \\
   &\quad\times \bigg[ \sinh^2\bm{S}_q
    + {\cal C}^2(t)\,\bm{Z}_q^{\eta,1}(p_E^2)\,\frac{1}{\bm{N}_q^{\eta,1}(p_E^2)} 
    - {\cal S}^2(t)\,\frac{\bm{N}_q^{\eta,2}(p_E^2)-1}{\bm{N}_q^{\eta,2}(p_E^2)} 
    + \mbox{h.c.} \bigg] \bigg\} \,,
\end{aligned}     
\end{equation}
and analogously for $T_-(p_E^2) $. In the case of one generation, the above expression reduces to formula (\ref{T1res}), once we identify $k_1(\hat p_E)=1+Z_q(p_E^2)\,\coth^2 S_q$. For the general case of three generations, the result (\ref{Tplgeneral}) simplifies if we take the limit $\eta\to 0$, in which we recover the results shown in (\ref{T3gen}). Note that in this case the dependence on $t$ inside the square brackets in (\ref{Tplgeneral}) disappears, due to the identity ${\cal C}^2(t)-{\cal S}^2(t)=1$. Therefore, as already mentioned in Section~\ref{sec:4.5}, we would have obtained the same result by setting $t=t'=1^-$, as shown in (\ref{TRLnaive}).

\subsubsection*{Generalizations for the model with custodial symmetry}

The derivation of the propagator functions in the RS model with custodial symmetry proceeds in an analogous way. In fact, the only difference arises in the equations in (\ref{Kidef}), where $\bm{D}_{1,2}^Q$ and $\bm{D}_{2,1}^q$ must be replaced by 
\begin{equation}
   \left( \begin{array}{cc}
    \bm{D}_{1,2}^Q(\hat p_E,t) & 0 \\ 0 & \bm{D}_{3,4}^Q(\hat p_E,t) \end{array} \right) 
    \quad \mbox{and} \quad
   \left( \begin{array}{ccc}
    \bm{D}_{2,1}^{u^c}(\hat p_E,t) & 0 & 0 \\
    0 & \bm{D}_{4,3}^{\tau_1}(\hat p_E,t) & 0 \\
    0 & 0 & \bm{D}_{4,3}^{\tau_2}(\hat p_E,t) \end{array} \right)
\end{equation}
for up-type quarks, and analogously for down- and $\lambda$-type quarks, with patterns that can be read off from (\ref{eq85}). The appearance of the functions 
\begin{equation}
\begin{aligned}
   \bm{D}_3^A(\hat p_E,t) 
   &= I_{-\bm{c}_A+\frac12}(\epsilon\hat p_E)\,I_{\bm{c}_A-\frac12}(\hat p_E t)
    - I_{\bm{c}_A-\frac12}(\epsilon\hat p_E)\,I_{-\bm{c}_A+\frac12}(\hat p_E t) \,, \\
   \bm{D}_4^A(\hat p_E,t) 
   &= I_{-\bm{c}_A+\frac12}(\epsilon\hat p_E)\,I_{\bm{c}_A+\frac12}(\hat p_E t)
    - I_{\bm{c}_A-\frac12}(\epsilon\hat p_E)\,I_{-\bm{c}_A-\frac12}(\hat p_E t)
\end{aligned}
\end{equation}
gives rise to the ratios $\bm{R}_A^{(-)}(\hat p_E)=\bm{D}_3^A(\hat p_E,1)/\bm{D}_4^A(\hat p_E,1)$ defined in (\ref{Rminus}).

\section{Case of a bulk-Higgs field}
\label{app:bulkHiggs}
\renewcommand{\theequation}{B.\arabic{equation}}
\setcounter{equation}{0}

This section intents to relate an RS model with a scalar sector in the bulk, in which the Higgs field and its vev have profiles that are strongly peaked near the IR brane, to the RS model with a brane-localized Higgs sector. Our discussion will follow the expositions given in \cite{Cacciapaglia:2006mz,Archer:2012qa}, but we will generalize these results in some aspects. 

\subsubsection*{Definition of the model}

Using the orbifold coordinate $x_5\equiv r\phi$, the action for the Higgs sector reads
\begin{equation}\label{eqn:HSaction}
   S_h = \int d^4x\int_{-r\pi}^{r\pi}\!dx_5\,e^{-4\sigma(\phi)}\,\Big[ 
    g^{MN} D_M\Phi^\dagger D_N\Phi - \mu^2\,|\Phi|^2 - V_{\rm UV}(\Phi)\,\delta(x_5) 
    - V_{\rm IR}(\Phi)\,\delta(|x_5|-r\pi) \Big] \,,   
\end{equation}
where $\mu$ provides a bulk mass for the scalar field, which can be tachyonic (see below). The potentials localized on the UV and IR branes determine the boundary conditions of the scalar fields and induce electroweak symmetry breaking. They are chosen to be
\begin{equation}
   V_{\rm UV}(\Phi) = M_{\rm UV}\,|\Phi|^2 \,, \qquad
   V_{\rm IR}(\Phi) = - M_{\rm IR}\,|\Phi|^2 + \lambda_{\rm IR}\,|\Phi|^4 \,,
\end{equation}
with mass dimensions $[M_{\rm UV}]=[M_{\rm IR}]=1$ and $[\lambda_{\rm IR}]=-2$. The dimensionful parameters in the 5D action naturally scale with appropriate powers of $M_{\rm Pl}$, and we find it useful to introduce dimensionless ${\cal O}(1)$ parameters by the rescalings
\begin{equation}
   m_{\rm UV}\equiv\frac{M_{\rm UV}}{2k} \,, \qquad
   m_{\rm IR}\equiv\frac{M_{\rm IR}}{2k} \,, \qquad
   \lambda\equiv\frac{\lambda_{\rm IR}\,k}{4r} \,.
\end{equation}

We now change variables from $\phi$ to $t=\epsilon\,e^{\sigma(\phi)}$ and express the scalar doublet $\Phi$ in the form
\begin{equation}\label{bulkPhi}
   \Phi(x,t) = \frac{t}{\epsilon\sqrt r}
    \begin{pmatrix} - i\varphi^+(x,t) \\ 
     \frac{1}{\sqrt2} \left[ v(t) + h(x,t) + i\varphi_3(x,t) \right]
    \end{pmatrix} ,
\end{equation}
where $v(t)$ denotes the profile of the Higgs vev along the extra dimension, $h(x,t)$ is the 5D physical Higgs scalar after electroweak symmetry breaking, and $\varphi^+(x,t)$, $\varphi_3(x,t)$ are 5D Goldstone bosons. For the following analysis we do not consider the Goldstone fields any further (unitary gauge). Integrating by parts, the Lagrangian corresponding to the action $S_h=\int d^4x\,{\cal L}_h(x)$ in (\ref{eqn:HSaction}) can be rewritten in the form
\begin{equation}\label{Lh}
\begin{aligned}
   {\cal L}_h(x)
   &= \frac{2\pi}{L} \int_{\epsilon}^1\!\frac{dt}{t}\,\bigg\{ 
    \frac12\,\partial_\mu h(x,t)\,\partial^\mu h(x,t) \\
   &\qquad\mbox{}+ \frac{M_{\rm KK}^2}{2} \bigg[ \frac{v(t)+2h(x,t)}{t}
    \left( t^2\partial_t^2 + t\partial_t - \beta^2 \right) \frac{v(t)}{t} 
    + \frac{h(x,t)}{t} \left( t^2\partial_t^2 + t\partial_t - \beta^2 \right) 
    \frac{h(x,t)}{t} \bigg] \bigg\} \\
   &\quad\mbox{}- \frac{\pi M_{\rm KK}^2}{L}\,\bigg\{\!
    \left[ \frac{v(t)+2h(x,t)}{t^2}\,\partial_t\left[ t\,v(t) \right] 
    + \frac{h(x,t)}{t^2}\,\partial_t\left[ t\,h(x,t) \right] \right]_{t=\epsilon^+}^{1^-} 
    + \frac{m_{\rm UV}}{\epsilon^2}\,\big[ v(\epsilon) + h(x,\epsilon) \big]^2 \\
   &\hspace{2.5cm}\mbox{}- m_{\rm IR}\,\big[ v(1) + h(x,1) \big]^2 
    + \frac{\lambda}{M_{\rm KK}^2}\,\big[ v(1) + h(x,1) \big]^4 \bigg\} \,,
\end{aligned}
\end{equation}
where $\beta=\sqrt{4+\mu^2/k^2}$. Requiring that the terms linear or quadratic in $h(x,t)$ cancel on the UV and IR branes yields the boundary conditions\footnote{These conditions can also be derived by integrating the field equations over infinitesimal intervals about the branes.}
\begin{equation}\label{UVIRBCs}
\begin{aligned}
   \partial_t\left[ t\,v(t) \right]_{t=\epsilon^+} 
   &= m_{\rm UV}\,v(\epsilon) \,, &\quad
    \partial_t\left[ t\,v(t) \right]_{t=1^-} 
    &= m_{\rm IR}\,v(1) - \frac{2\lambda}{M_{\rm KK}^2}\,v^3(1) \,, \\
   \partial_t\left[ t\,h(x,t) \right]_{t=\epsilon^+} 
   &= m_{\rm UV}\,h(x,\epsilon) \,, &\quad
    \partial_t\left[ t\,h(x,t) \right]_{t=1^-} 
    &= m_{\rm IR}\,h(x,1) - \frac{6\lambda}{M_{\rm KK}^2}\,v^2(1)\,h(x,1) \,.
\end{aligned}
\end{equation}
The notation $\epsilon^+$ and $1^-$ indicates that the orbifold fixed points must be approached from the appropriate sides.

\subsubsection*{Profile of the Higgs vacuum expectation value}

By means of the variational principle with respect to $v(t)$, one obtains the equation 
\begin{equation}\label{eqn:diffv}
   \left( t^2\partial_t^2 + t\partial_t - \beta^2 \right) \frac{v(t)}{t} = 0 \,,
    \qquad \mbox{with} \quad
   \beta^2 = 4 + \frac{\mu^2}{k^2}
\end{equation}
which ensures that the tadpole terms in the Lagrangian (\ref{Lh}) cancel out. We then obtain
\begin{equation}\label{Lh2}
\begin{aligned}
   {\cal L}_h(x)
   &= \frac{2\pi}{L} \int_{\epsilon}^1\!\frac{dt}{t}\,\bigg[ 
    \frac12\,\partial_\mu h(x,t)\,\partial^\mu h(x,t) 
    + \frac{M_{\rm KK}^2}{2}\,\frac{h(x,t)}{t}
    \left( t^2\partial_t^2 + t\partial_t - \beta^2 \right) \frac{h(x,t)}{t} \bigg] \\
   &\quad\mbox{}- \frac{\pi}{L}\,\lambda\,\Big[
    - v^4(1) + 4v(1)\,h^3(x,1) + h^4(x,1) \Big] \,.
\end{aligned}
\end{equation}
The general solution to the differential equation (\ref{eqn:diffv}) subject to the boundary conditions (\ref{UVIRBCs}) is
\begin{equation}\label{vtsol}
   v(t) = N_v \left( t^{1+\beta} - r_v\,t^{1-\beta} \right) , \qquad \mbox{with} \quad
   r_v = \epsilon^{2\beta}\,\frac{2+\beta-m_{\rm UV}}{2-\beta-m_{\rm UV}} \,, 
\end{equation}
and
\begin{equation}
   N_v^2 = \frac{M_{\rm KK}^2}{2\lambda}\,
    \frac{(m_{\rm IR}-2-\beta)-r_v\,(m_{\rm IR}-2+\beta)}{\left(1-r_v\right)^3} \,.
\end{equation}

Before proceeding, let us first discuss which values the parameter $\beta$ can take. Motivated by the observation that the energy-momentum flux in a pure anti-de~Sitter space without an IR brane (which corresponds to taking the limit $r\to\infty$) vanishes at the boundary only if the 5D scalar field obeys the Breitenlohner-Friedman bound $\mu^2>-4k^2$ \cite{Breitenlohner:1982jf}, one usually assumes that $\beta$ must be a real positive number, even though not necessarily larger than~2. Unless $\beta$ is very close to zero, it follows that the coefficient $r_v\propto\epsilon^{2\beta}$ in (\ref{vtsol}) is extremely small and can be set to zero for all practical purposes. The only exception would be the region where $t\sim\epsilon$ is very near the UV brane, which however is irrelevant for our analysis here. It follows that
\begin{equation}\label{vtsol2}
   v(t) = v(1)\,t^{1+\beta} \,, \qquad \mbox{with} \quad
   v(1) = M_{\rm KK}\,\sqrt{\frac{m_{\rm IR}-2-\beta}{2\lambda}} \,.
\end{equation}
The requirement that the Higgs vev be a real number imposes an upper bound on the parameter $\beta$, since $\lambda>0$ is required by vacuum stability. We thus obtain the allowed range
\begin{equation}\label{brange}
   0<\beta<m_{\rm IR}-2 \,.
\end{equation}

We proceed to relate the parameter $v(1)$ to the physical value $v_{\rm SM}$ of the Higgs vev in the SM. After electroweak symmetry breaking, the mass terms for the $W$ and $Z$ bosons are generated by the 5D Lagrangian 
\begin{equation}
   S_m = \int d^4x\,\frac{2\pi}{L}\int_\epsilon^1\!\frac{dt}{t}\,\frac{v^2(t)\,g_5^2}{4} 
   \left[ W_\mu^+(x,t)\,W^{-\mu}(x,t) + \frac{1}{2\cos^2\theta_W}\,Z_\mu(x,t)\,Z^\mu(x,t) \right] ,
\end{equation}
where the 5D gauge coupling $g_5$ is related to the gauge coupling $g$ of the SM by $g=g_5/\sqrt{2\pi r}$ \cite{Davoudiasl:1999tf}. Introducing the KK decomposition
\begin{equation}
   Z^\mu(x,t) = \frac{1}{\sqrt r}\,\sum_{n=0}^\infty\,Z_n^\mu(x)\,\chi_n^Z(t) \,,
\end{equation}
and similarly for the $W$ bosons, and using that the zero-mode profiles are flat, $\chi_n^Z(t)=1/\sqrt{2\pi}$ up to higher-order terms in $v^2/M_{\rm KK}^2$ \cite{Davoudiasl:1999tf}, we can identify
\begin{equation}\label{vSMv1}
   v_4^2 \equiv \frac{2\pi}{L} \int_\epsilon^1\,\frac{dt}{t}\,v^2(t)
   = \frac{\pi}{L}\,\frac{v^2(1)}{1+\beta} \,,
\end{equation}
where once again we neglect terms suppressed by powers of $\epsilon$. It follows that
\begin{equation}\label{vtfinal}
   v(t) = v_4\,\sqrt{\frac{L}{\pi}\,(1+\beta)}\,\,t^{1+\beta} \,.
\end{equation}
The parameter $v_4$ coincides with the parameter $v$ used elsewhere in this paper. At lowest order in an expansion in powers of $v^2/M_{\rm KK}^2$, it coincides with the SM parameter $v_{\rm SM}$ as defined, e.g., via the value of the Fermi constant. Higher-order corrections to the relation $v_{\rm SM}=v_4$ could be calculated by solving the differential equations for the profiles of the gauge-boson zero modes in the presence of the Higgs vev.

\subsubsection*{Profiles for the Higgs boson and its KK excitations}

We now proceed to study the eigenvalue problem for the physical Higgs boson and its KK excitations. We write the KK decomposition of the 5D Higgs field as 
\begin{equation}\label{hKKdec}
   h(x,t) = \sum_{n=0}^\infty\,h_n(x)\,\chi_n(t) \,,
\end{equation}
where the zero mode $h(x)\equiv h_0(x)$ corresponds to the SM Higgs boson. The profile functions obey the orthonormality condition
\begin{equation}
   \frac{2\pi}{L} \int_\epsilon^1\!\frac{dt}{t}\,\chi_m(t)\,\chi_n(t) = \delta_{mn} \,,
\end{equation}
which ensures that the kinetic terms in the effective 4D Lagrangian are canonically normalized. In order to obtain canonical mass terms from the Lagrangian (\ref{Lh2}), we must impose the equation of motion
\begin{equation}
   \left( t^2\partial_t^2 + t\partial_t + t^2 x_n^2 - \beta^2 \right) \frac{\chi_n(t)}{t} = 0 \,,
\end{equation}
where $x_n=m_n/M_{\rm KK}$ denote the masses of the KK scalar bosons in units of $M_{\rm KK}$. The general solution to this equation is a linear combination of Bessel functions,
\begin{equation}
   \chi_n(t) = N_n\,t\,\big[ J_\beta(x_n t) - r_n Y_\beta(x_n t) \big] \,,
\end{equation}
where the boundary condition on the UV brane in (\ref{UVIRBCs}) once again implies that $r_n\propto\epsilon^{2\beta}$ is extremely small and can be set to zero for all practical purposes, since we are not interested in the region where $t\sim\epsilon$. We then obtain
\begin{equation}\label{chin}
   \chi_n(t) = \sqrt{\frac{L}{\pi}}\,
    \frac{t\,J_\beta(x_n t)}{\sqrt{J_\beta^2(x_n)-J_{\beta+1}(x_n)\,J_{\beta-1}(x_n)}} \,.
\end{equation}
The boundary condition on the IR brane gives rise to the eigenvalue equation, which determines the masses of the scalar modes. We find
\begin{equation}\label{xndet}
   \frac{x_n J_{\beta+1}(x_n)}{J_\beta(x_n)} = 2(m_{\rm IR}-2-\beta)\equiv 2\delta \,.
\end{equation}
It follows from this equation that even the zero mode (the SM Higgs boson) has a mass that is naturally of order the KK scale $M_{\rm KK}$, which empirically cannot be less than a few TeV. This is the little hierarchy problem, which as mentioned in the Introduction is not addressed in RS scenarios. In order to obtain a realistic Higgs mass $m_h\ll M_{\rm KK}$, we must assume that 
\begin{equation}\label{tune}
   \delta = m_{\rm IR}-2-\beta \ll 1 \,.
\end{equation}
Once this is done, it is straightforward to obtain a formula for the zero-mode mass in a power series in $\delta$. We find
\begin{equation}
   x_0^2 = \frac{m_h^2}{M_{\rm KK}^2}
   = 4(1+\beta)\,\delta \left[ 1 - \frac{\delta}{2+\beta}
    + \frac{2\delta^2}{\left(2+\beta\right)^2(3+\beta)} + \dots \right] .
\end{equation}
Assuming $M_{\rm KK}=2$\,TeV, for example, implies that $(1+\beta)\,\delta\approx 10^{-3}$, which corresponds to a fine-tuning of 1 in 1000. For the zero-mode profile, it is now straightforward to obtain an expansion in powers of $x_0^2$. The leading terms are given by
\begin{equation}\label{chi0fin}
   \chi_0(t) = \sqrt{\frac{L}{\pi}\,(1+\beta)}\,\,t^{1+\beta}
    \left[ 1 - \frac{x_0^2}{4} \left( \frac{t^2}{1+\beta} - \frac{1}{2+\beta} \right) 
    + \dots \right] .
\end{equation}

Dropping the irrelevant constant proportional to $v^4(1)$, the Higgs Lagrangian (\ref{Lh2}) can now be written as
\begin{equation}\label{Lhfin}
\begin{aligned}
   {\cal L}_h(x)
   &= \sum_n \bigg[ \frac12\,\partial_\mu h_n(x)\,\partial^\mu h_n(x) 
    - \frac{m_n^2}{2}\,h_n^2(x) \bigg] \\
   &\quad\mbox{}- v_4\,\frac{4L}{\pi}\,(1+\beta)^2\,\lambda \sum_{l,m,n}\,
    \xi_l\,\,\xi_m\,\xi_n\,h_l(x)\,h_m(x)\,h_n(x) \\
   &\quad\mbox{}- \frac{L}{\pi}\,(1+\beta)^2\,\lambda \sum_{k,l,m,n}
    \xi_k\,\xi_l\,\,\xi_m\,\xi_n\,h_k(x)\,h_l(x)\,h_m(x)\,h_n(x) \,,
\end{aligned}
\end{equation}
where $\xi_n\equiv\chi_n(1)/\sqrt{\frac{L}{\pi}\,(1+\beta)}$. From (\ref{chi0fin}) we find $\xi_0\approx 1$ for the zero mode, while (\ref{chin}) and (\ref{xndet}) imply that $\xi_n\approx\pm1/\sqrt{1+\beta}$ for the KK excitations. We proceed to relate the parameter $\lambda$ to the physical value $\lambda_4$ of the Higgs self coupling. The relevant terms in the SM Lagrangian are
\begin{equation}\label{Lsm}
   {\cal L}_{\rm SM} \ni - \frac{m_h^2}{2}\,h^2 - v_{\rm SM}\,\lambda_{\rm SM}\,h^3
    - \frac{\lambda_{\rm SM}}{4}\,h^4 \,,
\end{equation} 
where $m_h^2=2\lambda_{\rm SM}\,v_{\rm SM}^2$. Matching either one of these terms with the corresponding term in (\ref{Lhfin}), we obtain at leading order
\begin{equation}\label{lam4rel}
   \lambda_{\rm SM} = \lambda_4 = \frac{4L}{\pi}\,(1+\beta)^2\,\lambda 
   = \lambda_{\rm IR}\,k^2\,(1+\beta)^2 \,. 
\end{equation}
The relation between $\lambda_{\rm SM}$ and $\lambda_4$ receives higher-order corrections in $v^2/M_{\rm KK}^2$, which are calculable in the model and depend on which of the three couplings in (\ref{Lsm}) is used to perform the matching. 

\subsubsection*{Yukawa interactions}

We finally consider the Yukawa couplings of the scalar field to the fermions. In the model with a brane-localized Higgs sector, which we have considered for most of this work, one has in analogy with~(\ref{Lhqq})
\begin{equation}\label{LYbrane}
   - {\cal L}_Y^{\rm brane}(x) = \sum_{q=u,d}\,\int_\epsilon^1\!dt\,
    \frac{v\,\delta_v^\eta(t-1)+h(x)\,\delta_h^\eta(t-1)}{\sqrt2}\,
    \bar{\cal Q}_L(t,x)\,\frac{k}{2}\,
    \bigg( \begin{array}{cc} 0 & \bm{Y}_q^{5D} \\ 
           \bm{Y}_q^{5D\dagger} & 0 \end{array} \bigg)\,{\cal Q}_R(t,x) + \mbox{h.c.} \,,
\end{equation}
where the 5D Yukawa matrices $\bm{Y}_q^{5D}$ have mass dimension $-1$. In the model with a bulk-Higgs field, we have instead
\begin{equation}
   - {\cal L}_Y^{\rm bulk}(x) = \sum_{q=u,d}\,\int_\epsilon^1\!dt\,
    \frac{v(t)+\sum_n h_n(x)\,\chi_n(t)}{\sqrt2}\,
    \bar{\cal Q}_L(t,x)\,\frac{1}{\sqrt r}\,
    \bigg( \begin{array}{cc} 0 & \bm{Y}_{q,\rm bulk}^{5D} \\ 
           \bm{Y}_{q,\rm bulk}^{5D\dagger} & 0 \end{array} \bigg)\,{\cal Q}_R(t,x) 
    + \mbox{h.c.} \,,
\end{equation}
where the 5D Yukawa matrices $\bm{Y}_{q,\rm bulk}^{5D}$ now have mass dimension $-1/2$. In order to match the two expression onto each other, we must rewrite the functions $v(t)$ from (\ref{vtfinal}) and $\chi_0(t)$ from (\ref{chi0fin}) in terms of functions with unit area, which can be mapped onto the normalized distributions $\delta_v^\eta(t-1)$ and $\delta_h^\eta(t-1)$. We obtain
\begin{equation}
\begin{aligned}
   v(t) &= v_4\,\sqrt{\frac{L}{\pi}}\,\frac{\sqrt{1+\beta}}{2+\beta}\,
    \delta_v^{1/\beta}(t-1) \,, \\
   \chi_0(t) &= \sqrt{\frac{L}{\pi}}\,\frac{\sqrt{1+\beta}}{2+\beta}\,
    \left[ 1 + \frac{\beta\,x_0^2}{4(1+\beta)(2+\beta)(4+\beta)} + \dots \right]
    \delta_h^{1/\beta}(t-1) \,, 
\end{aligned}
\end{equation}
with
\begin{equation}\label{profilefuns}
\begin{aligned}
   \delta_v^{1/\beta}(t-1) &= (2+\beta)\,t^{1+\beta} \,, \\ 
   \delta_h^{1/\beta}(t-1) &= (2+\beta)\,t^{1+\beta} 
    \left[ 1 - \frac{x_0^2}{4(1+\beta)} \left( t^2 - \frac{2+\beta}{4+\beta} \right) 
    + \dots \right] \,.
\end{aligned}
\end{equation}
Here $1/\beta$ plays the role of the regulator $\eta$ in (\ref{LYbrane}). Using the quark bilinear terms as a reference, the corresponding matching relations between the two Yukawa matrices read
\begin{equation}\label{b33}
   \bm{Y}_q \equiv \frac{k}{2}\,\bm{Y}_q^{5D} 
   = \frac{\sqrt{k\,(1+\beta)}}{2+\beta}\,\bm{Y}_{q,\rm bulk}^{5D} \,.
\end{equation}
The quantities on the left-hand side of the equation are the dimensionless Yukawa matrices introduced in (\ref{Lhqq}), whose elements are assumed to be random numbers bounded in magnitude by $y_*$. If one used the $hq\bar q$ couplings instead, the above relation would receive corrections of ${\cal O}(x_0^2)$.

\subsubsection*{Limit of a narrow bulk-Higgs field}

We are finally in a position to study the limit $\beta\gg 1$, in which the profile functions in (\ref{profilefuns}) become strongly localized near the IR brane. We can then identify $1/\beta$ with the width of the Higgs profile, which plays the role of the regulator $\eta$ in our brane-Higgs scenario. The Yukawa matrices of the bulk-Higgs model must then be identified with $\bm{Y}_q\leftrightarrow\sqrt{k/\beta}\,\bm{Y}_{q,\rm bulk}^{5D}\approx(k/\sqrt{\mu})\,\bm{Y}_{q,\rm bulk}^{5D}$. It would be inappropriate to conclude from this relation that the Yukawa matrices $\bm{Y}_q$ vanish in the limit $\beta\to\infty$. Rather, one should consider the dimensionless Yukawa couplings as fixed quantities, which are related to the observed masses and mixing angles of the SM quarks by means of relations derived in \cite{Casagrande:2008hr}. It then follows that the dimensionful Yukawa matrices $\bm{Y}_{q,\rm bulk}^{5D}$ must scale with $\sqrt{\beta/k}\approx\sqrt{\mu}/k$ (see also the discussion in~\cite{Azatov:2009na}).

Finally, since $t$ is pushed near 1, we conclude from (\ref{profilefuns}) that
\begin{equation}
   \frac{\delta_h^{1/\beta}(t-1)}{\delta_v^{1/\beta}(t-1)} 
   = 1 + {\cal O}\bigg( \frac{m_h^2}{\beta^2 M_{\rm KK}^2} \bigg) \,,
\end{equation}
as was claimed near the beginning of Section~\ref{sec:prop}.

Taking the limit of very large $\beta$ is not particularly natural, since $\beta=\sqrt{4+\mu^2/k^2}$ is naturally of ${\cal O}(1)$. For large $\beta$, we have the double hierarchy
\begin{equation}
   \frac{1}{r}\ll k\ll\mu\approx\frac{M_{\rm IR}}{2} \,, \qquad \mbox{or} \quad
   \frac{10}{r}\sim k\sim\frac{\mu}{\beta} \,.
\end{equation}
Large $\beta$ can be achieved by taking $k$ significantly smaller than the Planck scale (and $1/r$ yet smaller by an order of magnitude), or by assuming that $\mu$ and $M_{\rm IR}$ are significantly larger than $M_{\rm Pl}$. The first possibility appears more plausible. Note that for large $\beta$ relation (\ref{lam4rel}) implies that $\lambda_4\approx\lambda_{\rm IR}\,\mu^2$, indicating that increasing $\beta$ by lowering the curvature parameter $k$ does not affect the relation between $\lambda_4$ and $\lambda_{\rm IR}$ in a significant way.

\section{Case of two different Yukawa matrices}
\label{app:Y1Y2}
\renewcommand{\theequation}{C.\arabic{equation}}
\setcounter{equation}{0}

We briefly discuss the generalization of our results to the case where the two Yukawa couplings in~(\ref{gdef}), involving products of $Z_2$-even and $Z_2$-odd fermion profiles, are associated with different Yukawa matrices, such that
\begin{equation}\label{C.1}
   g_{mn}^u = \frac{\sqrt 2\pi}{L\epsilon} \int_\epsilon^1\!dt\,\delta^\eta(t-1)\,
    \Big[ a_m^{(U)\dagger}\,\bm{C}_m^{(Q)}(t)\,\bm{Y}_u^C\,\bm{C}_n^{(u)}(t)\,a_n^{(u)}
    + a_m^{(u)\dagger}\,\bm{S}_m^{(u)}(t)\,\bm{Y}_u^{S\,\dagger} \bm{S}_n^{(Q)}(t)\,a_n^{(U)}
    \Big] \,.
\end{equation}
At the level of the gluon fusion amplitude (\ref{ampl}), the above modification is implemented by the substitution 
\begin{equation}
   \frac{1}{\sqrt2}\,
    \bigg( \begin{array}{cc} 0 & \bm{Y}_q \\ 
           \bm{Y}_q^\dagger & 0 \end{array} \bigg) 
   \to \frac{1}{\sqrt2}\,
    \bigg( \begin{array}{cc} 0 & \bm{Y}_q^C \\ 
           \bm{Y}_q^{S\,\dagger} & 0 \end{array} \bigg) P_R
    + \frac{1}{\sqrt2}\,
    \bigg( \begin{array}{cc} 0 & \bm{Y}_q^S \\ 
           \bm{Y}_q^{C\,\dagger} & 0 \end{array} \bigg) P_L \,.    
\end{equation}
This generalization is only allowed if the Higgs boson is localized on the IR brane. For a bulk-Higgs field, it is forbidden by 5D Lorentz invariance, since $i\gamma_5$ is one of the 5D Dirac matrices $\gamma^a$. 

The equations of motion (\ref{coupledeqs}) for the propagator functions must now be generalized to 
\begin{equation}\label{newcoupled}
\begin{aligned}
   p^2 \bm{\Delta}_{LL}^q(t,t';-p^2)
    - M_{\rm KK} \left( \frac{\partial}{\partial t} + {\cal M}_q(t) \right) 
    \bm{\Delta}_{RL}^q(t,t';-p^2)
   &= \delta(t-t') \,, \\
   \bm{\Delta}_{RL}^q(t,t';-p^2)
    - M_{\rm KK} \left( - \frac{\partial}{\partial t} + {\cal M}_q^\dagger(t) \right) 
    \bm{\Delta}_{LL}^q(t,t';-p^2)
   &= 0 \,,
\end{aligned}
\end{equation}
where
\begin{equation}
   {\cal M}_q(t) 
   = \frac{1}{t}\,\bigg(\! \begin{array}{cc} \bm{c}_Q & \,\,\,0 \\ 
                        0 & -\bm{c}_q \end{array} \!\bigg) 
    + \frac{v}{\sqrt 2 M_{\rm KK}}\,\delta_v^\eta(t-1)\,
    \bigg( \begin{array}{cc} 0 & \bm{Y}_q^C \\ 
           \bm{Y}_q^{S\,\dagger} & 0 \end{array} \bigg)
\end{equation}
replaces the generalized mass matrix in (\ref{Mqdef}). The coupled set of first-order differential equations in (\ref{newcoupled}) can be combined to yield the second-order equation
\begin{equation}\label{new2nd}
   \left[ \frac{\partial^2}{\partial t^2} - {\cal M}_q(t)\,{\cal M}_q^\dagger(t)
    - \frac{d{\cal M}_q^\dagger(t)}{dt} + 
    \!\left( {\cal M}_q(t) - {\cal M}_q^\dagger(t) \right)\! \frac{\partial}{\partial t}
    - \hat p_E^2 \right] \bm{\Delta}_{LL}^q(t,t';-p^2)
   = \frac{1}{M_{\rm KK}^2}\,\delta(t-t') \,.
\end{equation}
In the bulk region $t<1-\eta$, where the profile $\delta_v^\eta(t-1)$ of the Higgs vev vanishes and the mass matrix is hermitian, this equation reduces to the original equation (\ref{2ndorder}). However, its structure becomes much more complicated for $t>1-\eta$. We have not succeeded to derive the general solution in that region. 

In the case of infinitesimal $\eta$ (at fixed $p^2$), however, it is consistent to only keep the terms in (\ref{newcoupled}) that are enhanced by $1/\eta$ for $1-\eta<t<1$. Taking $t'<1-\eta$ in the bulk region, squaring the resulting differential operators, and adopting the Higgs profile given in (\ref{simplebox}), we thus need to solve
\begin{equation}
\begin{aligned}
   \left[ \frac{\partial^2}{\partial t^2} - \frac{v^2}{2M_{\rm KK}^2\eta^2} 
    \bigg( \begin{array}{cc} \bm{Y}_q^C \bm{Y}_q^{S\dagger} & 0 \\ 
           0 & \bm{Y}_q^{S\,\dagger} \bm{Y}_q^C \end{array} \bigg) \right] 
    \bm{\Delta}_{RL}^q(t,t';-p^2) &= 0 + \dots \,, \\
   \left[ \frac{\partial^2}{\partial t^2} - \frac{v^2}{2M_{\rm KK}^2\eta^2} 
    \bigg( \begin{array}{cc} \bm{Y}_q^S \bm{Y}_q^{C\dagger} & 0 \\ 
           0 & \bm{Y}_q^{C\,\dagger} \bm{Y}_q^S \end{array} \bigg) \right] 
    \bm{\Delta}_{LL}^q(t,t';-p^2) &= 0 + \dots \,,
\end{aligned}
\end{equation}
where the dots denote subleading terms. The solutions to these equations involve hyperbolic trigonometric functions, whose arguments contain the matrices
\begin{equation}
   \bm{X}_q = \frac{v}{\sqrt2 M_{\rm KK}} \sqrt{\bm{Y}_q^C\bm{Y}_q^{S\,\dagger}} \,, \qquad
   \bar{\bm{X}}_q = \frac{v}{\sqrt2 M_{\rm KK}} \sqrt{\bm{Y}_q^{S\,\dagger}\bm{Y}_q^C}
\end{equation}
and their hermitian conjugates. It is then not difficult to show that, in the limit $\eta\to 0$, the boundary conditions given in (\ref{eq56}) still hold, provided we use $\bm{X}_q$ as defined here instead of the original definition in (\ref{Xqdef}), and $\tilde{\bm{Y}}_q$ as shown in (\ref{newdefs}) instead of the original definition in (\ref{Ytildef}). Solving the bulk equations of motion for the propagator functions with these boundary conditions, we recover our previous solutions with the substitutions just described.

\section{Perturbativity bounds on the Yukawa couplings}
\label{app:perturbativity}
\renewcommand{\theequation}{D.\arabic{equation}}
\setcounter{equation}{0}

One can impose an upper bound on the size of the 5D Yukawa couplings by requiring that the Yukawa interactions remain perturbative up to the cutoff of the RS model under consideration (see e.g.\ \cite{Csaki:2008zd,Cacciapaglia:2006mz}). In 5D language, NDA estimates of the one-loop corrections to the Yukawa interactions in a model with brane-localized Higgs sector hint at a quadratic divergence. One thus obtains a condition of the form \cite{Ponton:2012bi}
\begin{equation}\label{ybound}
   c_g \left( \frac{|Y_q^{\rm 5D}|}{\sqrt2} \right)^2 \frac{l_4}{l_5^2}\,M_{\rm Pl}^2
   = \frac{c_g |Y_q|^2}{18\pi^4} \left( \frac{\Lambda_{\rm TeV}}{M_{\rm KK}} \right)^2
   \stackrel{!}{<} 1 \,,
\end{equation}
where $|Y_q^{\rm 5D}|=2|Y_q|/k$ sets the scale of the dimensionful 5D Yukawa couplings, $l_4=16\pi^2$ and $l_5=24\pi^3$ are appropriate 4D and 5D phase-space factors, $M_{\rm Pl}$ is the physical UV cutoff of the RS model, and in the last step we have used that $\Lambda_{\rm TeV}=M_{\rm Pl}\epsilon$ and $M_{\rm KK}=k\epsilon$. The coefficient $c_g$ accounts for the multiplicity of fermion generations and is chosen such that $c_g=1$ for the case of one generation. In general, for $N_g$ fermion generations, its value $c_g=2N_g-1$ is determined by the relation
\begin{equation}
   \big\langle \! \left( \bm{Y}_q\bm{Y}_q^\dagger\bm{Y}_q \right)_{ij} \big\rangle
   = (2N_g-1)\,|Y_q|^2 \left( \bm{Y}_q \right)_{ij} ,
\end{equation} 
which holds in the sense of an expectation value for a large sample of anarchic, complex random matrices. It is instructive to reproduce condition (\ref{ybound}) by employing a 4D picture in terms of KK modes, where the quadratic behavior on the cutoff arises from a double sum over the $N_{\rm KK}$ levels of states with masses below the cutoff $\Lambda_{\rm TeV}$ \cite{Csaki:2008zd}. This leads to the condition 
\begin{equation}\label{ybound2}
   c_g \left( \frac{|Y_q|}{\sqrt2} \right)^2 \frac{1}{l_4}\,N_{\rm KK}^2
   \approx \frac{c_g |Y_q|^2}{32\pi^4} \left( \frac{\Lambda_{\rm TeV}}{M_{\rm KK}} \right)^2
   \stackrel{!}{<} 1 \,,
\end{equation}
where we have used that the masses of the KK modes are determined by the zeroes of some Bessel functions, such that the states in the $N^{th}$ KK level have masses approximately given by $N\pi M_{\rm KK}$ (valid for large $N$), and hence $N_{\rm KK}\approx\Lambda_{\rm TeV}/(\pi M_{\rm KK})$. The two estimates in (\ref{ybound}) and (\ref{ybound2}) differ by a harmless ${\cal O}(1)$ factor but are parametrically equivalent (including factors of $\pi$) as NDA estimates. Employing (\ref{eq87}) and solving for $y_*$, we find the condition $y_*<y_{\rm max}$, with the upper bounds $y_{\rm max}=(6\pi^2/\sqrt{c_g})\,M_{\rm KK}/\Lambda_{\rm TeV}$ derived from (\ref{ybound}) and $y_{\rm max}=(8\pi^2/\sqrt{c_g})\,M_{\rm KK}/\Lambda_{\rm TeV}$ derived from (\ref{ybound2}). Assuming as before that $\Lambda_{\rm TeV}\sim 10 M_{\rm KK}$, one obtains $y_{\rm max}\approx 2.6$ in the first case and $y_{\rm max}\approx 3.5$ in the second. These estimates are somewhat more refined than those presented elsewhere in the literature (because we include the dependence on $N_g$), but they are compatible with the conventional choice $y_{\rm max}=3$ adopted in most phenomenological analyses of RS models. Using the more stringent upper bound derived from (\ref{ybound}), and assuming that the Yukawa couplings are not much smaller than the values given by the perturbativity bound, we can rewrite condition (\ref{ceffesti}) in the form
\begin{equation}
   c_{\rm eff} \ll \frac{3\pi^2}{2}\,\frac{N_g^2}{2N_g-1} \approx 27 \,,
\end{equation}
which is now independent of the value of the ratio $M_{\rm KK}/\Lambda_{\rm TeV}$. This argument shows that, even if the UV completion of the RS model is strongly coupled and $c_{\rm eff}={\cal O}(1)$, the contributions from higher-dimensional operators are expected to be numerically much smaller than the KK loop effects, provided that the Yukawa couplings are not much smaller than the perturbativity bounds.

Repeating the same argument for the case of an RS model in which the Higgs sector lives in the bulk, we obtain from relation (\ref{b33}) in Appendix~\ref{app:bulkHiggs} the condition
\begin{equation}
   c_g \left( \frac{|Y_q^{\rm 5D}|}{\sqrt2} \right)^2 \frac{1}{l_5}\,M_{\rm Pl}
   = \frac{c_g |Y_q|^2}{48\pi^3}\,\frac{(2+\beta)^2}{1+\beta}\,
    \frac{\Lambda_{\rm TeV}}{M_{\rm KK}} 
   \stackrel{!}{<} 1 \,,   
\end{equation}
which translates into $y_*<y_{\rm max}$ with $y_{\rm max}=\sqrt{96\pi^3/c_g}\,\frac{\sqrt{1+\beta}}{2+\beta}\,\sqrt{M_{\rm KK}/\Lambda_{\rm TeV}}$. Here $\beta\sim 1/\eta$ is related to the width of the Higgs profile. Note that in the bulk-Higgs case the suppression in the ratio $M_{\rm KK}/\Lambda_{\rm TeV}$ is parametrically weaker than in the case of a brane-localized Higgs field. In practice, with $\Lambda_{\rm TeV}\sim 10 M_{\rm KK}$, this effect is not too important, however. Even for a very broad bulk Higgs with $\beta\to 0$, we obtain $y_{\rm max}\approx 3.9$, which is of the same order as the bound in the brane-Higgs case. In the present work we are only interested in a narrow bulk-Higgs scenario, for which $\eta=1/\beta\ll 1$ is a small parameter (see Table~\ref{tab:models}). We can then simplify $y_{\rm max}=\sqrt{96\pi^3/c_g}\,\sqrt{\eta M_{\rm KK}/\Lambda_{\rm TeV}}\approx 7.7\sqrt\eta$. This formula can only be trusted as long as $\eta\gtrsim M_{\rm KK}/\Lambda_{\rm TeV}\approx 0.1$. For smaller $\eta$, the relevant bound is that found in the brane-Higgs case, $y_{\rm max}\approx 2.6$. From a practical point of view, there is no significant difference between the two bounds.

\end{appendix}

%\newpage

\end{document}